\documentclass{aa}  
\usepackage{graphicx}
\usepackage{txfonts}
\usepackage[colorlinks=true,citecolor=blue,linkcolor=blue,urlcolor=cyan]{hyperref}
\usepackage{pdflscape}
\newcommand{\nodata}{}

\begin{document} 


\title{Structure and kinematics of the Taurus star-forming region from Gaia-DR2 and VLBI astrometry}

\author{P.~A.~B.~Galli \inst{1}
\and
L.~Loinard \inst{2,3}
\and
H.~Bouy \inst{1}
\and
L.~M.~Sarro \inst{4}
\and 
G.~N.~Ortiz-Le{\'o}n \inst{5}
\and
S.~A.~Dzib \inst{5}
\and
J.~Olivares \inst{1}
\and
M.~Heyer \inst{6}
\and
J.~Hernandez \inst{7}
\and
C.~Rom\'an-Z\'u\~niga \inst{7}
\and
M.~Kounkel\inst{8}
\and
K.~Covey \inst{8}
}

\institute{
Laboratoire d'Astrophysique de Bordeaux, Univ. Bordeaux, CNRS, B18N, All\'ee Geoffroy Saint-Hilaire, 33615 Pessac, France
\and
Instituto de Radioastronom\'ia y Astrof\'isica, Universidad Nacional Aut\'onoma de M\'exico, Apartado Postal  3-72, Morelia 58089, M\'exico
\and
Instituto de Astronom\'ia, Universidad Nacional Aut\'onoma de M\'exico, Apartado Postal 70-264, 04510 Ciudad de M\'exico, M\'exico
\and
Depto. de Inteligencia Artificial , UNED, Juan del Rosal, 16, 28040 Madrid, Spain
\and
Max Planck Institut f{\"u}r Radioastronomie, Auf dem H{\"u}gel 69, D-53121, Bonn, Germany
\and
Department of Astronomy, University of Massachusetts, Amherst, MA 01003, USA
\and
Instituto de Astronom\'ia, Universidad Nacional Aut\'onoma de M\'exico, Unidad Acad\'emica en Ensenada, Ensenada 22860, M\'exico
\and
Department of Physics and Astronomy, Western Washington University, 516 High St, Bellingham, WA 98225, USA.
}

\date{Received  ; accepted  }

 \abstract
{}
{We take advantage of the second data release of the Gaia space mission and the state-of-the-art astrometry delivered from very long baseline interferometry observations to revisit the structure and kinematics of the nearby Taurus star-forming region.}
{We apply a hierarchical clustering algorithm for partitioning the stars in our sample into groups (i.e., clusters) that are associated with the various molecular clouds of the complex, and derive the distance and spatial velocity of individual stars and their corresponding molecular clouds.}
{We show that the  molecular clouds are located at different distances and confirm the existence of important depth effects in this region reported in previous studies. For example, we find that the L~1495 molecular cloud is located at $d=129.9^{+0.4}_{-0.3}$~pc, while the filamentary structure connected to it (in the plane of the sky) is at $d=160.0^{+1.2}_{-1.2}$~pc. We report B~215 and L~1558 as the closest ($d=128.5^{+1.6}_{-1.6}$~pc) and most remote ($d=198.1^{+2.5}_{-2.5}$~pc) substructures of the complex, respectively. The median inter-cloud distance is 25~pc and the relative motion of the  subgroups is on the order of a few km/s. We find no clear evidence for expansion (or contraction) of the Taurus complex, but signs of the potential effects of a global rotation. Finally, we compare the radial velocity of the stars with the velocity of the underlying $^{13}$CO molecular gas and report a mean difference of $0.04\pm0.12$~km/s (with r.m.s. of 0.63~km/s) confirming that the stars and the gas are tightly coupled. }
{}


\maketitle

\section{Introduction}\label{section1}

The Taurus-Auriga star-forming region (or simply Taurus) is one of the most intensively studied regions of low-mass star formation and an ideal laboratory for  observing young stellar objects (YSOs) from the most embedded sources at the early stages of evolution (i.e., protostars) to disk-free stars that are actively forming planets   \citep[see, e.g.,][]{Kenyon2008}. Previous studies suggest that Taurus hosts a few hundred  YSOs spread over a large area on the sky of about $15\times15$~deg  \citep{Esplin2014,Esplin2017}. The sky-projected spatial distribution shows that the stars are not randomly distributed but clustered in small groups  and overdense structures in and around the different star-forming clouds and filaments of the region \citep{Gomez1993,Joncour2017,Joncour2018}. The morphology and kinematics of these gaseous clouds and filaments have been clearly characterized in recent years based on CO surveys and extinction maps \citep[see, e.g.,][]{Ungerechts1987,Cambresy1999,Dame2001,Dobashi2005,Goldsmith2008}, and increasing progress is being made to constrain the three-dimensional structure and stellar kinematics of the individual clouds. However, until recently many studies have been hampered by the lack of accurate data for a significant number of stars, in particular stellar distances and spatial velocities, which could provide us with valuable information about the star formation history in this region. 

Distances to individual stars are, in general, derived from trigonometric parallaxes; until recently   there were very few parallaxes    for Taurus stars. \citet{Bertout1999} used the trigonometric parallaxes of 17~stars from the \textit{Hipparcos} catalog \citep{ESA1997} and estimated the distances to three groups in this sample, $125^{+21}_{-16}$, $140^{+16}_{-13}$, and $168^{+42}_{-28}$~pc, which are roughly associated with the central, northern, and southern clouds of the complex, respectively. The situation did not improve significantly with the first data release of the \textit{Gaia} space mission \citep[Gaia-DR1,][]{GaiaDR1}. The \textit{Tycho-Gaia Astrometric Solution} \citep[TGAS,][]{Lindegren2016} catalog provided trigonometric parallaxes for only 19 stars in Taurus that are obviously more precise than the \textit{Hipparcos} results for the same stars, but still represent a small fraction of the sample of known members. This sample is restricted to the brightest stars (i.e., $G< 12$~mag) and the parallaxes were nevertheless affected by systematic errors on the order of 0.3~mas \citep[see][]{Lindegren2016}.

A  major effort to determine the distance to individual stars in the Taurus region was successfully undertaken using very long baseline interferometry \citep[VLBI,][]{Lestrade1999,Loinard2007,Torres2007,Torres2009,Torres2012}. In recent years the Gould's Belt Distances Survey \citep[GOBELINS,][]{Loinard2011} has targeted a number of YSOs in nearby star-forming regions to deliver state-of-the-art trigonometric parallaxes and proper motions \citep[see][]{OrtizLeon2018b,OrtizLeon2018a,Kounkel2017,OrtizLeon2018}. In one paper in this series, \citet{Galli2018} measured the trigonometric parallaxes of 18 stars in Taurus with  precision ranging  from 0.3$\%$ to 5$\%$. The resulting distances suggest that the various molecular clouds of the complex are located at different distances and reveal the existence of significant depth effects in this region. For example, the Lynds~1531 and 1536 molecular clouds (hereafter L~1531 and L~1536) were reported to be the closest ($d=126.6\pm1.7$~pc) and most remote   ($d=162.7\pm0.8$~pc) structures of the complex, respectively. The VLBI astrometry combined with published radial velocities yielded a one-dimensional velocity dispersion of about 2-3~km/s among the various clouds in Taurus. This  is significantly lower than the value of 6~km/s used by \citet{Bertout2006} to derive kinematic distances based on the convergent point method. Such a discrepancy could arise, for example  from the   internal motions within the complex, indicating that more study is clearly warranted in this regard. 

The small number of sources with complete data  in the sample (proper motion, parallax, and radial velocity)  compared to the number of known members prevented \citet{Galli2018} from investigating in more detail the three-dimensional structure and kinematics of the various subgroups. In this context, the recently published second data release of the \textit{Gaia} space mission \citep[Gaia-DR2,][]{GaiaDR2} offers a unique opportunity to revisit the previous analysis with a much more significant number of stars and the same level of astrometric precision obtained from VLBI observations. For example, Gaia-DR2 increases the number of Taurus stars with available astrometry by a factor of more than 20  compared to its predecessor Gaia-DR1 including the faintest members at $G\simeq20$~mag and having smaller systematic errors on the trigonometric parallaxes  of about 0.1~mas on global scales \citep{Luri2018}.

In a recent paper \citet{Luhman2018} used Gaia-DR2 data to refine the census of Taurus stars, to identify new candidates with similar properties of known members, and to determine the shape of the initial mass function (IMF). The revised sample of members shows that the older population of stars ($>10$~Myr) which was proposed to be associated with this region in other studies \citep[see, e.g.,][]{Kraus2017,Zhang2018} has no physical relationship with the Taurus molecular clouds, and the Taurus IMF resembles other star-forming regions (e.g., IC~348 and the Orion Nebula Cluster). We  incorporated this updated census of stars in our analysis and  we present here   our discussion   of the structure and kinematics of the region.

This paper is organized as follows. In Section~\ref{section2} we describe the sample of stars used in this study for our analysis and in Section~\ref{section3} we compare the VLBI and Gaia-DR2 astrometry for the stars in common between the two projects in the Taurus region. In Section~\ref{section4} we present our methodology based on hierarchical clustering for partitioning the stars in our sample into different groups with similar properties, for rejecting outliers in the sample, and for defining the subsamples of stars that are associated with the various molecular clouds of the Taurus complex. In Section~\ref{section5} we present our results for   the distance and spatial velocity of individual stars and subgroups derived from Bayesian inference using the most precise astrometric and spectroscopic data available to date and   the existence of internal motions, expansion, and rotation effects in the complex, and we  compare the stellar velocities with the kinematics of the underlying gaseous clouds. Finally, we summarize our results and conclusions in Section~\ref{section6}.

\section{Sample}\label{section2}

To construct our initial sample of Taurus stars, we begin by compiling known YSOs and new candidates associated with this region that have been previously identified in the literature. Several studies in the literature have proposed different lists of Taurus stars \citep[see, e.g.,][]{Joncour2017,Kraus2017,Zhang2018,Luhman2018}, but the recent study performed by \citet{Luhman2018} shows that the samples of stars proposed by \citet{Kraus2017} and \citet{Zhang2018} are older (> 10~Myr) and show kinematic properties that are inconsistent with membership in Taurus. We  therefore restricted our sample of stars to the lists given by \citet{Joncour2017} and \citet{Luhman2018}. We combined the sample of 338 stars from \citet{Joncour2017} with the lists of known members (438 stars) and new candidates (62~stars) given by \citet{Luhman2018}. The resulting sample consists of 519 stars after removing the sources in common between the two surveys. Multiple systems are counted as one single source in our sample unless they were resolved in these studies or by the Gaia satellite (as described below).

We proceeded as follows to access the astrometric measurements in Gaia-DR2 for our targets and avoid erroneous cross-identifications. Gaia-DR2 provides cross-matched tables with a number of external catalogs. First, we use the unique source identifier from the 2MASS catalog \citep{Cutri2003} given in the original tables used to construct our sample and cross-match our list of source identifiers with  the \texttt{TMASS\_BEST\_NEIGHBOUR} table provided by the Gaia archive\footnote{see \href{http://gea.esac.esa.int/archive/}{http://gea.esac.esa.int/archive/}}. This procedure returns the unique Gaia-DR2 source identifier that corresponds to our target, the number of sources in the 2MASS catalog that match the Gaia source, and the number of Gaia sources that have the same source as best-neighbor.  We find a direct one-to-one relationship for most sources in our sample, which confirms that they have been correctly identified. We note that three sources (2MASS~J04210934+2750368, 2MASS~J04400174+2556292, and 2MASS~J05122759+2253492) which were not resolved in previous surveys have more than one counterpart in Gaia-DR2. In such cases, we retain the two components of the system and count them as independent stars. Then, we use the resulting list of Gaia-DR2 identifiers to search our targets in the main catalog table (\texttt{GAIA\_SOURCE}) and retrieve the astrometric measurements that will be used in the forthcoming analysis.  We repeated this procedure for the 492 stars with known 2MASS identifiers in our sample and searched the remaining sources in Gaia-DR2 using their stellar positions and a search radius of 1\arcsec. Doing so, we found proper motions and trigonometric parallaxes for 411~stars from our initial sample.  

Radial velocities in Gaia-DR2 are available for only 34 stars in our sample, so we searched the CDS/SIMBAD databases \citep{Wenger2000} to access more radial velocity measurements. Our search in the literature, which was as exhaustive as possible, was based on \citet{Wilson1953}, \citet{Hartmann1986}, \citet{Hartmann1987}, \citet{HBC}, \citet{Reipurth1990}, \citet{Duflot1995}, \citet{Mathieu1997}, \citet{Wichmann2000}, \citet{White2003}, \citet{Muzerolle2003}, \citet{Gontcharov2006}, \citet{Torres2006}, \citet{Kharchenko2007}, \citet{Scelsi2008}, \citet{Nguyen2012}, and \citet{Kraus2017}. In addition, we also used the more recent measurements collected with the Apache Point Observatory Galactic Evolution Experiment (APOGEE) spectrograph \citep{Kounkel2019}. In the case of multiple radial velocity measurements in the literature we took the most precise result as our final estimate. By combining these external sources with Gaia-DR2 we found radial velocities for a total of 248 stars. 

Table~\ref{tab1} lists the 519 stars in our initial sample with the data collected from the literature and the membership status of each star as derived from our forthcoming analysis (see Sect.~\ref{section4}). 

\begin{landscape}
\renewcommand\thetable{1} 
\begin{table}
\caption{Identifiers, positions (epoch=2000), proper motions, trigonometric parallaxes, and radial velocities for the 519 stars in our initial sample. The columns ``Cluster'' and ``Member'' provide the cluster membership from the analysis presented in Sect.~\ref{section4.2} (zeros denote the outliers in the sample) and the final membership status (0 = probable outlier, 1 = confirmed member) of each star in the corresponding cluster (see Sect.~\ref{section4.3}). The last three columns indicate whether the star is included in one of the original tables from the literature used to construct our initial sample of stars (1 = included, 0 = not included). They refer to the samples of \citet{Joncour2017} and Tables 1 (members) and 6 (candidate members) of \citet{Luhman2018}, respectively. (This table will be available in its entirety in machine-readable form.) }\label{tab1}
\scriptsize{
\begin{tabular}{lllccccccccccccc}
\hline\hline
2MASS Identifier&Gaia-DR2 Identifier&Other Identifier& $\alpha$&$\delta$&$\mu_{\alpha}\cos\delta$ & $\mu_{\delta}$&$\varpi$&Source&$V_{r}$&Ref.&Cluster&Member&\multicolumn{3}{c}{Table}\\
&&& (h:m:s) &($^{\circ}$ $^\prime$ $^\prime$$^\prime$)  & (mas/yr) & (mas/yr)&(mas)&&(km/s)&&&&\multicolumn{3}{c}{(literature)}\\
\hline\hline
2MASS J04034930+2610520& Gaia DR2 162535413750345856 & HBC358A+B+C & 04 03 49.32& 26 10 52.0& \nodata& \nodata& \nodata& \nodata& \nodata&\nodata& 0 & 0 & 1 & 0 & 0 \\
2MASS J04034997+2620382& Gaia DR2 162541942104406784 & XEST06-006 & 04 03 49.99& 26 20 38.4& $ 14.212 \pm 0.311 $& $ -19.389 \pm 0.221 $& $ 6.866 \pm 0.208 $& GaiaDR2 & \nodata&\nodata& 0 & 0 & 1 & 1 & 0 \\
2MASS J04035084+2610531& Gaia DR2 162535345034688768 & HBC359 & 04 03 50.83& 26 10 53.0& $ 18.855 \pm 0.159 $& $ -29.927 \pm 0.123 $& $ 7.781 \pm 0.129 $& GaiaDR2 & $ 14.2 \pm 2.0 $&4& 0 & 0 & 1 & 0 & 0 \\
2MASS J04043936+2158186& Gaia DR2 53092775104124288 & HBC360 & 04 04 39.36& 21 58 18.5& $ 3.399 \pm 0.405 $& $ -15.695 \pm 0.226 $& $ 8.220 \pm 0.237 $& GaiaDR2 & \nodata&\nodata& 0 & 0 & 1 & 0 & 0 \\
2MASS J04043984+2158215& Gaia DR2 53092775104123776 & HBC361 & 04 04 39.84& 21 58 21.4& $ 5.102 \pm 0.394 $& $ -15.118 \pm 0.221 $& $ 8.183 \pm 0.252 $& GaiaDR2 & $ 16.2 \pm 2.0 $&4& 0 & 0 & 1 & 0 & 0 \\
2MASS J04044307+2618563& \nodata& IRAS04016+2610 & 04 04 43.08& 26 18 56.5& \nodata& \nodata& \nodata& \nodata& \nodata&\nodata& 0 & 0 & 1 & 1 & 0 \\
2MASS J04053087+2151106& Gaia DR2 53098994216894208 & HBC362 & 04 05 30.89& 21 51 10.8& $ 4.271 \pm 0.189 $& $ -14.224 \pm 0.126 $& $ 7.969 \pm 0.128 $& GaiaDR2 & $ 14.8 \pm 2.0 $&4& 0 & 0 & 1 & 0 & 0 \\
2MASS J04053214+2733139& Gaia DR2 163511608280866816 & \nodata& 04 05 32.16& 27 33 13.5& $ 11.426 \pm 0.196 $& $ -25.316 \pm 0.135 $& $ 7.606 \pm 0.149 $& GaiaDR2 & \nodata&\nodata& 0 & 0 & 0 & 0 & 1 \\
2MASS J04064263+2902014& Gaia DR2 164265220422661760 & \nodata& 04 06 42.65& 29 02 01.2& $ 11.027 \pm 0.320 $& $ -17.797 \pm 0.162 $& $ 6.004 \pm 0.168 $& GaiaDR2 & \nodata&\nodata& 0 & 0 & 0 & 0 & 1 \\
2MASS J04064443+2540182& Gaia DR2 162259814291151616 & \nodata& 04 06 44.45& 25 40 18.0& $ 14.009 \pm 0.261 $& $ -18.784 \pm 0.159 $& $ 6.536 \pm 0.159 $& GaiaDR2 & \nodata&\nodata& 0 & 0 & 0 & 1 & 0 \\
2MASS J04065134+2541282& Gaia DR2 162260226608005120 & V1195Tau & 04 06 51.37& 25 41 28.6& $ 12.368 \pm 0.272 $& $ -18.119 \pm 0.133 $& $ 6.415 \pm 0.138 $& GaiaDR2 & $ 16.01 \pm 0.20 $&15& 0 & 0 & 0 & 1 & 0 \\
2MASS J04065364+2540368& Gaia DR2 162259951730099968 & \nodata& 04 06 53.66& 25 40 36.6& $ 14.318 \pm 0.263 $& $ -18.857 \pm 0.157 $& $ 6.666 \pm 0.157 $& GaiaDR2 & \nodata&\nodata& 0 & 0 & 0 & 0 & 1 \\
2MASS J04080782+2807280& Gaia DR2 163926914439144320 & \nodata& 04 08 07.82& 28 07 28.2& $ -0.118 \pm 0.252 $& $ -10.310 \pm 0.157 $& $ 4.435 \pm 0.146 $& GaiaDR2 & \nodata&\nodata& 0 & 0 & 1 & 0 & 0 \\
2MASS J04105425+2501266& \nodata& \nodata& 04 10 54.25& 25 01 26.6& \nodata& \nodata& \nodata& \nodata& \nodata&\nodata& 0 & 0 & 0 & 1 & 0 \\
2MASS J04110081+2717163& \nodata& \nodata& 04 11 00.81& 27 17 16.3& \nodata& \nodata& \nodata& \nodata& \nodata&\nodata& 0 & 0 & 0 & 1 & 0 \\
2MASS J04124068+2438157& Gaia DR2 150073682105601408 & \nodata& 04 12 40.71& 24 38 15.4& $ 14.258 \pm 0.180 $& $ -18.838 \pm 0.151 $& $ 6.869 \pm 0.131 $& GaiaDR2 & \nodata&\nodata& 0 & 0 & 0 & 1 & 0 \\
2MASS J04131414+2819108& Gaia DR2 163246832135164544 & LkCa1 & 04 13 14.14& 28 19 10.9& $ 8.371 \pm 0.171 $& $ -24.403 \pm 0.137 $& $ 7.791 \pm 0.115 $& GaiaDR2 & $ 9.57 \pm 0.12 $&15& 7 & 1 & 1 & 1 & 0 \\
2MASS J04132722+2816247& Gaia DR2 163233981593016064 & Anon1 & 04 13 27.24& 28 16 25.0& $ 7.440 \pm 0.185 $& $ -23.835 \pm 0.145 $& $ 7.367 \pm 0.121 $& GaiaDR2 & $ 21.91 \pm 0.28 $&17& 7 & 1 & 1 & 1 & 0 \\
2MASS J04135328+2811233& Gaia DR2 163229544890946944 & IRAS04108+2803A & 04 13 53.28& 28 11 23.3& $ 11.323 \pm 2.530 $& $ -22.677 \pm 1.889 $& $ 6.936 \pm 1.013 $& GaiaDR2 & $ 15.32 \pm 0.27 $&17& 0 & 0 & 1 & 1 & 0 \\
2MASS J04135471+2811328& \nodata& IRAS04108+2803B & 04 13 54.72& 28 11 33.0& \nodata& \nodata& \nodata& \nodata& \nodata&\nodata& 0 & 0 & 1 & 1 & 0 \\
\hline\hline

\end{tabular}
\tablefoot{References for radial velocities: (1)~\citet{Wilson1953}, (2)~\citet{Hartmann1986}, (3)~\citet{Hartmann1987}, (4)~\citet{HBC}, (5)~\citet{Reipurth1990}, (6)~\citet{Duflot1995}, (7)~\citet{Mathieu1997}, (8)~\citet{Wichmann2000}, (9)~\citet{White2003}, (10)~\citet{Muzerolle2003}, (11)~\citet{Gontcharov2006}, (12)~\citet{Torres2006}, (13)~\citet{Kharchenko2007}, (14)~\citet{Scelsi2008}, (15)~\citet{Nguyen2012}, (16)~\citet{Kraus2017}, (17)~\citet{Kounkel2019} and (18)~\citet{GaiaDR2}. }
}
\end{table}
\end{landscape}
\clearpage

\section{Gaia-DR2 and VLBI astrometry in Taurus}\label{section3}

\begin{figure}
\begin{center}
\includegraphics[width=0.49\textwidth]{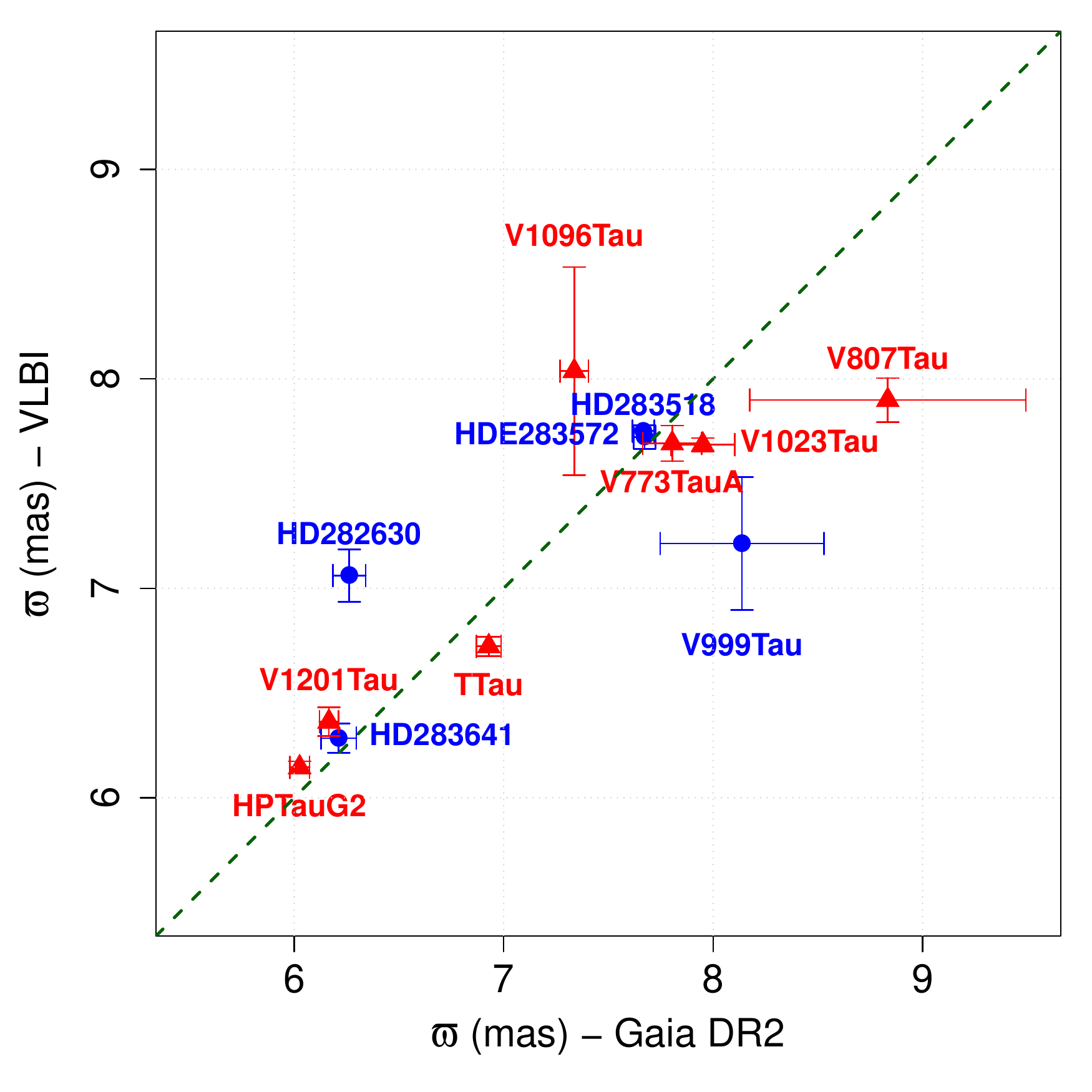}
\caption{
\label{fig1_compPLX}
Comparison of the trigonometric parallaxes obtained from the GOBELINS project \citep{Galli2018} and Gaia-DR2. Blue circles and red triangles indicate the stars that have been modeled as single and binary (multiple) sources for the VLBI astrometry, respectively. The green dashed line indicates perfect correlation between the measurements. 
}
\end{center}
\end{figure}

\begin{figure*}[!htp]
\begin{center}
\includegraphics[width=0.48\textwidth]{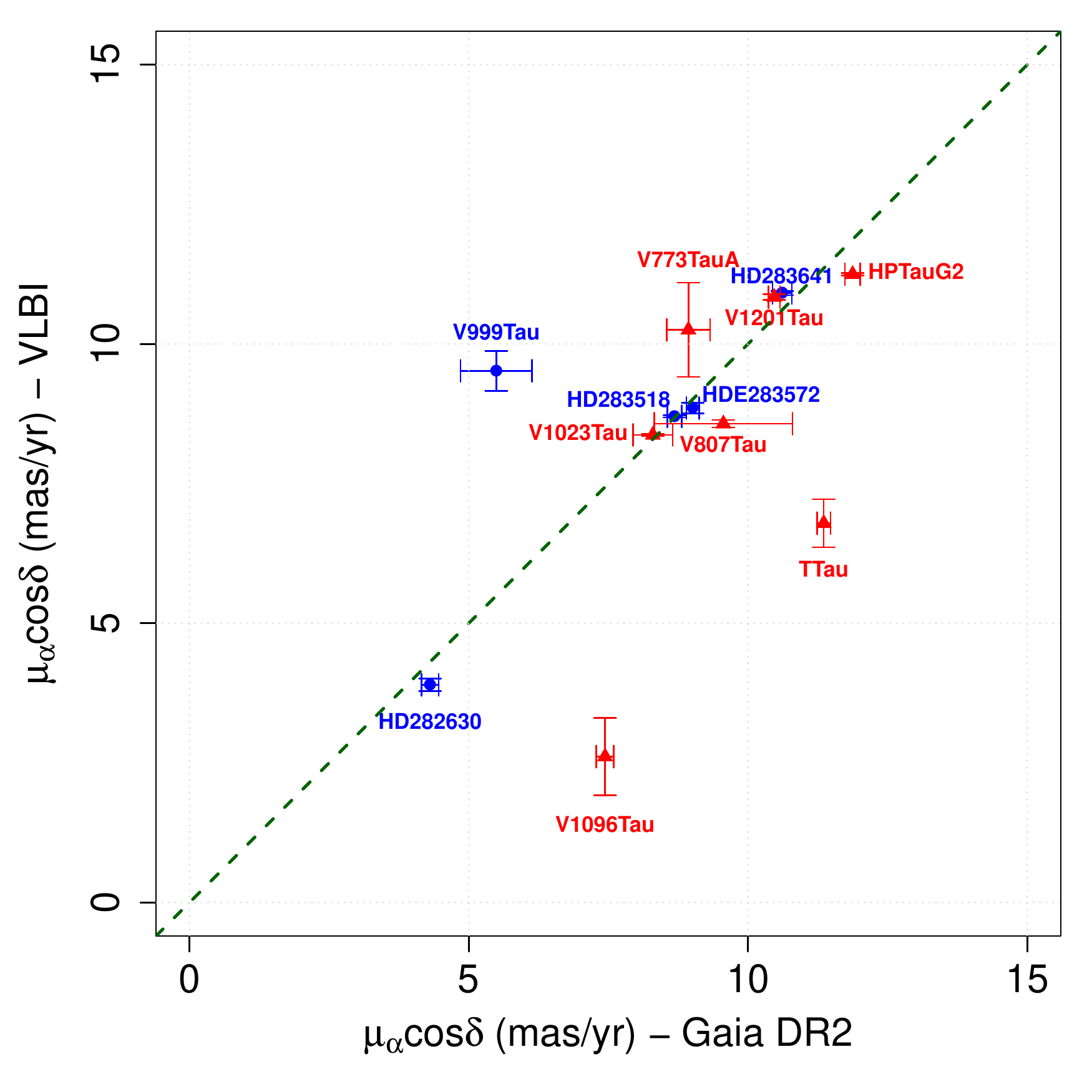}
\includegraphics[width=0.48\textwidth]{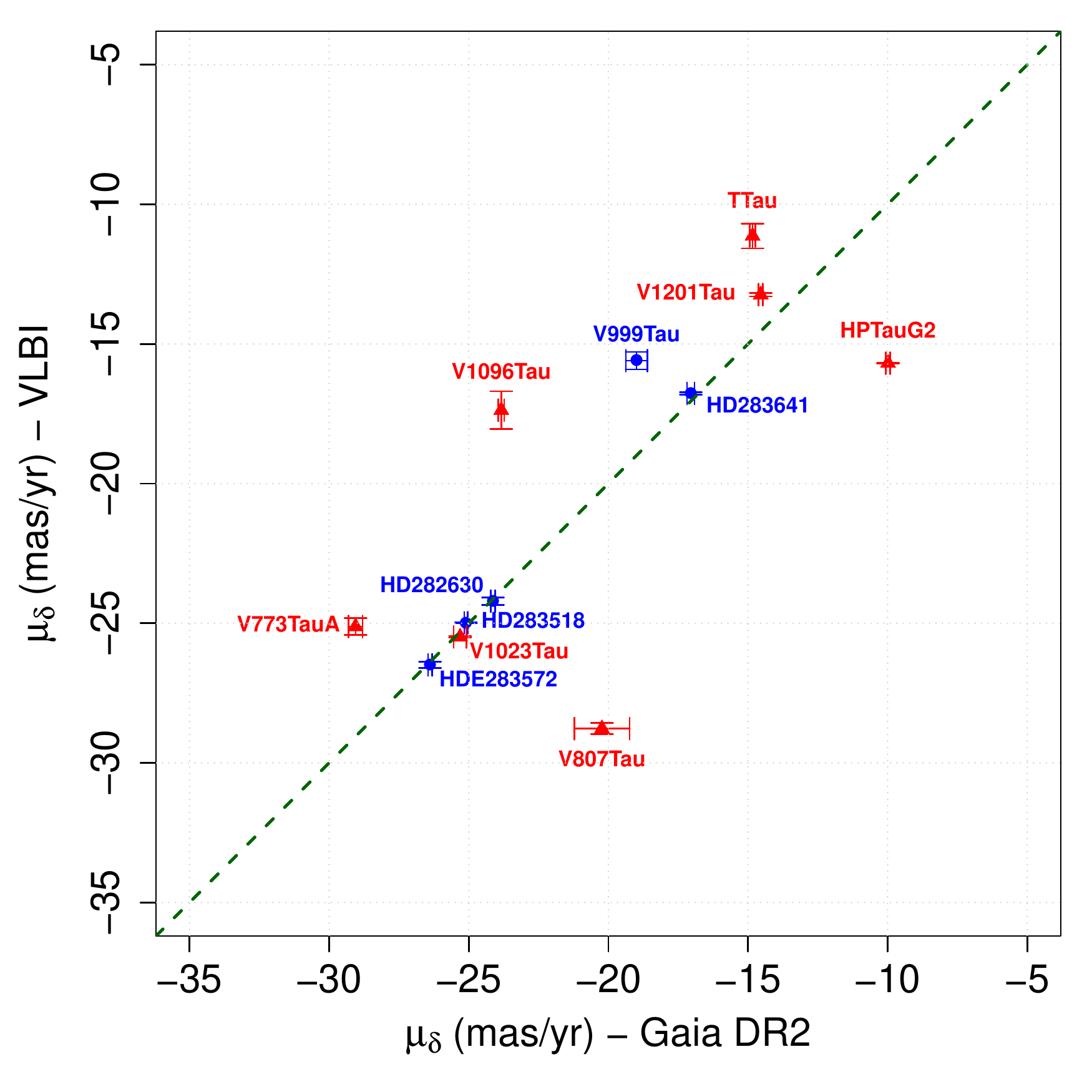}
\caption{
\label{fig2_comp_pmRA_pmDEC}
Comparison of the proper motions in right ascension \textit{(left panel)} and declination \textit{(right panel)} obtained from the GOBELINS project \citep{Galli2018} and Gaia-DR2. Blue circles and red triangles indicate the stars that have been modeled as single and binary (multiple) sources for the VLBI astrometry, respectively. The green dashed line indicates perfect correlation between the two measurements. 
}
\end{center}
\end{figure*}

In a recent study, \citet{Galli2018} derived trigonometric parallaxes and proper motions of 18 YSOs in the Taurus region based on multi-epoch VLBI radio observations as part of the GOBELINS project (see Sect.~\ref{section1}). In the following we exclude V1110~Tau from the discussion because it is more likely to be a foreground field star \citep[see discussion in Sect.~4.10 of][]{Galli2018} and we count the V1096~Tau~A-B  binary system as one source. We note that 12~YSOs from their sample are also included in Gaia-DR2. Figures~\ref{fig1_compPLX} and \ref{fig2_comp_pmRA_pmDEC} illustrate the comparison of trigonometric parallaxes and proper motions for the stars in common. The mean difference (Gaia-DR2 minus VLBI) and r.m.s. of the trigonometric parallaxes between the two projects are $0.035\pm0.152$~mas and $0.526$~mas, respectively. The same comparison in proper motions yields a mean difference of $0.455\pm0.682$~mas/yr and $-0.382\pm1.194$~mas/yr, and the r.m.s. of $2.410$~mas/yr and $4.154$~mas/yr, respectively, in right ascension and declination. 

Although these numbers provide valuable information for evaluating the consistency (or discrepancy) between VLBI and Gaia-DR2 results, two points are worth mentioning here regarding this comparison. First, the trigonometric parallaxes and proper motions derived from VLBI astrometry for two stars in common with Gaia-DR2 (V999~Tau and HD~282630) have been determined based on a small number of observational epochs. These results are therefore less precise and accurate  compared to the other stars in the VLBI sample \citep[see Sect.~4.3 of][]{Galli2018}. Second, the astrometric solutions delivered by Gaia-DR2 assume a model with uniform space motion of the stars so that non-linear  motions caused by binarity (and multiplicity) of the source have not been taken into account. \citet{Galli2018} performed a dedicated analysis of the binaries in the VLBI sample and solved for the full orbital motion of these systems with a sufficient number of detections. This explains the discrepancy observed between the two projects for such systems (see also Figures~\ref{fig1_compPLX} and \ref{fig2_comp_pmRA_pmDEC}). 

For the reasons discussed above we  decided to prioritize the trigonometric parallaxes and proper motions based on VLBI astrometry for both single stars and binaries. In the specific cases of V999~Tau, HD~282630, and the V1096~Tau~A-B binary system we prefer to use Gaia-DR2 data because of the small number of observations and the large errors produced in the VLBI solution for these specific sources \citep[see also Sects.~4.1 and 4.3 of ][]{Galli2018}. Thus, if we exclude V999~Tau, HD~282630, and V1096~Tau~A-B from the comparison, the mean difference and r.m.s. of the trigonometric parallaxes between the two projects becomes $0.111\pm0.115$~mas and 0.365~mas, respectively. The former is consistent with the systematic errors of about 0.1~mas that exist in the trigonometric parallaxes of the Gaia-DR2 catalog \citep[see][]{Lindegren2018}. One possibility to explain this discrepancy for the sample of stars under analysis is indeed the different source modeling  used in each project. For example, if we remove binaries and multiple systems from this comparison the mean difference between VLBI and Gaia-DR2 results drops to $-0.069\pm0.010$~mas (with r.m.s. of 0.086~mas).  

By combining the recently published Gaia-DR2 catalog with the state-of-the-art VLBI astrometry delivered by the GOBELINS project  in the Taurus region, we have a sample of 415 stars with measured trigonometric parallaxes and proper motions. We use the VLBI results obtained by \citet{Galli2018} for 13 stars and Gaia-DR2 data for the remaining 402 stars in this list. The astrometry reference used for each star in this study is indicated in Table~\ref{tab1}. 

One important point to mention about Gaia-DR2, which is the main source of data used in our study, is the presence of systematic errors in the catalog. They depend on the position, magnitude, and color  of each source, but they are believed to be limited on global scales to 0.1~mas in parallaxes and 0.1~mas/yr in proper motion \citep[see, e.g.,][]{Luri2018}. We added these numbers in quadrature to the random errors given in the Gaia-DR2 catalog for each star. This procedure is likely to overestimate the parallaxes and proper motion uncertainties for some stars in our sample, but the parameters that result from these observables (e.g., distance and spatial velocity) will take this effect into account when propagating the errors. We  also corrected the Gaia-DR2 parallaxes by the zero-point shift of $-0.030$~mas that is present in the published data, as reported by the Gaia collaboration \citep[see, e.g.,][]{Lindegren2018}, although the final impact of this correction in our distances is not significant due to the close proximity of the Taurus star-forming region.   

\section{Analysis}\label{section4}

One of the main objectives of the current study is to compare the properties of the various star-forming clouds in Taurus. In the following we describe our methodology for partitioning the stars in our sample into different groups that roughly define the clouds in the complex. We use the term ``cluster'' throughout this section to refer to the grouping of stars with similar properties that result from our clustering analysis, and we warn the reader that the terminology used here is not related to the astronomical context (i.e., star clusters).  

\subsection{Selection criteria}\label{section4.1}

Our sample of YSOs in Taurus compiled from the literature contains 415 stars with measured trigonometric parallaxes and proper motions. However, some of these sources are spread well beyond the molecular clouds of the complex. Thus, we restrict our sample to the general region of the main star-forming clouds in Taurus which roughly spans the following range of Galactic coordinates:   $166^{\circ}\leq l\leq 180^{\circ}$, $-18^{\circ}\leq b\leq -6^{\circ}$ for the central and northern clouds, and  $176^{\circ}\leq l\leq 183^{\circ}$, $-22^{\circ}\leq b\leq -17^{\circ}$ for the southernmost clouds of the complex. This reduces our sample to 388~stars. 

As explained in the previous section, we are using the  astrometric solution from Gaia-DR2 for most sources in this study. The Gaia-DR2 catalog is unprecedented for the quality and quantity of astrometric measurements, but it still contains some spurious solutions that need to be filtered for an optimal usage of the data \citep[see, e.g.,][]{Arenou2018}. We proceeded as follows to obtain an astrometrically clean sample of stars. First, we select only the sources with \texttt{visibility\_periods\_used}~$>8$, as suggested by \citet{Babusiaux2018}. This removes ten~stars from the sample with observations that are not spread out in time and result in poor astrometric solutions. Second, we adopt the renormalized unit weight error (RUWE) of the source as a goodness-of-fit statistic to remove poor astrometric solutions (i.e., $RUWE>1.4$).\footnote{see technical note  \href{https://www.cosmos.esa.int/web/gaia/dr2-known-issues}{GAIA-C3-TN-LU-LL-124-01} for more details} This procedure flags 94 sources in our sample, and rejecting them yields the astrometrically clean sample of 284 stars that we use in the forthcoming analysis. 

\subsection{Clustering analysis}\label{section4.2}

Mode association clustering is a non-parametric statistical approach used for clustering analysis that    finds the modes of a kernel-based estimate of the density of points in the input space and   groups the data points associated with the same modes into one cluster with arbitrary shape \citep[see][for more details]{Li2007}. Clustering by mode identification requires only the bandwidth $\sigma$ of the kernel to be defined. When the bandwidth increases, the density of points becomes smoother and more points are assigned to the same cluster. Thus, a hierarchy of clusters can be constructed in a bottom-up manner by gradually increasing the bandwidth of the kernel functions and treating the modes acquired from the preceding (smaller) bandwidths as new input  to be clustered. Hierarchical Mode Association Clustering (HMAC) has the advantage of elucidating the relationship (and hierarchy) among the various clusters in the sample when compared to other commonly used clustering algorithms, for example  $k$-means \citep{MacQueen1967} and DBSCAN \citep{DBSCAN}. It is used in this study to investigate the structure of the Taurus molecular cloud complex and to reveal important clues to the star formation process in this region. 
 
In the forthcoming analysis we use the \texttt{Modalclust} package \citep{Modalclust} which implements the HMAC algorithm in R programming language. We run HMAC from the \textit{phmac} routine using a number of smoothing levels (i.e., bandwidths) defined as described below, and use the \textit{hard.hmac} function to access the cluster membership of each star at a given clustering level. We construct our dataset for the clustering analysis with HMAC using only the five astrometric parameters $(\alpha,\delta,\mu_{\alpha}\cos\delta,\mu_{\delta},\varpi)$. Many stars in our sample are still lacking radial velocity measurements, thus they will be included only in a subsequent discussion (see Sect.~\ref{section5}) to refine our results. In the first step of our analysis we  rescaled the five astrometric parameters so that the resulting distributions have zero mean and unit variance. We obtain the same results using rescaled and non-rescaled parameters, and we therefore decided to work with the non-rescaled astrometry as given in the original sources.     

The hierarchical clustering is performed in a bottom-up manner using a sequence of bandwidths $\sigma_{1}<\sigma_{2} < ... < \sigma_{L}$ (in all dimensions) that need to increase by a sufficient amount to drive the merging of the existing clusters at level~$l$ with $l=1,2,...,L$, where $L$ is the highest level and merges the full sample into a single cluster. We construct the sequence of bandwidths $\sigma_{l}$ as described below. The smallest bandwidth $\sigma_{1}$ that is associated with the lowest level is defined based on the uncertainties of the astrometric parameters used in our analysis. The median errors in the stellar positions (right ascension and declination), proper motions (right ascension and declination), and parallax for the sample of 284~stars are, respectively, 0.093~mas, 0.054~mas, 0.224~mas/yr, 0.162~mas/yr, and 0.142~mas. We take the maximum value among the median uncertainties listed before as the bandwidth for the lowest level (i.e., $\sigma_{1}$ = 0.224). Then, we proceed as follows for the higher levels~($l>1$). We compute the covariance matrix of the clusters at level $l$ (after removing the outliers, see Sect.~\ref{section4.3}), and take the smallest variance observed among all clusters in this level as the bandwidth for the following level. If the new bandwidth does not produce cluster mergers in the next level, we use the second smallest variance and repeat the procedure until at least one merger is produced. This procedure is repeated for all clustering levels until all clusters (and outliers) are clustered into the only existing mode at level $L$. 

\begin{figure*}[!p]
\begin{center}
\includegraphics[width=1.0\textwidth]{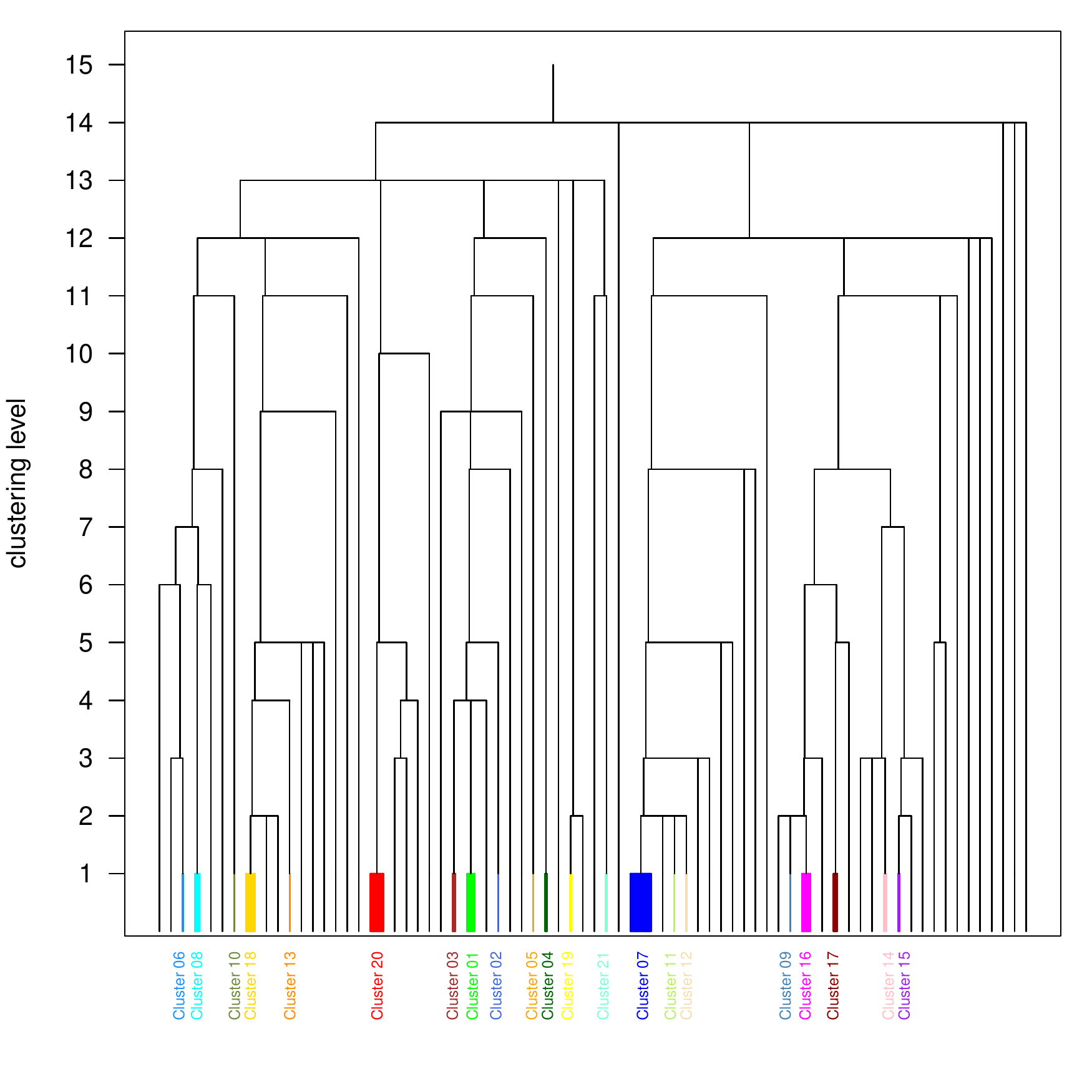}
\caption{
\label{fig_dendrogram}
Hierarchical tree (or dendrogram) obtained with HMAC for the sample of 284~stars. The different colors indicate the various clusters at the lowest clustering level of the tree. Outliers that result directly from the HMAC analysis are shown in black.    
}
\end{center}
\end{figure*}

\begin{figure*}[!p]
\begin{center}
\includegraphics[width=1.0\textwidth]{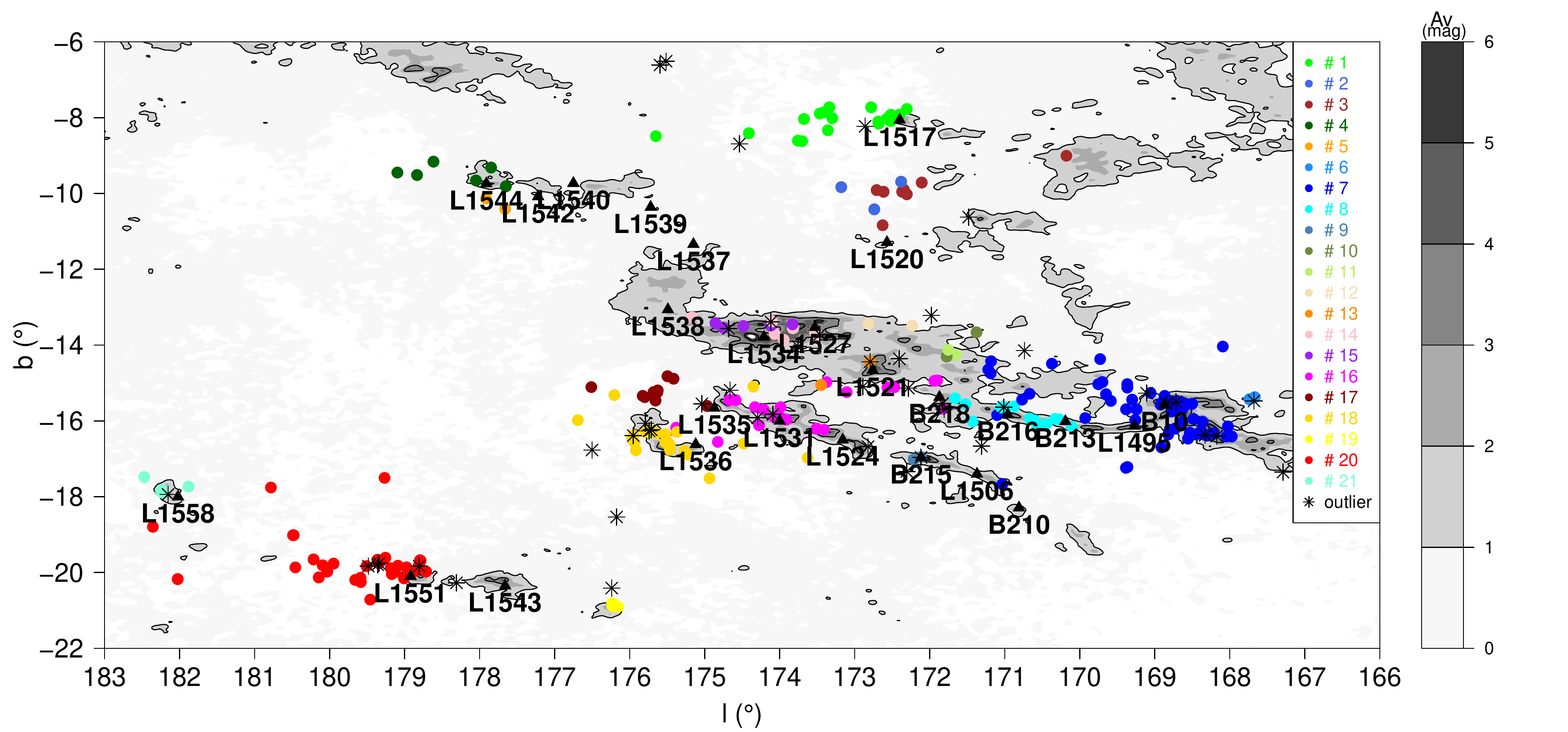}
\caption{
\label{fig_extinction_map}
Location of the 284 stars in our sample overlaid on the extinction map from \citet{Dobashi2005} in Galactic coordinates. The different colors represent the sources that belong to the 21 clusters identified at the lowest level of the hierarchical tree obtained with HMAC. Outliers that result directly from the HMAC analysis are shown as  black asteriks. The position of the most prominent clouds \citep{Barnard1927,Lynds1962} are indicated in the diagram with black triangles.   
}
\end{center}
\vspace{1cm}
\end{figure*}

Figure~\ref{fig_dendrogram} shows the resulting hierarchical tree (or dendrogram) obtained with HMAC for the sample of 284~stars. It reveals the existence of 21 clusters at the lowest clustering level which we discuss in more detail in Sect.~\ref{section4.4} (see cluster membership for each star in Table~\ref{tab1}). We also note the existence of 48 clusters with one single data point, which we consider to be extreme outliers because they exhibit different (unique) properties  compared to the other clusters in this level. Table~\ref{tab_HMAC} summarizes the results obtained in each clustering level. In Figure~\ref{fig_extinction_map} we show that the various clusters obtained from HMAC are indeed associated with different molecular clouds of the Taurus complex. Although the existence of a few additional outliers that could not be identified by the current methodology is still apparent, HMAC has proven to be a useful tool to separate the stars that belong to the several molecular clouds which often exhibit arbitrary shapes and unclear boundaries. The robustness of our clustering results obtained with HMAC is tested in Appendix~\ref{appendixA} based on synthetic data and confirms the results presented in this section.

\begin{table*}[!p]
\renewcommand\thetable{2} 
\centering
\caption{Results obtained with HMAC for each clustering level.
\label{tab_HMAC}}
\begin{tabular}{ccccccccccccccc}
\hline
\hline
Level&1&2&3&4&5&6&7&8&9&10&11&12&13&14\\
\hline
\hline
$\sigma_{l}$&0.224&0.244&0.264&0.266&0.292&0.304&0.311&0.330&0.356&0.373&0.498&0.745&0.971&4.193\\
No. of clusters&21&19&18&17&16&15&13&12&12&12&9&6&1&1\\
No. of outliers&48&40&35&31&23&22&21&17&14&13&9&5&4&0\\
\hline\hline
\end{tabular}
\tablefoot{We provide the bandwidth, number of clusters, and outliers for each clustering level of the hierarchical tree in Figure~\ref{fig_dendrogram}.}
\end{table*}

\subsection{Removing outliers in the individual clusters} \label{section4.3}

HMAC has shown to be able to detect the most extreme outliers in our sample which have been grouped into clusters of one single data point. However, we still note the existence of a more dispersed population of stars in some clusters that clearly extends beyond the limits of the molecular clouds (see, e.g., cluster 7 in Fig.~\ref{fig_extinction_map}). In this section we revise the membership status of these sources and reject potential outliers in the individual clusters. In this context, we use the minimum covariance determinant \citep[MCD,][]{Rousseeuw1999} method that is a robust estimator of multivariate location and scatter efficient in outlier detection. 

Our dataset used for the clustering analysis is stored in an $n\times p$ data matrix $\mathbf{X}=(\mathbf{x_{1}}, \mathbf{x_{2}}, ..., \mathbf{x_{n}})^{t}$ with $\mathbf{x_{i}}=(x_{i1}, x_{i2}, ..., x_{ip})^{t}$ for the $i$-th observation, where $n$ is the number of stars in the sample and $p$ is the number of dimensions (variables) used in our analysis ($p=5$).  The MCD estimator searches the subset of $h$ observations (out of $n$) that returns the covariance matrix with the lowest determinant. The tolerance ellipse is defined based on the set of $p$-dimensional points whose MCD-based robust distances
\begin{equation}
RD(\mathbf{x})=\sqrt{(\mathbf{x}-\boldsymbol{\mu})^{t}\boldsymbol{\Sigma}^{-1}(\mathbf{x}-\boldsymbol{\mu})}
\end{equation} 
equals $\sqrt{\chi^{2}_{p,\alpha}}$. We denote $\boldsymbol{\mu}$ as the MCD estimate of location, $\boldsymbol{\Sigma}$ as the MCD covariance matrix, and $\chi^{2}_{p,\alpha}$ as the $\alpha$-quantile of the $\chi^{2}_{p}$ distribution. Here we use the value of $\alpha=0.975$ to construct the tolerance ellipse and identify outliers following the procedure described by \citet{Hubert2010}.  

We compute the robust distance of the stars in the clusters derived from the HMAC analysis and remove the outliers based on the cutoff threshold $\sqrt{\chi^{2}_{p,0.975}}$. This procedure is applied to all clusters in our sample with $h>p$ and repeated at each level of the hierarchical tree. The final membership status of each star is given in the last column of Table~\ref{tab1}. 

\subsection{Notes on the individual clusters}\label{section4.4}

\begin{figure*}[!htp]
\begin{center}
\includegraphics[width=0.33\textwidth]{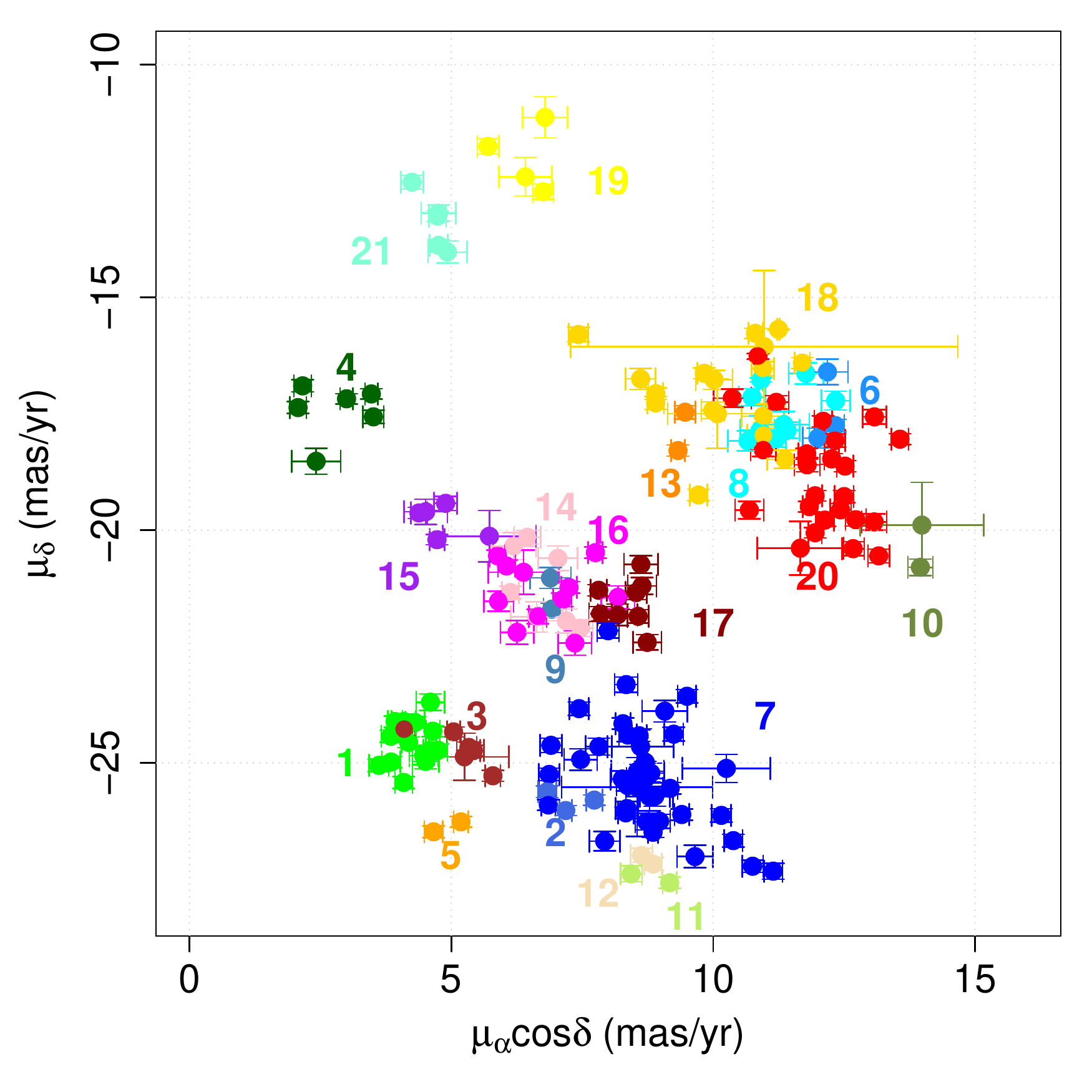}
\includegraphics[width=0.33\textwidth]{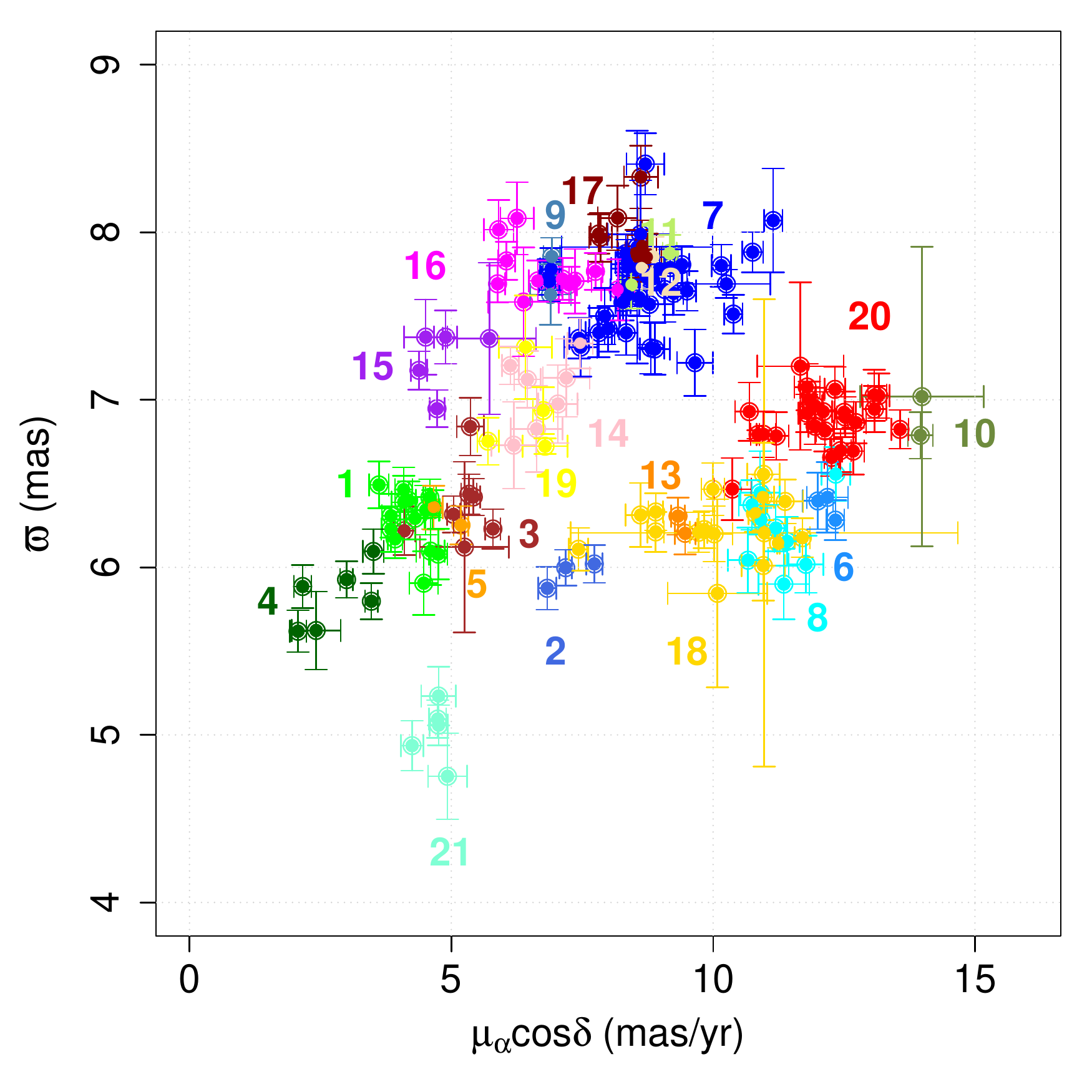}
\includegraphics[width=0.33\textwidth]{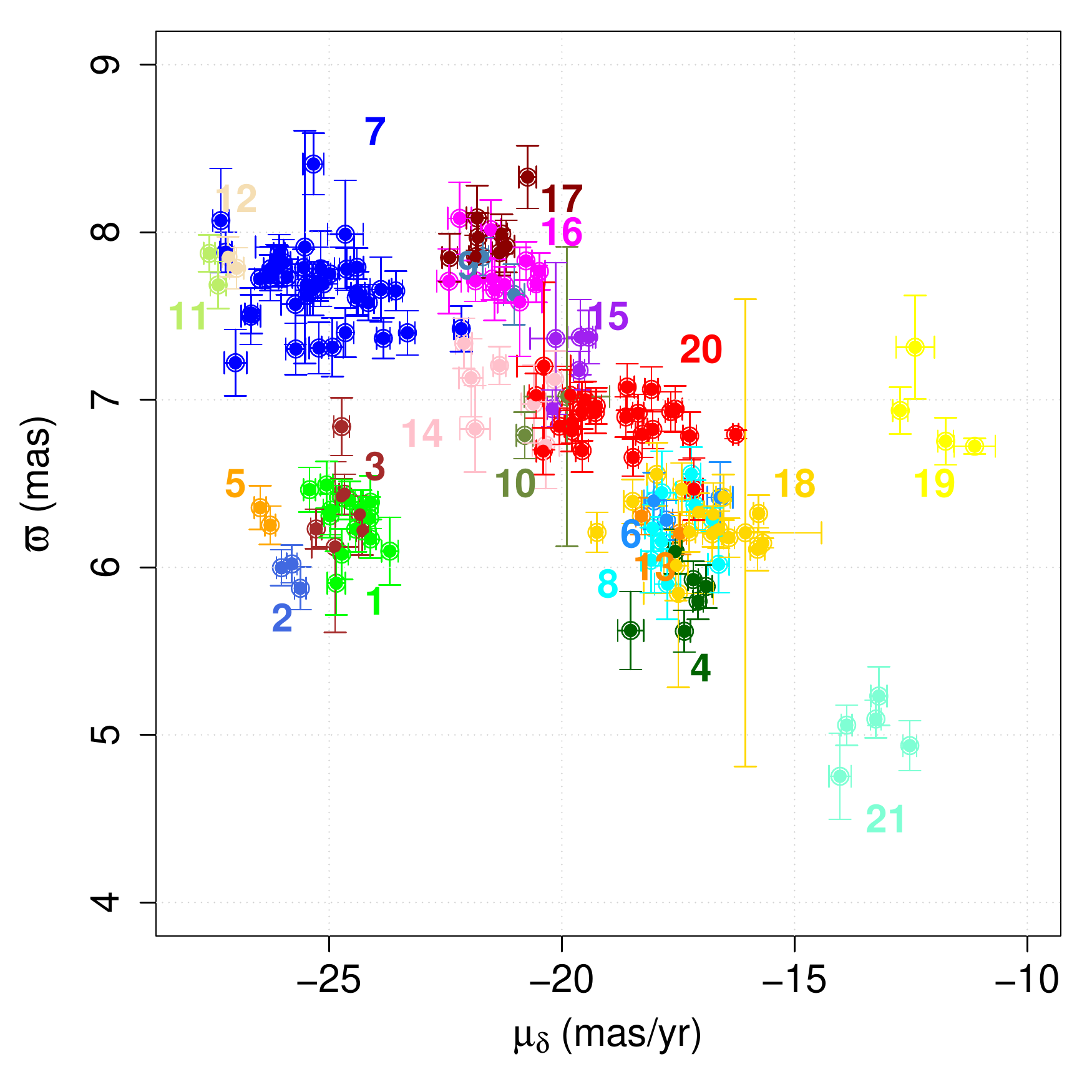}
\caption{
\label{fig_allClusters}
Clustering of the stars in the space of proper motions and parallaxes for the 21 clusters obtained with HMAC (after removing the outliers).  
}
\end{center}
\end{figure*}

In the following we discuss the individual clusters obtained with HMAC at the lowest level of the hierarchical tree. We present the clusters in order of ascending longitude and start with the northernmost clusters, as shown in Fig.~\ref{fig_extinction_map}. Figure~\ref{fig_allClusters} shows the proper motions and parallaxes of the stars in the various clusters, and Table~\ref{tab_allClusters} summarizes the cluster properties. 

\begin{figure*}
\begin{center}
\includegraphics[width=1.0\textwidth]{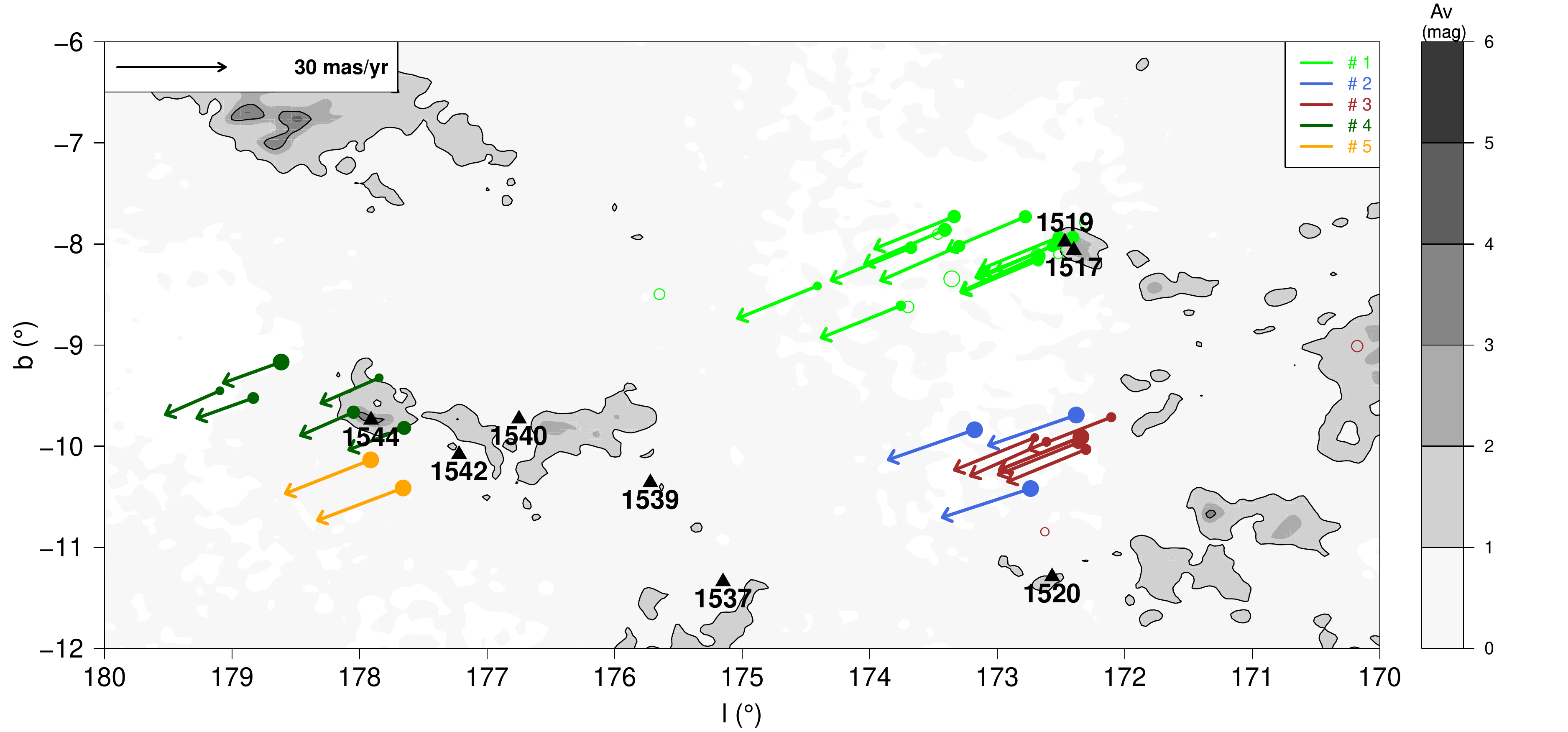}
\caption{
\label{fig_extinction_map1}
Location of the sources in clusters 1, 2, 3, 4, and 5 overlaid on the extinction map of \citet{Dobashi2005} in Galactic coordinates. The size of the symbols has been rescaled between the minimum and maximum parallaxes observed in each cluster to better distinguish between the closest (big symbols) and most remote (small symbols) members within each cluster. Filled and open symbols indicate, respectively, cluster members and outliers that have been removed in our analysis (see Sect.~\ref{section4.3}). The vectors indicate the stellar proper motions converted to the Galactic reference system (not corrected for  solar motion). The position of the most prominent clouds \citep{Barnard1927,Lynds1962} is indicated in the diagram with black triangles. 
}
\end{center}
\end{figure*}

\paragraph*{Cluster 1} is projected towards the northernmost molecular clouds of the complex, L~1517 and L~1519 (see Fig.~\ref{fig_extinction_map1}). It is interesting to note the existence of a more dispersed population of stars around these clouds with similar properties of the ``on-cloud'' population. We confirm that the mean parallax of the more dispersed stars ($\varpi=6.253\pm0.088$~mas) is in good agreement with the mean parallax of the on-cloud stars ($\varpi=6.302\pm0.046$~mas), and both values are consistent with the mean parallax of all stars in the cluster ($\varpi=6.281 \pm 0.045$~mas, see Table~\ref{tab_allClusters}). The proper motions of the two populations are also consistent within 1~mas/yr in both components.

\paragraph*{Clusters 2 and 3} overlap in the same sky region and they are not projected towards any cloud of the complex, as shown in Fig.~\ref{fig_extinction_map1}. Their parallaxes differ significantly (see Fig.~\ref{fig_allClusters} and Table~\ref{tab_allClusters}) which explains the clustering in separate groups. 

\paragraph*{Cluster 4} is a grouping of seven sources located in the northern part of the Taurus complex. Three of them (V836~Tau, CIDA~8, and CIDA~9B) are projected towards the molecular cloud L~1544 (see Fig.~\ref{fig_extinction_map1}), and their mean parallax ($\varpi=5.812\pm0.095$~mas) is consistent with the mean parallax ($\varpi=5.837\pm0.139$~mas) of the more dispersed cluster members (RX~J0507.2+2437, CIDA~12 and 2MASS~J05080709+2427123). 

\paragraph*{Cluster 5} contains only two stars (2MASS J05010116+2501413 and 2MASS J05023985+2459337), which are located south of L~1544 (see Fig.~\ref{fig_extinction_map1}). Despite the close proximity (in the plane of the sky) to cluster 4, the sources in the two groups exhibit different proper motions (see also Fig.~\ref{fig_allClusters}), which justifies the clustering in separate groups. Cluster 5 is therefore not associated with any molecular cloud of the complex. 

\begin{figure*}
\begin{center}
\includegraphics[width=1.0\textwidth]{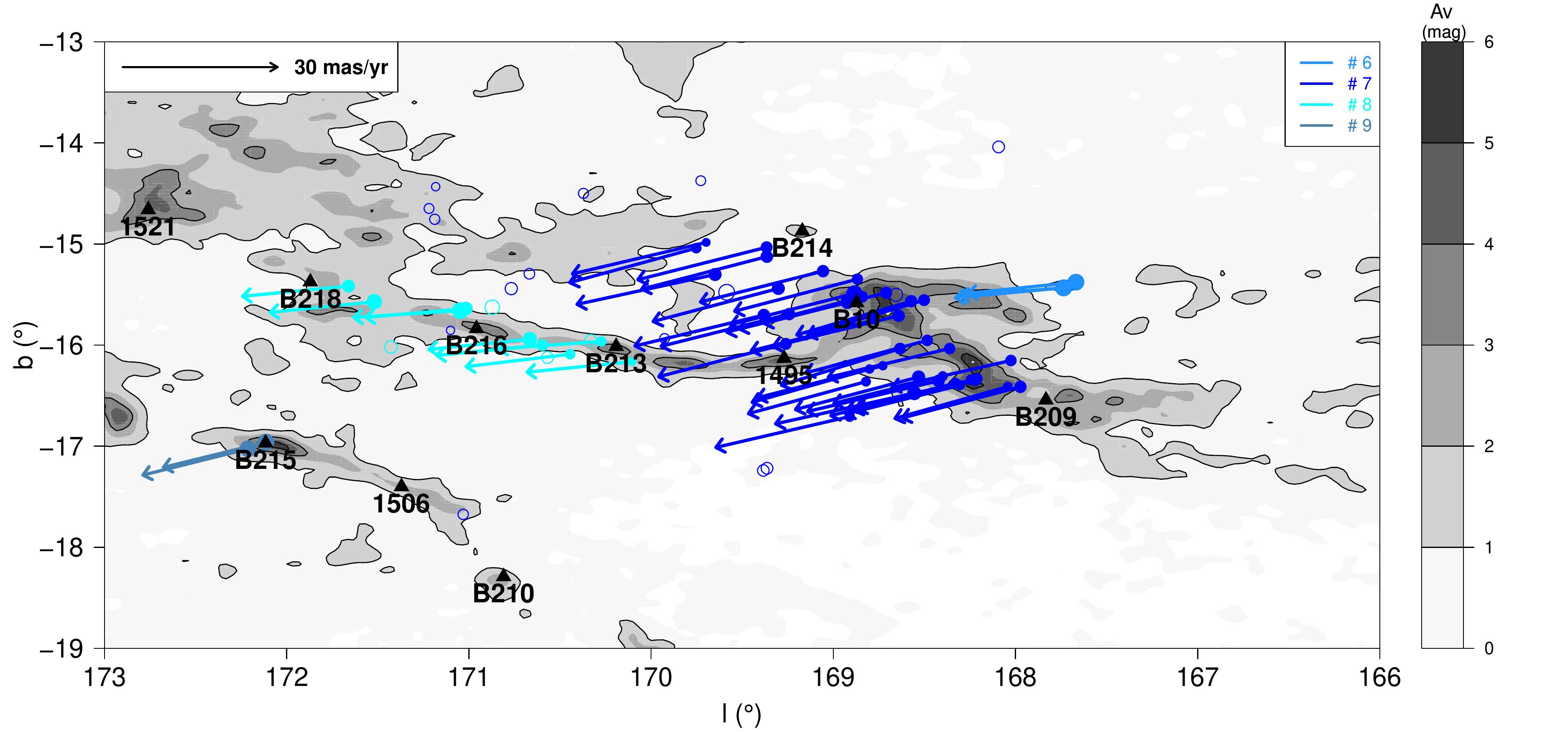}
\caption{
\label{fig_extinction_map2}
Same as Figure~\ref{fig_extinction_map1}, but for clusters 6, 7, 8, and 9.
}
\end{center}
\end{figure*}

\paragraph*{Cluster 6} consists of only three stars (2MASS J04154131+2915078, 2MASS J04154269+2909558, and 2MASS J04154278+2909597) projected towards a molecular cloudlet located northwest of L~1495 (hereafter L~1495~NW). Their parallaxes and proper motions differ significantly from the sources in L~1495 (i.e., cluster 7, see below) despite the close proximity in the plane of the sky (see Figs.~\ref{fig_allClusters} and \ref{fig_extinction_map2}). This suggests that L~1495~NW and L~1495 are different structures of the Taurus region. 

\paragraph*{Cluster 7} is the most populated cluster in our analysis (39 sources) and it is associated with the most prominent molecular cloud of the complex, namely  L~1495. The vast majority of stars in this cluster are located in the direction of the dense core B~10 of the cloud and many of the more dispersed sources in the vicinity of L~1495 have been flagged as outliers by the MCD estimator, as illustrated in Figure~\ref{fig_extinction_map2}. 

\paragraph*{Cluster 8}  is associated with the filamentary structure connected  (in the plane of the sky) to the central part of the L~1495 molecular cloud. \citet{Schmalzl2010} divided the  filament into five clumps (B ~211, B~213, B~216, B~217, and B~218) with  ranges of $169^{\circ}< l <172^{\circ}$ and $-16.2^{\circ}< b < -15.2^{\circ}$ (see Fig.~5 of their paper). Most of our sources in this cluster are located between B~213 and B~216 (see Fig.~\ref{fig_extinction_map2}), and we detect hints of a gradient in parallaxes along the filament from $l=170.1^{\circ}$ ($\varpi=5.900\pm0.210$~mas) to $l=171.0^{\circ}$ ($\varpi=6.557\pm0.162$~mas).  Figure~\ref{fig_allClusters} and Table~\ref{tab_allClusters} show that the parallaxes and proper motions of the sources in the filament and central part of L~1495 (i.e., cluster 7) are significantly different, which  confirms them as independent structures. This is also confirmed by the late merging of the two clusters in the hierarchical clustering, as shown in Figure~\ref{fig_dendrogram}. Interestingly, the stars in the filament exhibit parallaxes and proper motions that are more consistent with the sources in the L~1495~NW cloudlet despite the angular separation of a few degrees on the sky. 

\paragraph*{Cluster 9} includes two sources (FU~Tau~A and FF~Tau) which are projected towards the B~215 star-forming clump (see Fig.~\ref{fig_extinction_map2}). Their parallaxes are still consistent with the sources in L~1495 (cluster 7), but the proper motions are shifted by about 4~mas/yr in declination (see also Table~\ref{tab_allClusters}). 

\begin{figure*}
\begin{center}
\includegraphics[width=1.0\textwidth]{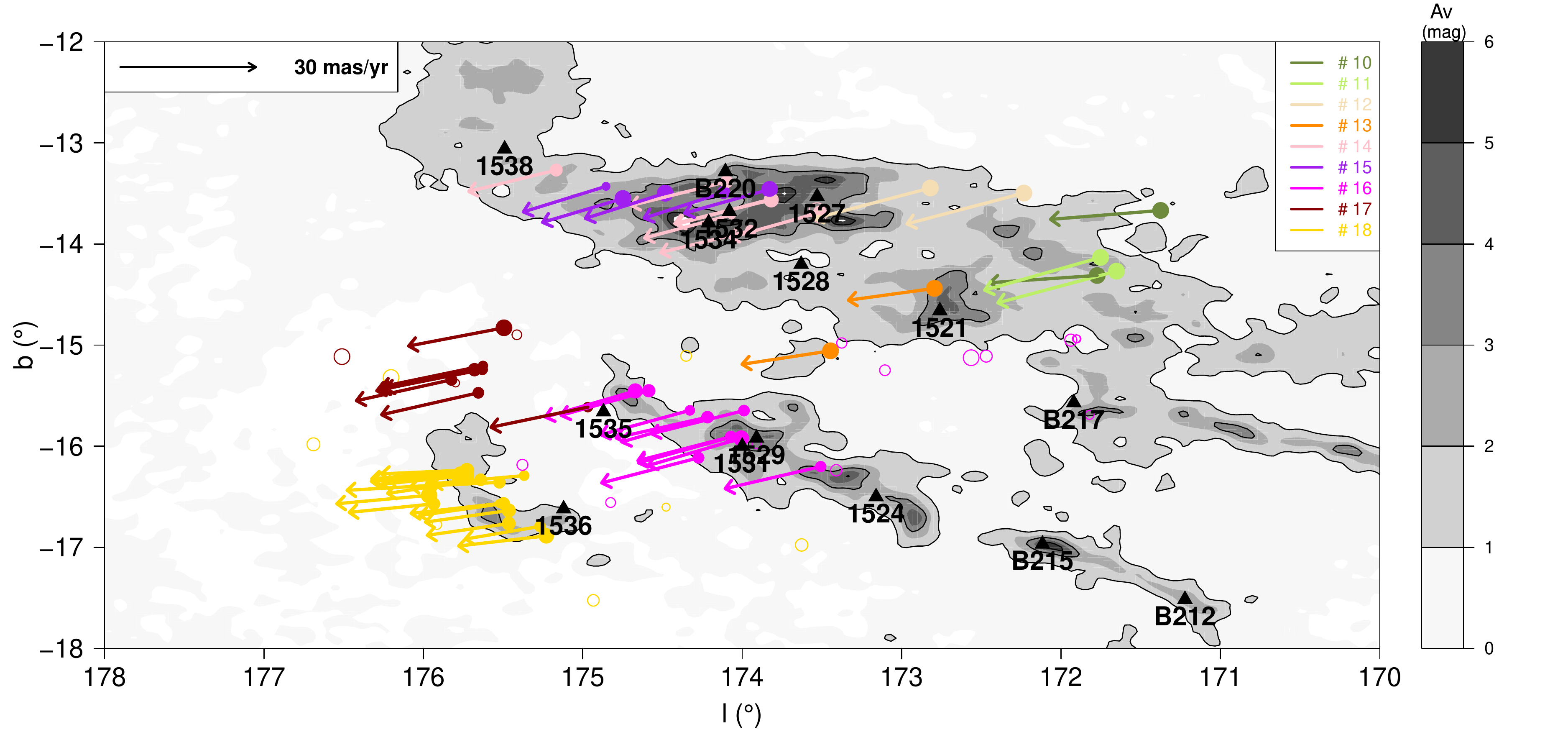}
\caption{
\label{fig_extinction_map3}
Same as Figure~\ref{fig_extinction_map1}, but for clusters 10, 11, 12, 13, 14, 15, 16, 17, and 18.
}
\end{center}
\end{figure*}

\paragraph*{Clusters 10} has the two stars with the largest proper motions (in right ascension) in the sample (2MASS~J04312669+2703188 and 2MASS~J04322873+2746117). They are separated by about 1$^{\circ}$ in the plane of the sky (see Fig.~\ref{fig_extinction_map3}) and they are not associated with any star-forming clump. The closest clusters in terms of similarity are clusters 6 and 8. Figure~\ref{fig_dendrogram} shows that the tree clusters merge at level 11 of the hierarchical tree to form one single group.   

\paragraph*{Clusters 11 and 12} are spread over 2 degrees in Galactic longitude and each of them contains two stars (see Fig.~\ref{fig_extinction_map3}). The two clusters exhibit similar proper motions and parallaxes   to the sources in cluster 7 (see Fig.~\ref{fig_allClusters}). This is confirmed by the early merging of these two clusters with cluster 7 at level 2 of the hierarchical tree to form one single group (see Fig.~\ref{fig_dendrogram}). 

\paragraph*{Cluster 13} includes only two sources (DL~Tau and IT~Tau~A). Their positions, proper motions, and parallaxes differ from the other clusters in the central region of the complex which justifies the clustering into a different group. IT~Tau is projected towards the molecular cloud L~1521 and DL~Tau is located in a different cloudlet separated by about 1$^{\circ}$ in the plane of the sky (see Fig.~\ref{fig_extinction_map3}), making it unclear whether this cluster is associated with any cloud of the complex. The small number of sources and their somewhat different sky positions led us to the decision to exclude it from our forthcoming discussion about the properties of the molecular clouds.  

\paragraph*{Clusters 14 and 15} are collectively discussed because they are both located in the Heiles Cloud~2 and overlap in the plane of the sky (see Fig.~\ref{fig_extinction_map3}). The sources in these two clusters are spread over the star-forming clumps L~1527, L~1532, L~1534, and B~220. Their parallax and proper motion values  are somewhat different (see Table~\ref{tab_allClusters}) and define a different locus in Fig.~\ref{fig_allClusters}. This indicates the existence of substructures in this cloud that we discuss in our forthcoming analysis using the three-dimensional spatial distribution of the stars (see Sect.~\ref{section5}).  

\paragraph*{Cluster 16} contains 11 stars spread over the molecular clouds L~1535, L~1529, L~1531, and L~1524 (see Fig.~\ref{fig_extinction_map3}). We note that the sources projected towards the various clouds associated with this cluster exhibit similar properties. For example, the sources projected towards the L~1535, L~1529, L~1531, and L~1524 molecular clouds have a mean parallax of $\varpi=7.922\pm0.105$~mas, $\varpi=7.743\pm0.052$~mas, and $\varpi=7.656\pm0.182$~mas, respectively, and they are   consistent among themselves. The proper motions of the various sources are consistent within 1-2~mas/yr. Thus, we discuss their properties collectively under the same group.

\paragraph*{Cluster 17} is a grouping of eight sources located north of L~1536 (cluster 18, see below) that is not projected towards any cloud in the complex (see Fig.~\ref{fig_extinction_map3}). Despite the close proximity (in the plane of the sky) to the L~1536 molecular cloud, we note that the two clusters define a different locus in the proper motion vector diagram (see Fig.~\ref{fig_allClusters}). 

\paragraph*{Cluster 18} is one of the most populated clusters in our sample and it contains 17 stars spread in and around the L~1536 molecular cloud. The most dispersed sources in this cluster have been flagged as outliers by the MCD estimator, as illustrated in Figure~\ref{fig_extinction_map3}.  

\begin{figure*}
\begin{center}
\includegraphics[width=1.0\textwidth]{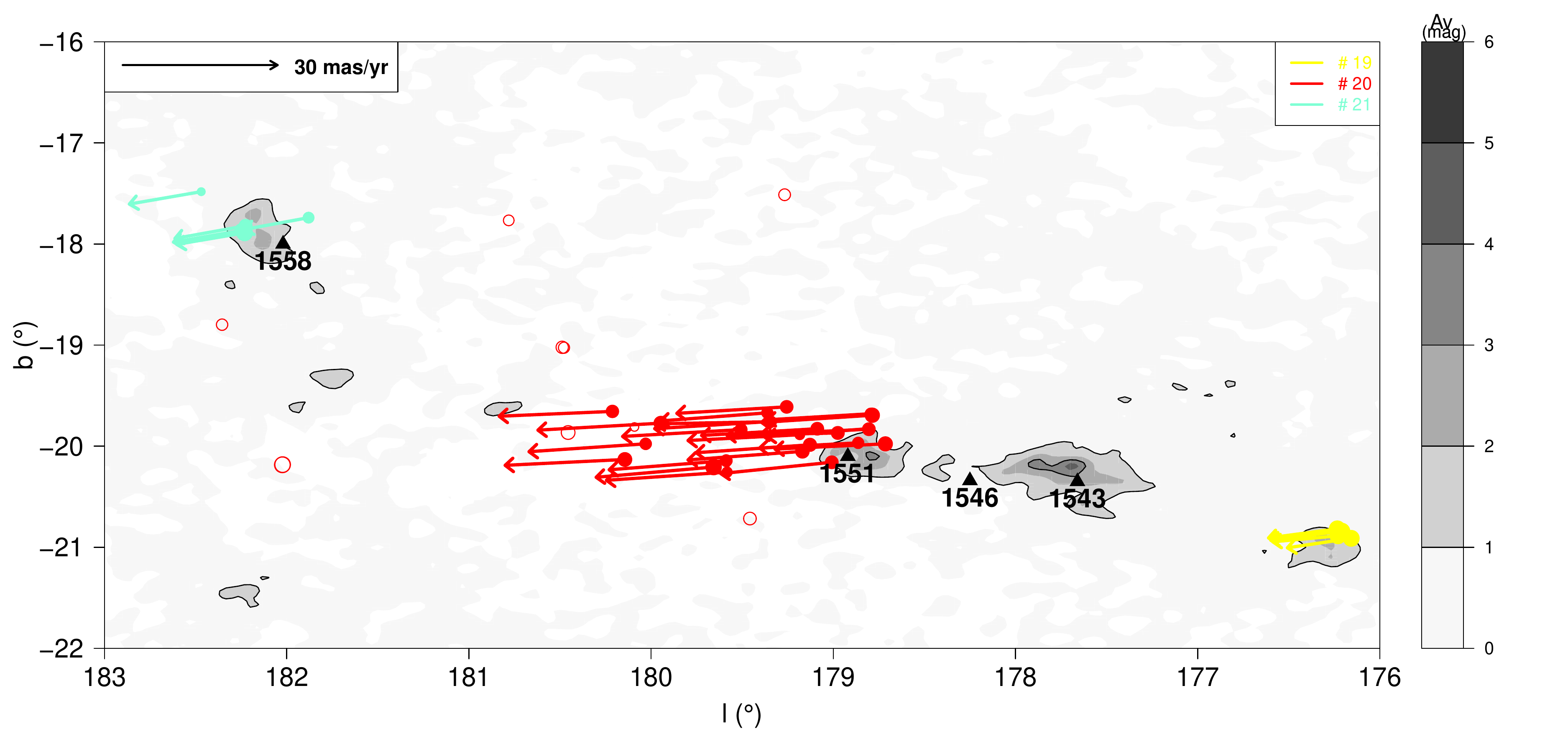}
\caption{
\label{fig_extinction_map4}
Same as Figure~\ref{fig_extinction_map1}, but for clusters 19, 20, and 21.
}
\end{center}
\end{figure*}

\paragraph*{Cluster 19} hosts four sources (T~Tau, IRAS~04187+1927, RX~J0422.1+1934, and 2MASS~J04221332+1934392) in a small cloud (hereafter the  T~Tau cloud) in the southern region of the Taurus complex (see Fig.~\ref{fig_extinction_map4}). We note in Fig.~\ref{fig_allClusters} that the sources in this cluster exhibit the smallest values for the proper motion component in declination among all the clusters in our sample.  

\paragraph*{Cluster 20} is the second most populated cluster in our analysis. It includes 25 sources that  are spread in and around the L~1551 molecular cloud (see Fig.~\ref{fig_extinction_map4}). The late merging of clusters 19 and 20 in the hierarchical tree (level 13, see Fig.~\ref{fig_dendrogram}) and the different proper motions (see Fig.~\ref{fig_allClusters}) suggest that L~1551 and the T~Tau cloud are indeed independent structures of the southern region of the Taurus complex.  

\paragraph*{Cluster 21} includes five sources projected towards the L~1558 molecular cloud (see Fig.~\ref{fig_extinction_map4}). As shown in Fig.~\ref{fig_allClusters} the sources in this cluster exhibit the smallest parallaxes making it the most distant cluster in our sample (see also discussion in Sect.~\ref{section5}).

\begin{landscape}
\renewcommand\thetable{3} 
\begin{table}
\caption{Properties of the clusters obtained with HMAC. }
\label{tab_allClusters}
\scriptsize{
\begin{tabular}{ccccccccccccccccl}
\hline
\hline
Cluster &$N_{0}$&$N$&$\alpha$&$\delta$&$l$&$b$&\multicolumn{3}{c}{$\mu_{\alpha}\cos\delta$}&\multicolumn{3}{c}{$\mu_{\delta}$}&\multicolumn{3}{c}{$\varpi$}&Molecular Clouds\\
&&&(h:m:s)&($^{\circ}$~\arcmin~\arcsec)&(deg)&(deg)&\multicolumn{3}{c}{(mas/yr)}&\multicolumn{3}{c}{(mas/yr)}&\multicolumn{3}{c}{(mas)}&\\
\hline
\hline
&&&&&&&Mean $\pm$ SEM&Median&SD&Mean $\pm$ SEM&Median&SD&Mean $\pm$ SEM&Median&SD&\\
\hline
1 & 21 & 14 & 04:57:04.1 & 30:05:31 & 173.0 & -8.0 & $ 4.249 \pm 0.095 $& 4.247 & 0.356 & $ -24.578 \pm 0.125 $& -24.644 & 0.470 & $ 6.281 \pm 0.045 $& 6.325 & 0.167 &  L~1517, L~1519 \\
2 & 3 & 3 & 04:49:17.7 & 29:05:52 & 172.8 & -10.0 & $ 7.249 \pm 0.262 $& 7.185 & 0.453 & $ -25.814 \pm 0.117 $& -25.800 & 0.203 & $ 5.965 \pm 0.045 $& 5.998 & 0.078 &  \nodata \\
3 & 9 & 7 & 04:48:29.6 & 29:25:15 & 172.4 & -9.9 & $ 5.187 \pm 0.200 $& 5.336 & 0.529 & $ -24.701 \pm 0.126 $& -24.732 & 0.334 & $ 6.369 \pm 0.089 $& 6.316 & 0.236 &  \nodata \\
4 & 6 & 6 & 05:06:04.2 & 25:01:35 & 178.3 & -9.5 & $ 2.773 \pm 0.263 $& 2.709 & 0.645 & $ -17.435 \pm 0.237 $& -17.273 & 0.580 & $ 5.825 \pm 0.076 $& 5.842 & 0.185 & L~1544 \\
5 & 2 & 2 & 05:01:50.5 & 25:00:37 & 177.8 & -10.3 & $ 4.923 \pm 0.112 $& 4.923 & 0.522 & $ -26.374 \pm 0.085 $& -26.374 & 0.214 & $ 6.304 \pm 0.087 $& 6.304 & 0.104 & \nodata \\
6 & 3 & 3 & 04:15:42.3 & 29:11:41 & 167.7 & -15.4 & $ 12.173 \pm 0.097 $& 12.183 & 0.168 & $ -17.458 \pm 0.436 $& -17.754 & 0.755 & $ 6.365 \pm 0.042 $& 6.397 & 0.073 & L~1495~NW \\
7 & 54 & 39 & 04:17:16.9 & 28:08:55 & 168.7 & -15.9 & $ 8.704 \pm 0.157 $& 8.614 & 0.981 & $ -25.308 \pm 0.179 $& -25.347 & 1.116 & $ 7.673 \pm 0.037 $& 7.692 & 0.233 & L~1495 \\
8 & 13 & 9 & 04:23:36.1 & 26:44:03 & 170.8 & -15.8 & $ 11.255 \pm 0.180 $& 11.196 & 0.541 & $ -17.482 \pm 0.184 $& -17.730 & 0.552 & $ 6.223 \pm 0.072 $& 6.235 & 0.215 &  B~213, B~216 \\
9 & 2 & 2 & 04:23:37.3 & 24:59:38 & 172.2 & -17.0 & $ 6.907 \pm 0.209 $& 6.907 & 0.023 & $ -21.362 \pm 0.129 $& -21.362 & 0.672 & $ 7.741 \pm 0.106 $& 7.741 & 0.226 & B~215 \\
10 & 2 & 2 & 04:31:57.7 & 27:24:45 & 171.6 & -14.0 & $ 13.973 \pm 0.602 $& 13.973 & 0.030 & $ -20.344 \pm 0.469 $& -20.344 & 0.906 & $ 6.903 \pm 0.452 $& 6.903 & 0.231 & \nodata \\
11 & 2 & 2 & 04:31:36.8 & 27:10:48 & 171.7 & -14.2 & $ 8.803 \pm 0.123 $& 8.803 & 0.736 & $ -27.480 \pm 0.105 $& -27.480 & 0.191 & $ 7.780 \pm 0.089 $& 7.780 & 0.189 & \nodata \\
12 & 2 & 2 & 04:36:26.7 & 27:03:04 & 172.5 & -13.5 & $ 8.743 \pm 0.133 $& 8.743 & 0.218 & $ -27.082 \pm 0.100 $& -27.082 & 0.186 & $ 7.820 \pm 0.086 $& 7.820 & 0.069 & \nodata \\
13 & 2 & 2 & 04:33:46.9 & 25:47:03 & 173.1 & -14.7 & $ 9.397 \pm 0.119 $& 9.397 & 0.135 & $ -17.878 \pm 0.093 $& -17.878 & 0.825 & $ 6.255 \pm 0.084 $& 6.255 & 0.103 & \nodata \\
14 & 9 & 7 & 04:40:22.3 & 25:50:19 & 174.1 & -13.6 & $ 6.728 \pm 0.194 $& 6.628 & 0.514 & $ -21.195 \pm 0.310 $& -21.340 & 0.820 & $ 7.046 \pm 0.081 $& 7.122 & 0.215 & Heiles Cloud 2: L~1527, L~1532, L~1534,  B~220 \\
15 & 5 & 5 & 04:41:38.0 & 25:38:26 & 174.4 & -13.5 & $ 4.847 \pm 0.237 $& 4.727 & 0.529 & $ -19.799 \pm 0.155 $& -19.630 & 0.347 & $ 7.247 \pm 0.084 $& 7.366 & 0.188 & Heiles Cloud 2: L~1527, L~1532, L~1534,  B~220 \\
16 & 23 & 11 & 04:33:12.4 & 24:19:57 & 174.2 & -15.8 & $ 6.799 \pm 0.236 $& 6.651 & 0.783 & $ -21.353 \pm 0.194 $& -21.444 & 0.642 & $ 7.768 \pm 0.046 $& 7.710 & 0.152 & L~1535, L~1529, L~1531, L~1524 \\
17 & 11 & 8 & 04:38:42.0& 23:38:14 & 175.6 & -15.3 & $ 8.368 \pm 0.131 $& 8.548 & 0.370 & $ -21.556 \pm 0.182 $& -21.570 & 0.516 & $ 7.984 \pm 0.057 $& 7.943 & 0.161 & \nodata \\
18 & 24 & 17 & 04:34:50.4 & 22:49:58 & 175.6 & -16.5 & $ 10.146 \pm 0.280 $& 10.082 & 1.156 & $ -16.994 \pm 0.238 $& -16.770 & 0.979 & $ 6.243 \pm 0.042 $& 6.212 & 0.171 & L~1536 \\
19 & 4 & 4 & 04:22:00.2 & 19:33:57 & 176.2 & -20.9 & $ 6.416 \pm 0.253 $& 6.586 & 0.506 & $ -12.006 \pm 0.355 $& -12.082 & 0.711 & $ 6.932 \pm 0.136 $& 6.845 & 0.272 & T~Tau cloud \\
20 & 34 & 24 & 04:32:59.4 & 17:57:15 & 179.3 & -19.9 & $ 12.068 \pm 0.168 $& 12.118 & 0.822 & $ -18.846 \pm 0.236 $& -18.943 & 1.157 & $ 6.880 \pm 0.032 $& 6.908 & 0.158 & L~1551 \\
21 & 5 & 5 & 04:47:09.1 & 17:05:31 & 182.2 & -17.8 & $ 4.686 \pm 0.114 $& 4.757 & 0.255 & $ -13.375 \pm 0.269 $& -13.254 & 0.602 & $ 5.015 \pm 0.081 $& 5.058 & 0.180 & L~1558 \\
\hline
\hline

\end{tabular}
\tablefoot{We list the initial number of members $N_{0}$, final number of members $N$ (after removing outliers, see Sect.~\ref{section4.3}), position (equatorial and Galactic coordinates), proper motion, trigonometric parallax, and the molecular clouds associated with each cluster. We also provide the mean, standard error of the mean (SEM), median and standard deviation (SD) values of the proper motion, and trigonometric parallax distributions of each cluster. The standard deviation given in the table represents the difference between the individual measurements when the cluster has only two stars. }
}
\end{table}
\end{landscape}
\clearpage

\subsection{Comparison of our results with those of  \citet{Luhman2018}}\label{section4.5}

In a recent study \citet{Luhman2018} divided a sample of Taurus stars into four populations with similar properties of proper motions, parallaxes, and photometry. Two points are worth mentioning here before comparing our results with that study. First, the two studies had  distinct objectives which explains the different strategy employed to explore the Gaia-DR2 data in the Taurus region. \citet{Luhman2018} performed an extensive analysis to improve the census of Taurus stars by refining the sample of known members and identifying new candidates. In this context, the sources were not filtered (as done in the present study) to minimize as much as possible the rejection of potential members of the Taurus region. On the other hand, we  decided to apply the RUWE selection criterion in the present study, which is a more conservative approach to filter the stars in the sample. This procedure is likely to remove some bona fide members of the Taurus region, but at the same time it minimizes the  number of stars with discrepant measurements in the sample due to a poor fit of the Gaia-DR2 astrometric solution or to non-membership. This was made necessary to derive more accurate distances and spatial velocities for the  subgroups, as we discuss in more detail in Section~\ref{section5}.  Second, the methodology used by \citet{Luhman2018} to identify the four populations of stars is based on a manual selection of the sources with similar properties rather than a clustering algorithm. For these reasons, the number of sources and the subgroups themselves identified in the two studies differ from each other, and the comparison between the two solutions is not straightforward. We proceeded as follows to compare the results given by the two studies.

\begin{figure}
\begin{center}
\includegraphics[width=0.48\textwidth]{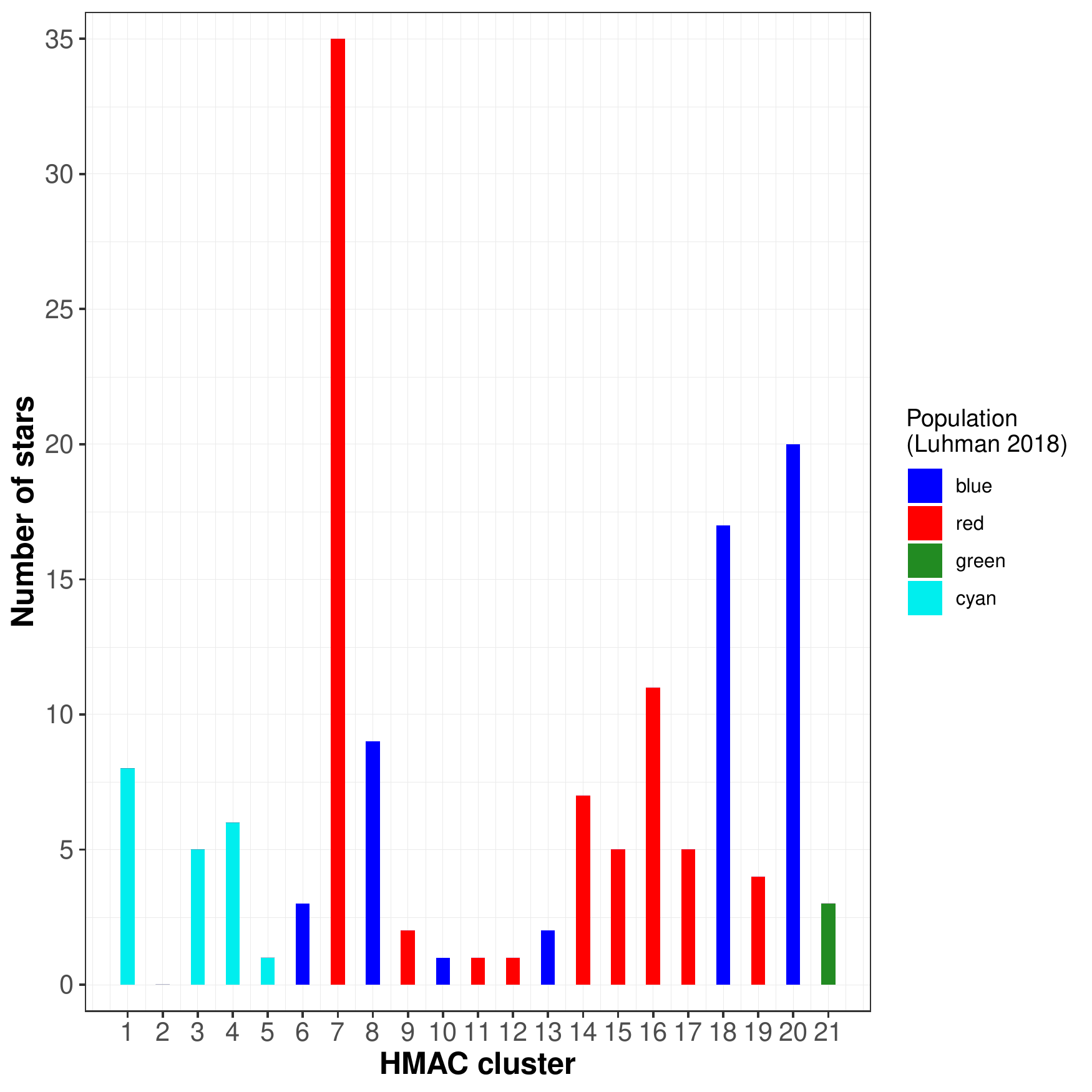}
\caption{Comparison of the HMAC clustering results obtained in this study with the four populations of Taurus stars (blue, red, green, and cyan) identified by \citet{Luhman2018} from the sample of known Taurus members (Table~1 of that paper). The bar chart indicates the number of stars of each population that are in common with the HMAC clusters. 
\label{comp_L18} 
}
\end{center}
\end{figure}

To begin with, we cross-matched our sample of cluster members with the list of stars from \citet{Luhman2018}. Figure~\ref{comp_L18} shows the distribution of Taurus stars in the four populations classified by \citet{Luhman2018} among the various clusters obtained in our clustering analysis with HMAC. We note that the HMAC clusters obtained in our analysis group only stars from one of the four populations discussed by \citet{Luhman2018} (i.e., we do not see a mix of populations in the various clusters). The clusters derived from the HMAC analysis that contain only a subset of the sample of stars given by \citet{Luhman2018}, due to the different selection criteria used to filter the Gaia-DR2 sources in each study, as explained before.

Another interesting point to mention is that the four populations of \citet{Luhman2018} are closely related to the HMAC clustering results that we obtained at level 12 of the hierarchical tree (see Fig.~\ref{fig_dendrogram}), as explained below. At this level we have six groups of clusters that include all  21 clusters discussed in Sect.~\ref{section4.4} (see also Table~\ref{tab_HMAC}). We label them as follows (from the left to  right in Fig.~\ref{fig_dendrogram}): Group~A (includes clusters 6, 8, 10, 13, and 18), Group~B (includes cluster 20), Group~C (includes clusters 1, 2, 3, 4, and 5), Group~D (includes cluster 19), Group~E (includes cluster 21), and Group~F (includes clusters 7, 9, 11, 12, 14, 15, 16, and 17). 
We note from Fig.~\ref{comp_L18} that groups~A, B and groups~D, F correspond to the blue and red populations, respectively. The position of the stars was not used by \citet{Luhman2018} to define the various populations, which explains why the red and blue populations are separated into several groups in the HMAC analysis. Group~C represents the cyan population (i.e., the northern clouds) and Group~E is associated with the green population. 

We note that 51 stars from the sample of 62 new candidate members given in Table~6 of \citet{Luhman2018} have been retained for the clustering analysis after applying the selection criteria described in Sect.~\ref{section4.1}. Twenty-six of them were selected in our analysis and assigned to clusters 1, 2, 3, 5, 7, 10, 11, 12, 17, 20, and 21. In particular, we note that cluster 2 is formed exclusively by new candidate members, which explains in Figure~\ref{comp_L18} the absence of known Taurus members of the four populations identified by \citet{Luhman2018}. In addition, we find 16 stars in the sample of \citet{Joncour2017} that are not included in the list of known members given by \citet{Luhman2018}. Only five of them satisfy our selection criteria described in Section~\ref{section4.1}, and all of them have been identified as outliers in the HMAC clustering analysis. Table~\ref{tab1} lists the membership status of each star in our sample given by \citet{Joncour2017} and \citet{Luhman2018} compared to the results obtained in this study. 

Thus, we conclude that our methodology based on the HMAC analysis is able to recover the four populations of Taurus stars that were previously identified by \citet{Luhman2018}, and that the two studies return consistent results with respect to the clustering of the stars in several substructures across the Taurus complex.

\section{Discussion}\label{section5}

\subsection{Distance and spatial velocity of Taurus stars}\label{section5.1}

In this section we convert the observables used in our clustering analysis (positions, proper motions, and parallaxes) into distances, three-dimensional positions, and spatial velocities to discuss the properties of the stars projected against the various molecular clouds in this region. The forthcoming discussion will be restricted to the 13 clusters listed in Table~\ref{tab_allClusters} that are associated with a molecular cloud of the complex, and hereafter we  use  the molecular cloud identifiers when refering to the individual clusters rather than the cluster numbering from the HMAC terminology.  

First, we convert the trigonometric parallaxes and proper motions of individual stars into distances and two-dimensional tangential velocities using Bayesian inference and following the online tutorials available in the Gaia archive \citep[see][]{Luri2018}. This procedure uses an exponentially decreasing space density prior for the distance with length scale $L=1.35$~kpc \citep{Bailer-Jones2015,Astraatmadja2016} and a beta function for the prior over speed.\footnote{see also  \href{https://github.com/agabrown/astrometry-inference-tutorials/blob/master/3d-distance/resources/3D_astrometry_inference.pdf}{GAIA-C8-TN-LU-MPIA-CBJ-081} for more details} This methodology takes the full covariance matrix of the observables into account to estimate our uncertainties on the final distances and tangential velocities of the stars. Then we use the resulting distances to compute the three-dimensional position $XYZ$ of the stars in a reference system that has its origin at the Sun, where $X$ points to the Galactic center, $Y$ points in the direction of Galactic rotation, and $Z$ points to the Galactic north pole to form a right-handed system.

\begin{figure}
\begin{center}
\includegraphics[width=0.42\textwidth]{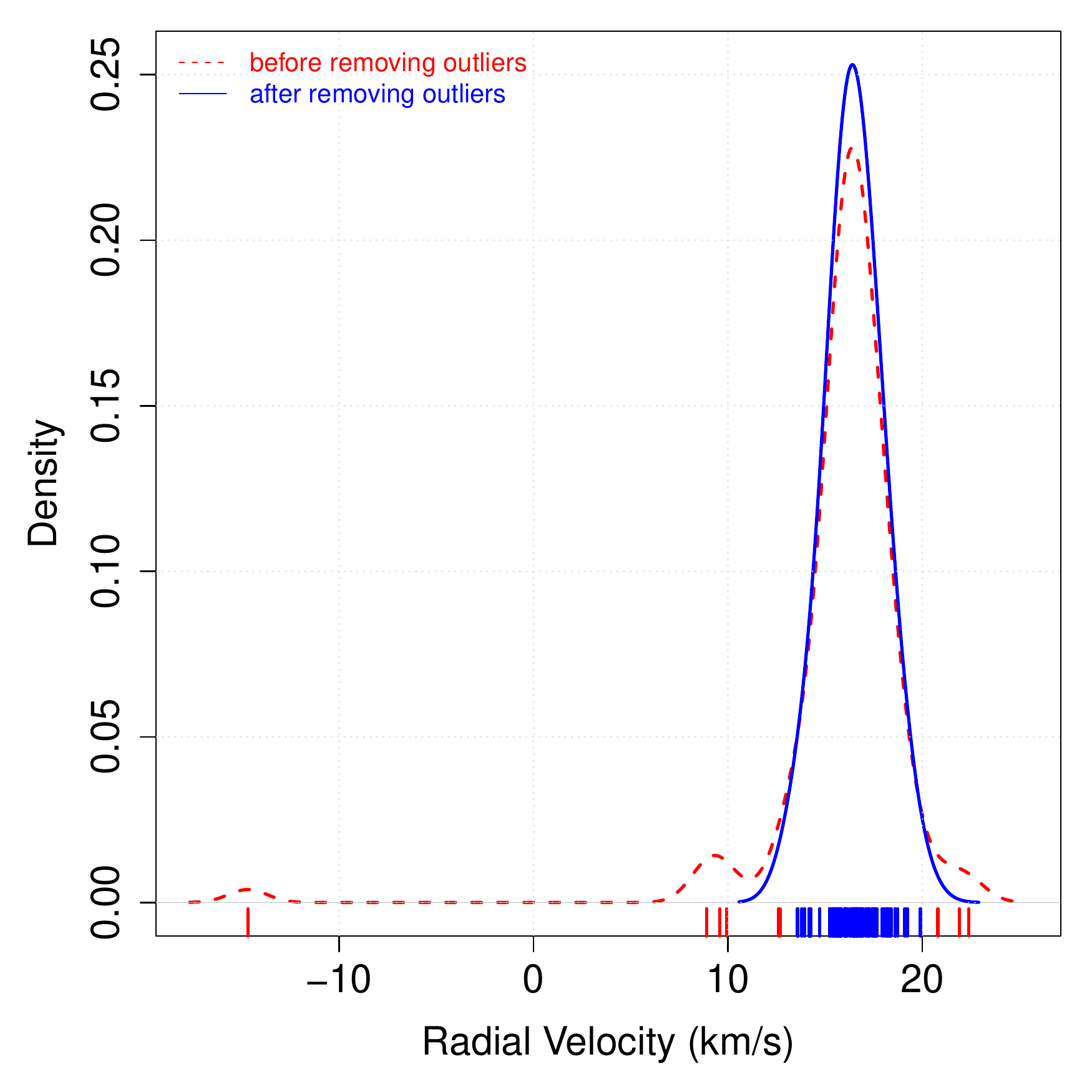}
\caption{Kernel density estimate (for a kernel bandwidth of 1~km/s) of the distribution of radial velocity measurements collected from the literature for 102 stars in the sample of cluster members obtained in our clustering analysis with HMAC. The tick marks in the horizontal axis indicate the individual measurements of each star. 
\label{dist_RV} 
}
\end{center}
\end{figure}

Second, we combine the resulting two-dimensional tangential velocities with the radial velocities collected from the literature (see Sect.~\ref{section2}) to derive the $UVW$ spatial velocity of the stars in the same reference system as described above and following the transformation outlined by \citet{Johnson1987}. We note that 102 stars among those that were confirmed as cluster members in our previous analysis have published radial velocity measurements in the literature. Figure~\ref{dist_RV} shows the distribution of radial velocities in our sample. We flag the radial velocities for ten~stars as outliers based on the interquartile range (IQR) criterium. These measurements lie over $1.5*IQR$ below the first quartile (Q1) or above the third quartile (Q3) of the distribution, and in many cases they are likely to be affected by binarity. Thus, we discard the radial velocities of LkCa~1, Anon~1, XEST~20-066, LkCa~3 (V1098~Tau), Hubble~4 (V1023~Tau), MHO~5, HD~28867, DQ~Tau, 2MASS~J04482128+2927120, and AB~Aur when computing the $UVW$ spatial velocities (but we still retain them as cluster members in the forthcoming discussion based on our previous results from Sect.~\ref{section4}). 

Table~\ref{tab_distStars} lists the individual distances, three-dimensional positions and spatial velocities for the 174 stars that were confirmed as cluster members (i.e., Member=1 in Table~\ref{tab1}). We also provide in this table the spatial velocities $uvw$ corrected for the velocity of the Sun relative to the local standard of rest (LSR) using the solar  motion of  $(U,V,W)_{\odot}=(11.10^{+0.69}_{-0.75},12.24^{+0.47}_{-0.47},7.25^{+0.37}_{-0.36})$~km/s from \citet{Schonrich2010}. The formal uncertainties on the distances and spatial velocities provided in this paper are computed from the 16\% and 84\% quantiles of the corresponding distributions, which roughly provide us with a 1$\sigma$ standard deviation. Although recent studies based on Bayesian inference \citep[e.g.,][]{Bailer-Jones2015} recommend  using a 90\% confidence interval (e.g., 5\% and 95\% quantiles) we  decided to proceed as explained above to better compare our results with previous studies that only present  the 1$\sigma$ uncertainty on their results. 

We provide in Tables~\ref{tab_distClusters} and \ref{tab_velClusters} the distance estimate and the mean spatial velocity of the stars projected towards the molecular clouds represented in our sample. The Bayesian distance for each cloud is computed from the individual parallaxes and their uncertainties based on the online tutorials for inference of cluster distance available in the Gaia archive \citep[see also][]{Luri2018} and using the same prior over the distance as before. We also  list  the distances obtained by the more common approach of simply inverting the mean parallax of each molecular cloud. In the latter case it is important to take into account the  possible systematic errors in the Gaia-DR2 parallaxes that largely dominate our sample. Although we have already included the systematic errors of 0.1~mas in the uncertainties of Gaia-DR2 parallaxes (as described in Sect.~\ref{section3}), this effect is still present in the mean parallaxes listed in Table~\ref{tab_allClusters} in the sense that averaging the individual parallaxes of cloud members will not reduce the final uncertainties below the 0.1~mas level. We note that the uncertainties on the mean parallaxes given in Table~\ref{tab_allClusters} are much smaller than the the systematic error of 0.1~mas for most clusters (i.e., molecular clouds) in our sample. In these cases we used 0.1 mas as the uncertainty for the mean parallax to estimate the (asymmetric) uncertainties in the distances derived from the inversion method.

\begin{figure*}
\begin{center}
\includegraphics[width=0.25\textwidth]{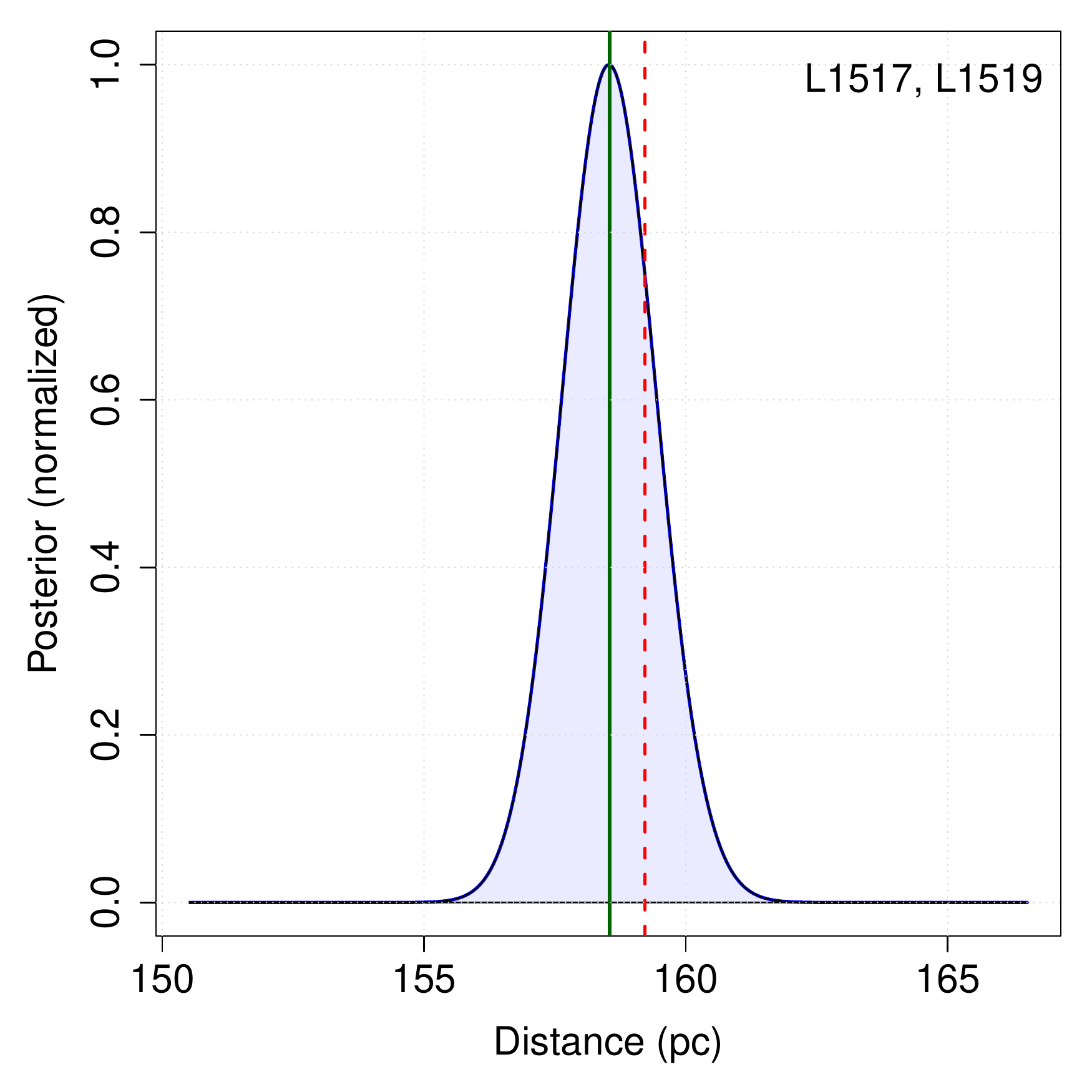}
\includegraphics[width=0.25\textwidth]{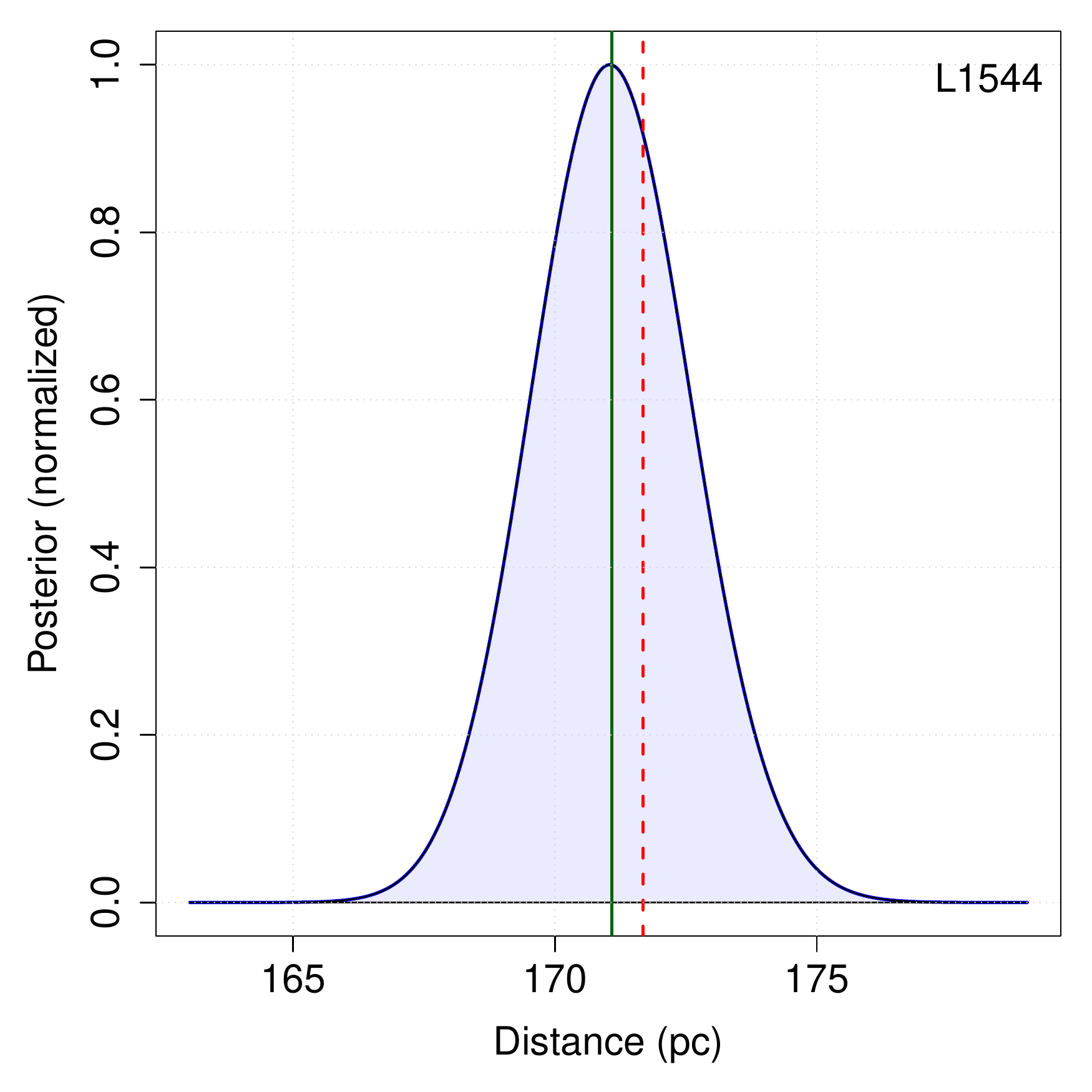}
\includegraphics[width=0.25\textwidth]{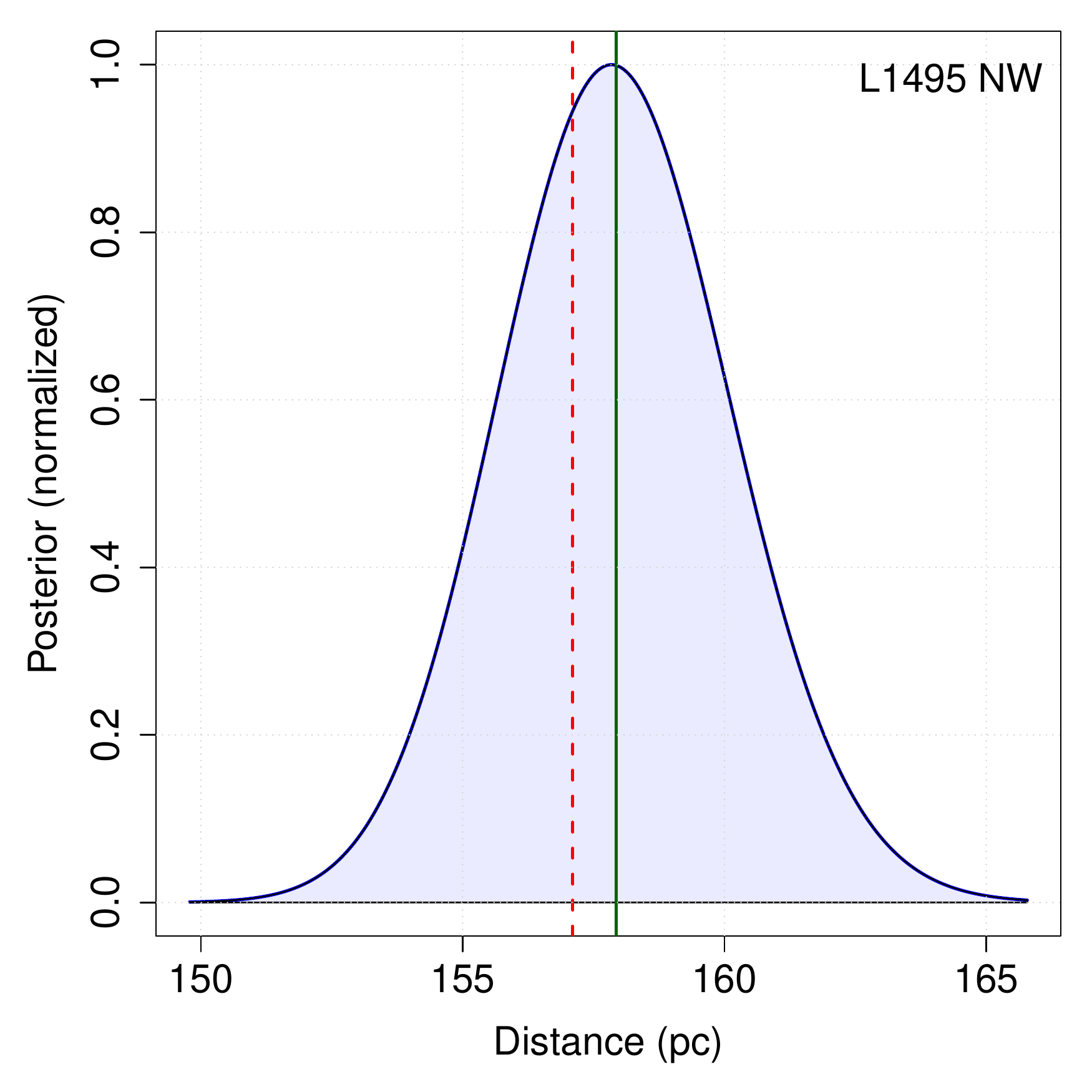}
\includegraphics[width=0.25\textwidth]{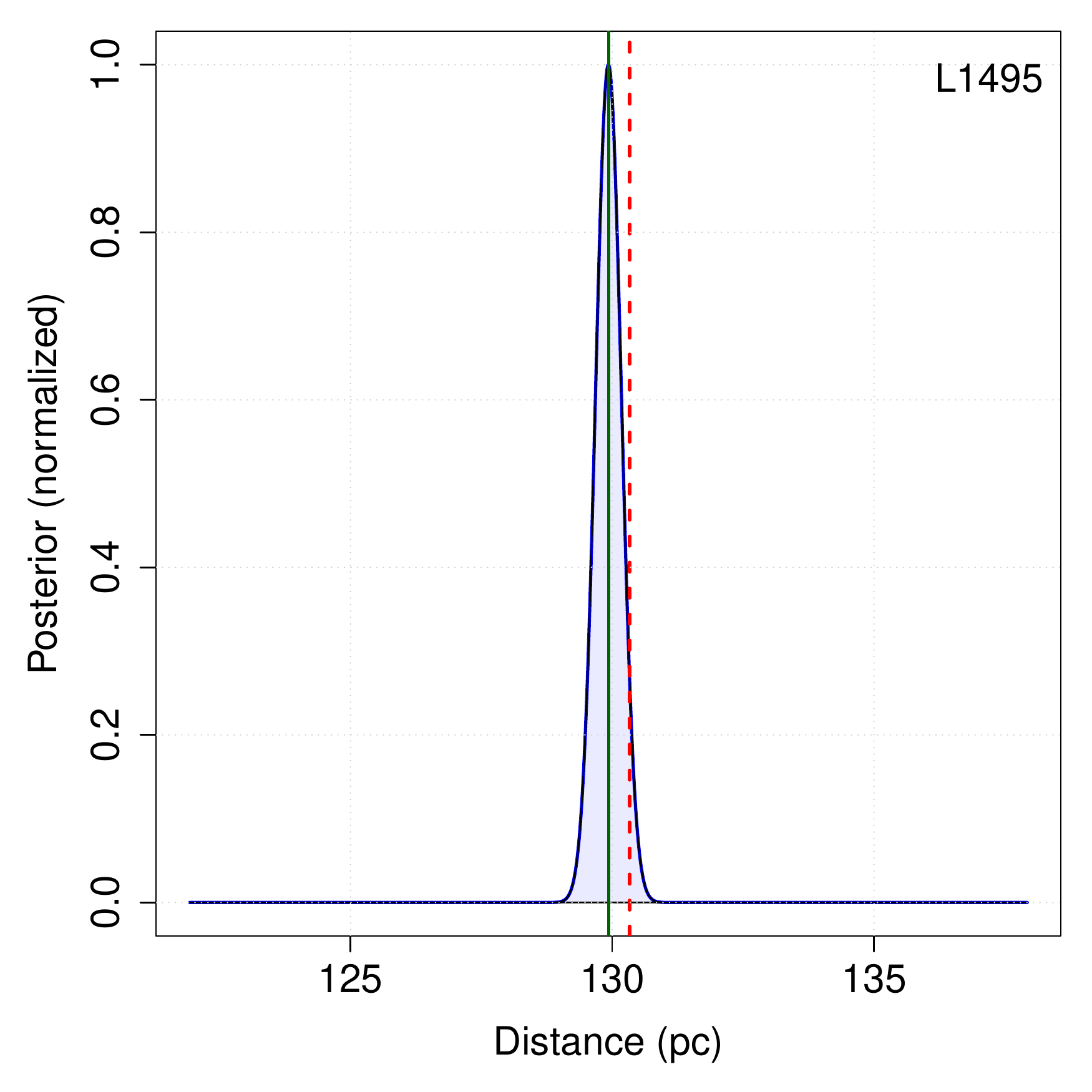}
\includegraphics[width=0.25\textwidth]{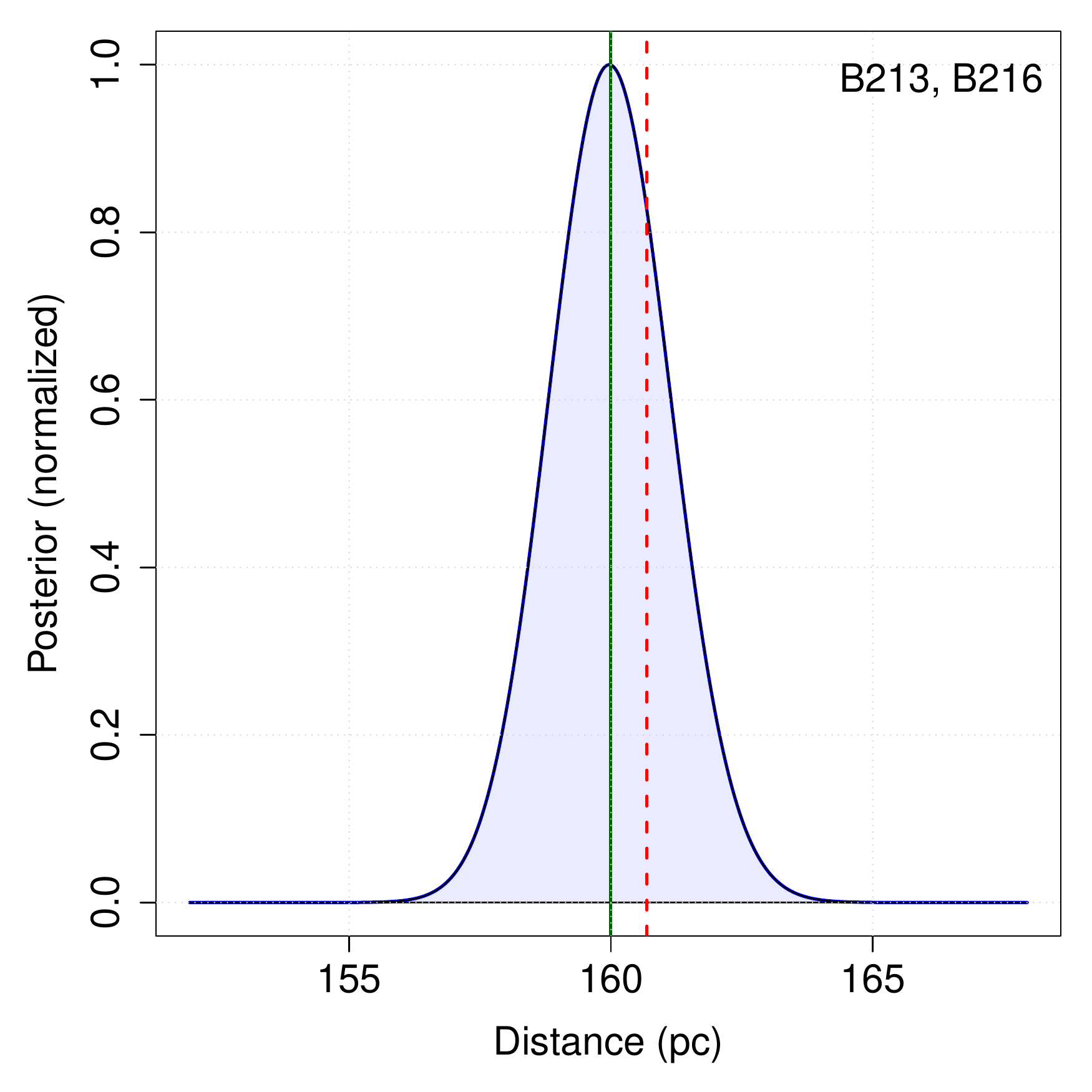}
\includegraphics[width=0.25\textwidth]{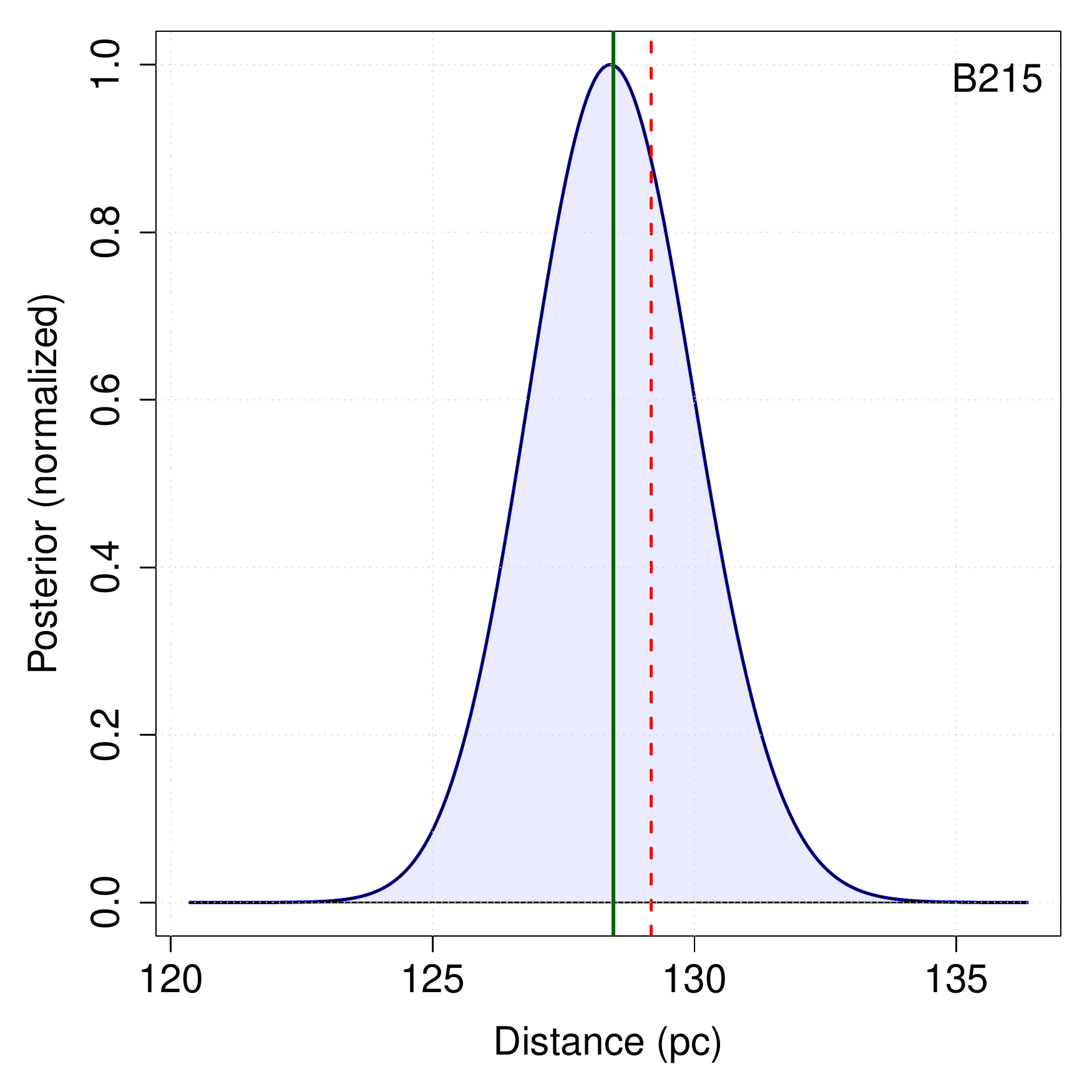}
\includegraphics[width=0.25\textwidth]{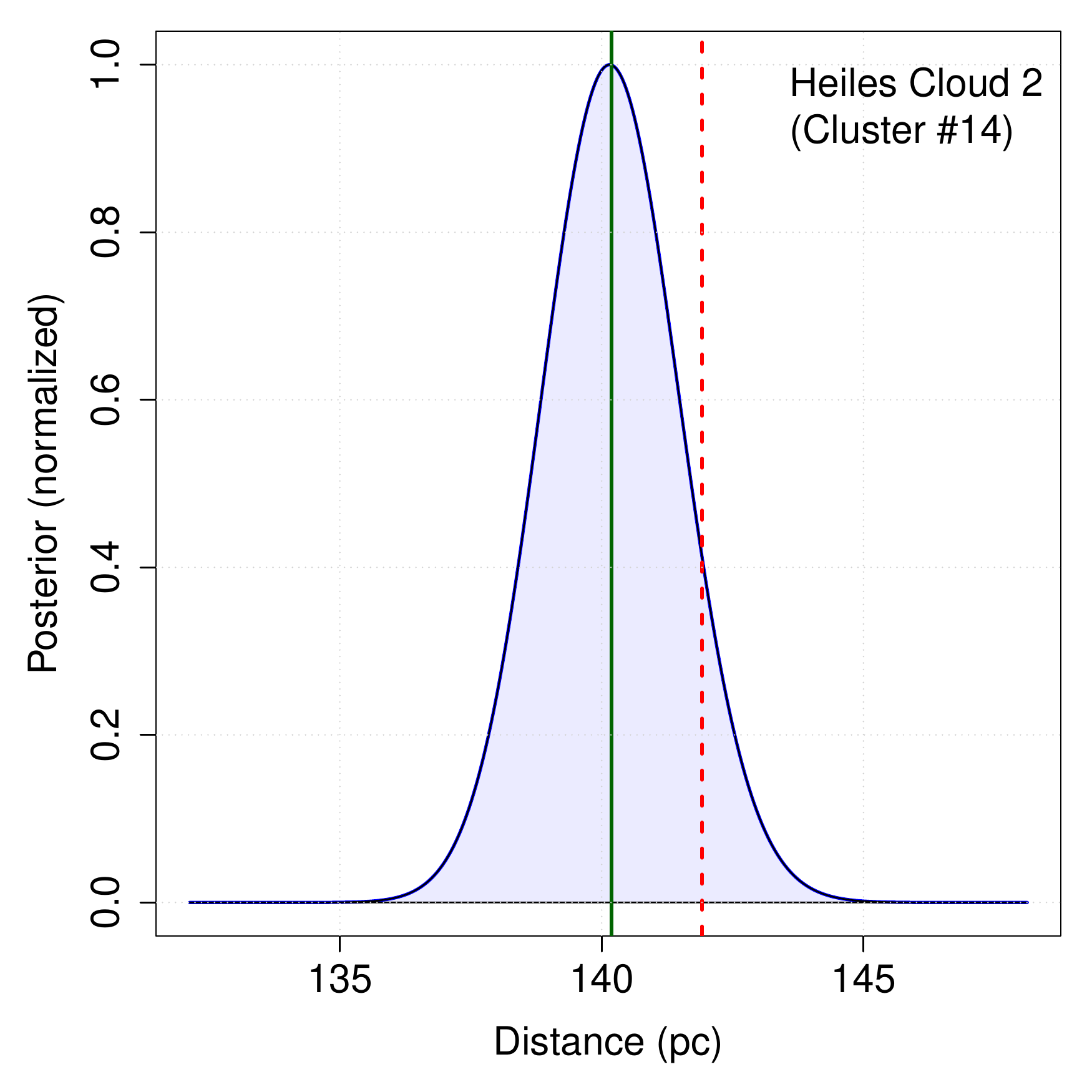}
\includegraphics[width=0.25\textwidth]{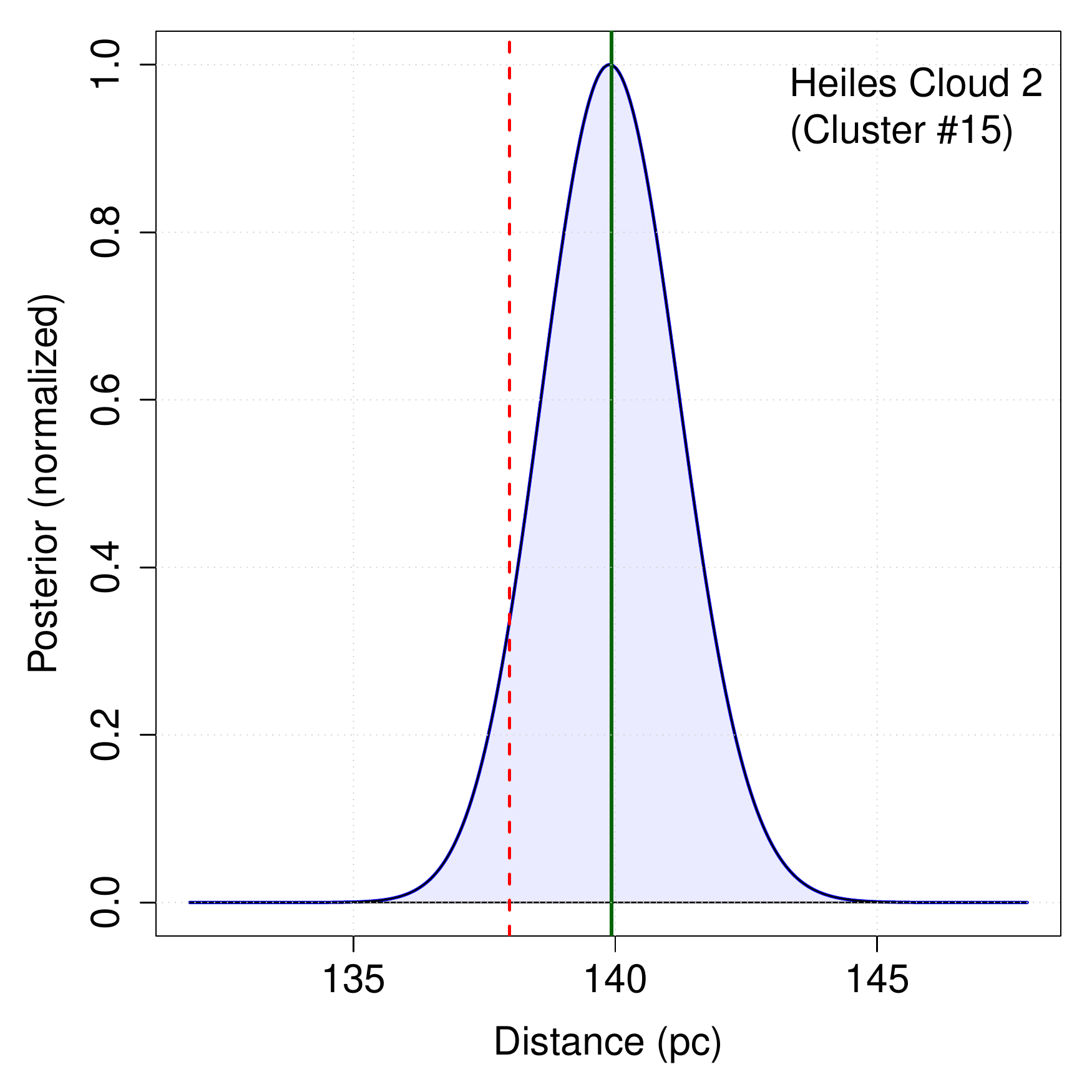}
\includegraphics[width=0.25\textwidth]{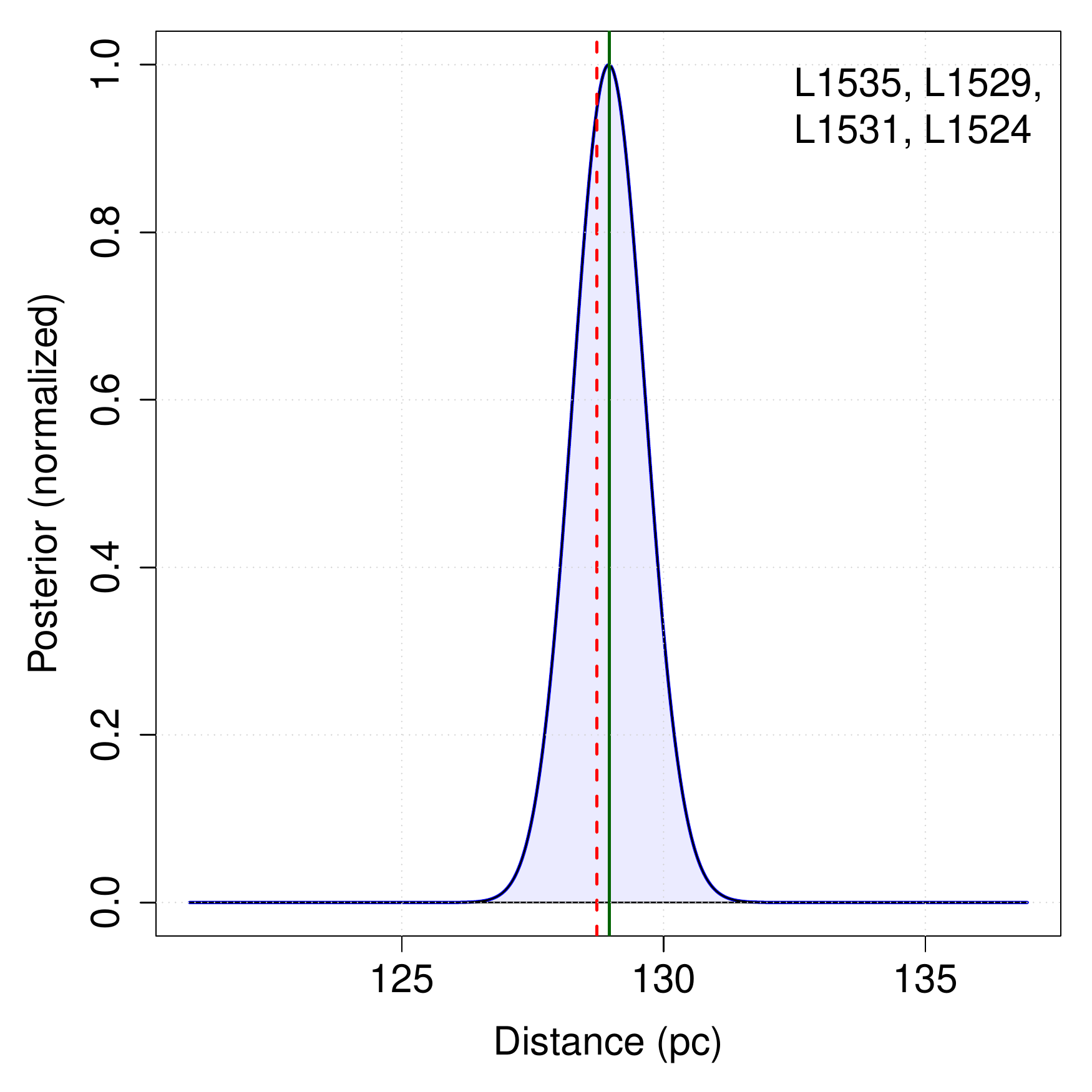}
\includegraphics[width=0.25\textwidth]{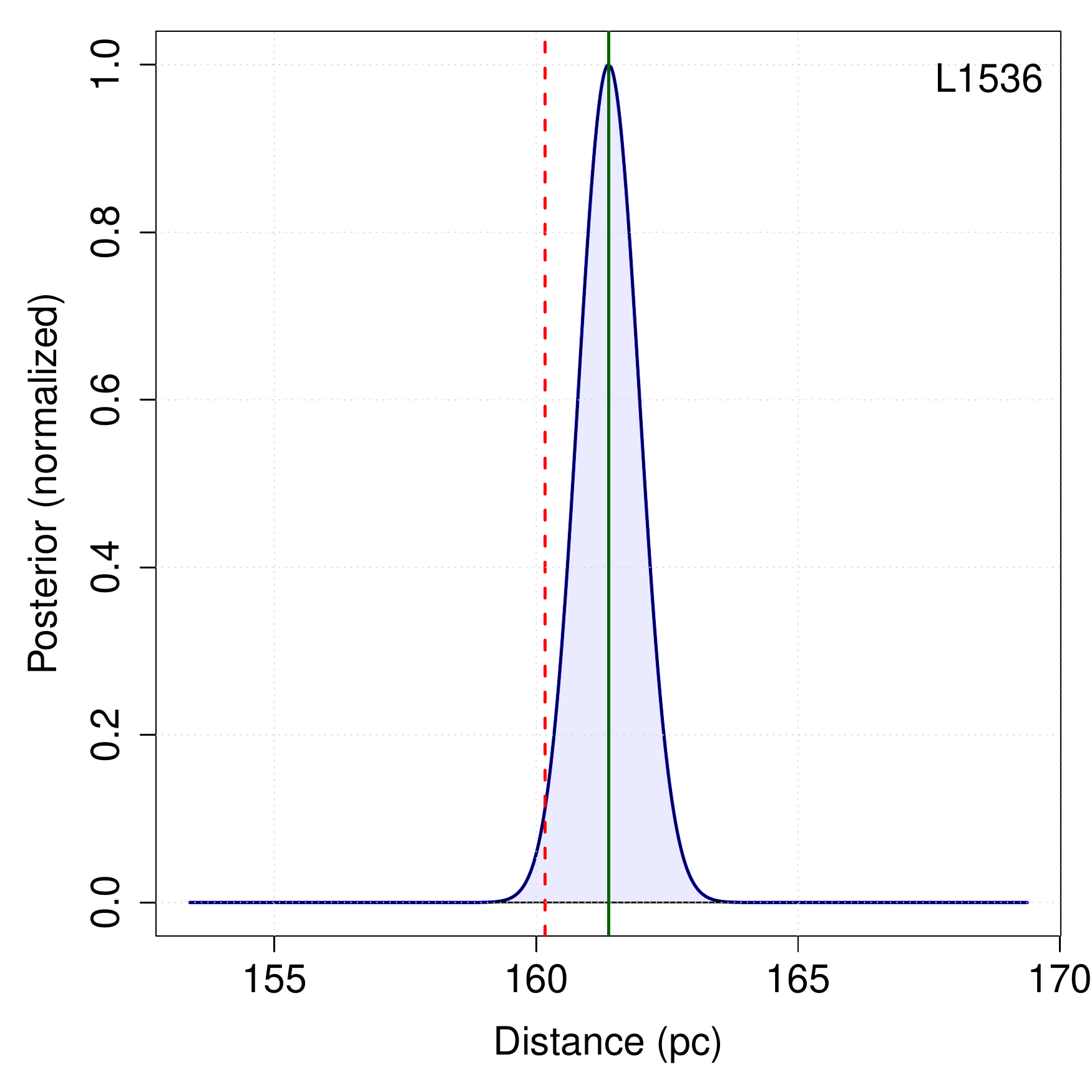}
\includegraphics[width=0.25\textwidth]{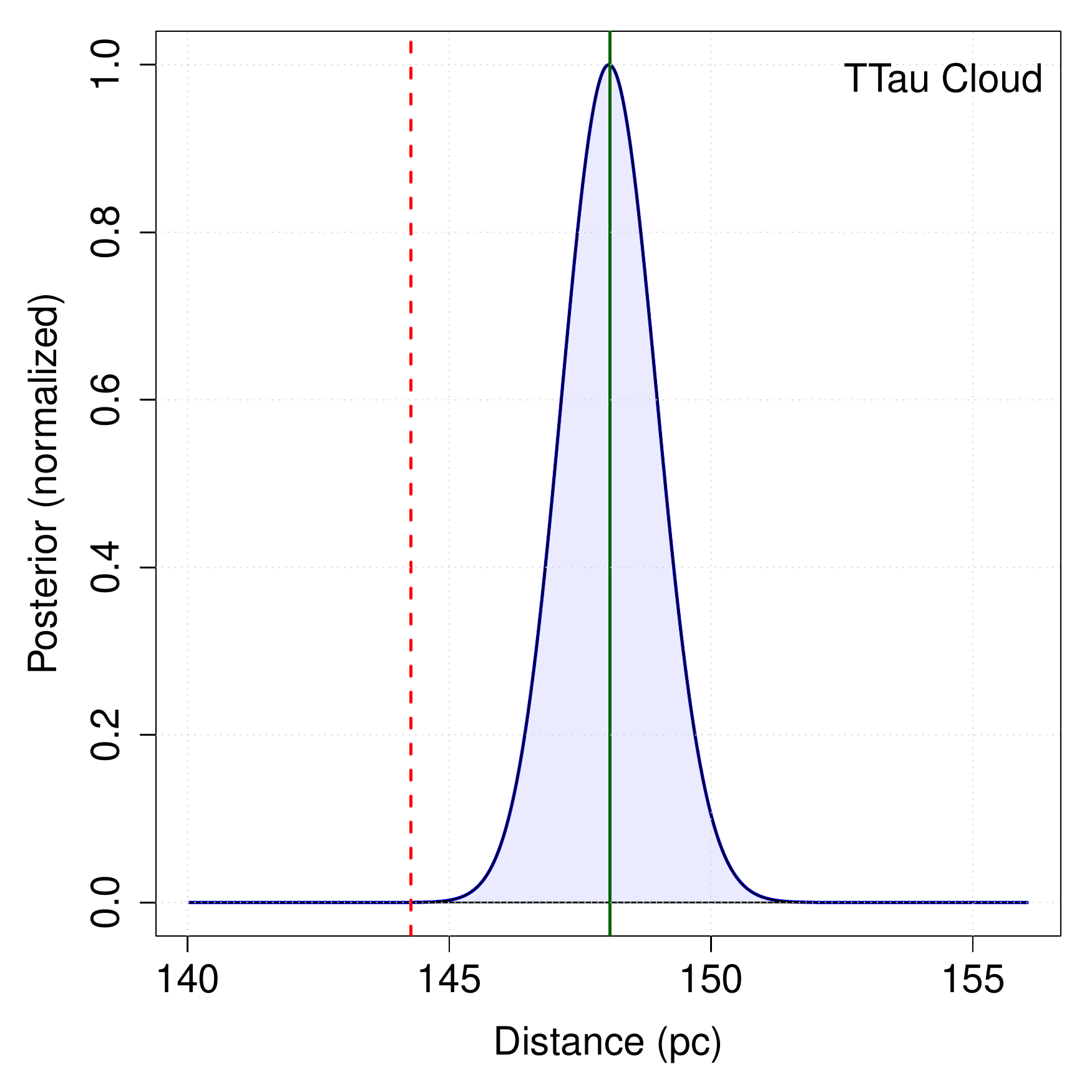}
\includegraphics[width=0.25\textwidth]{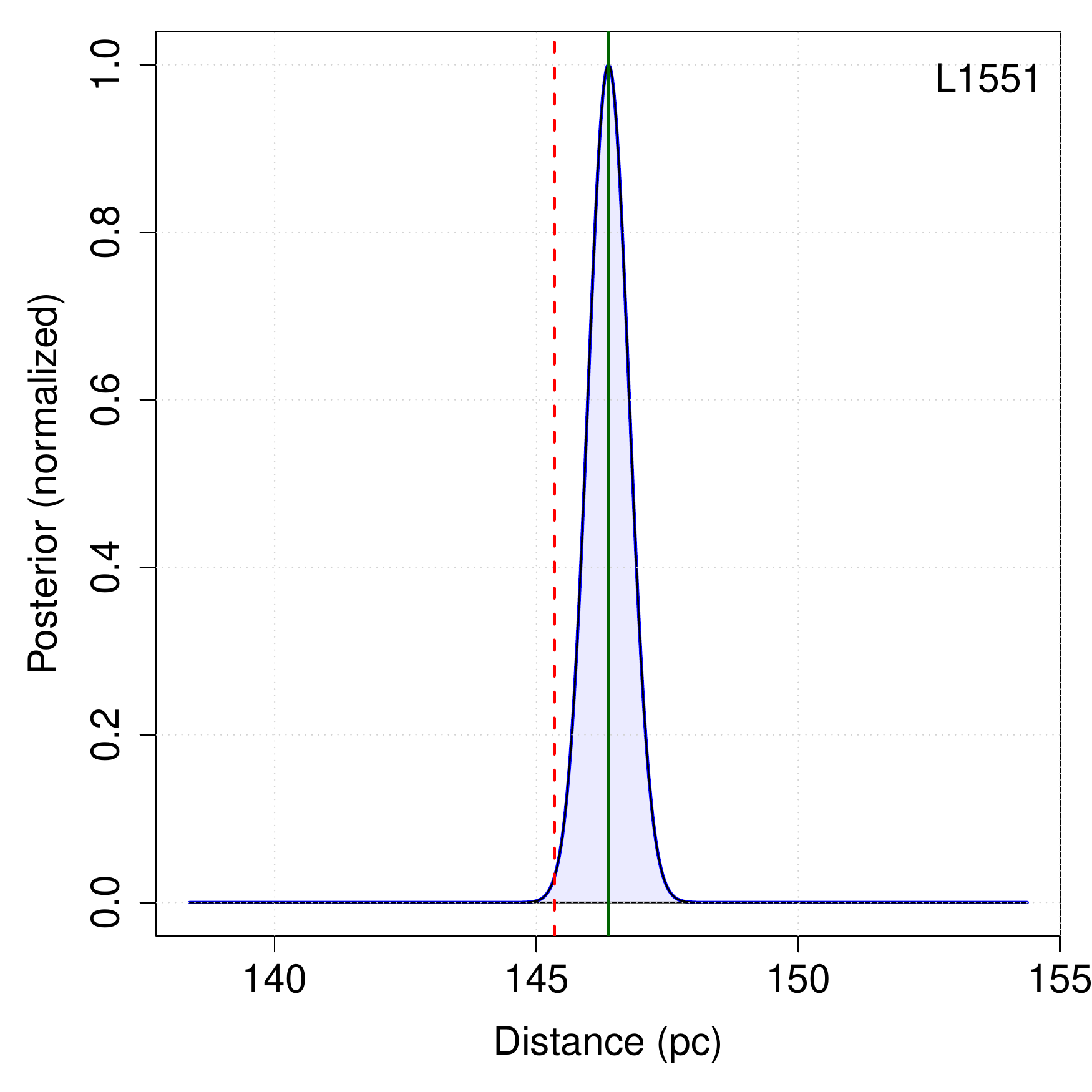}
\includegraphics[width=0.25\textwidth]{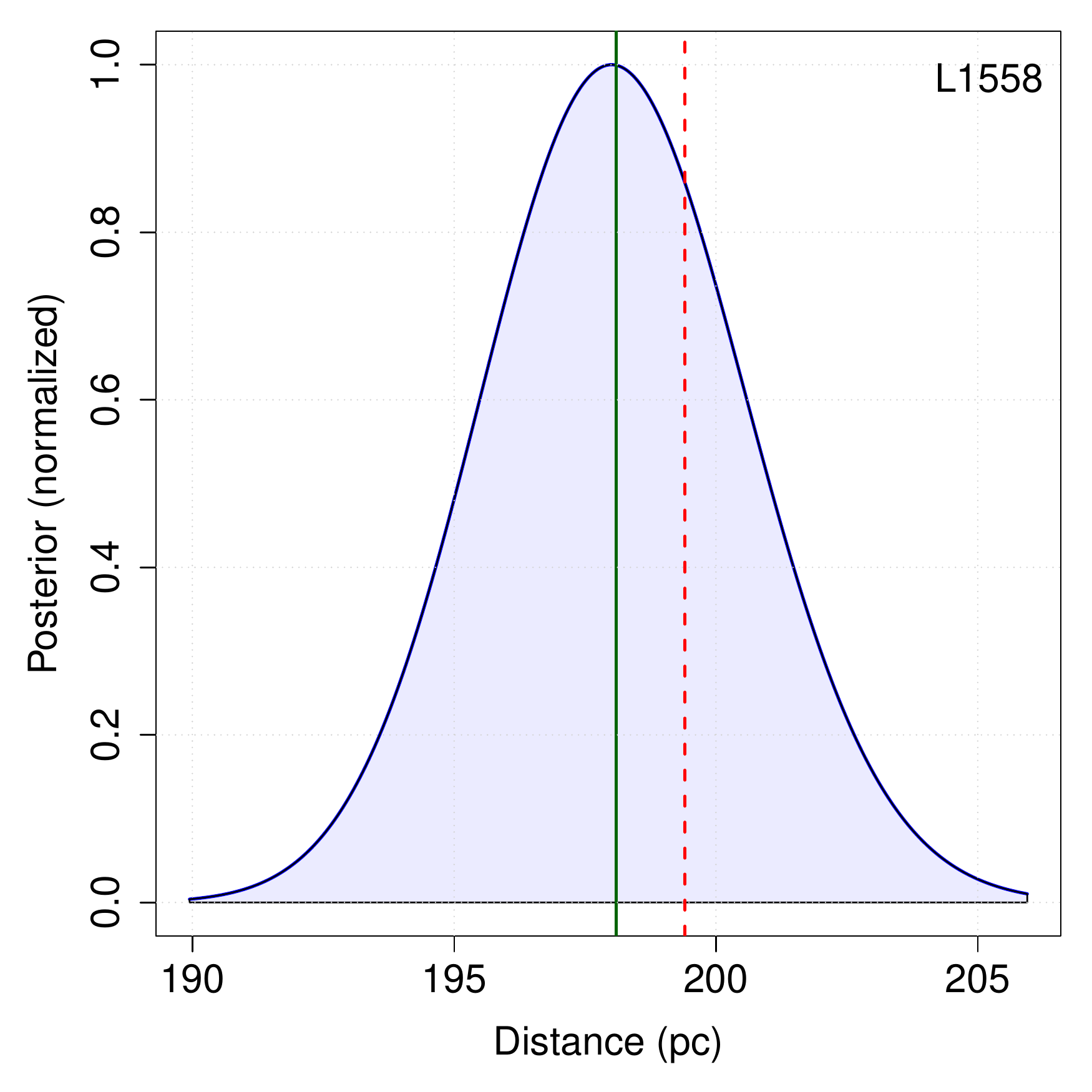}
\caption{Posterior probability density function of the distance to the various molecular clouds of the Taurus complex.  The solid and dashed lines indicate, respectively, the distances obtained from the Bayesian approach (see Sect.~\ref{section5}) and  by inverting the mean parallax. 
\label{distance_pdf} 
}
\end{center}
\end{figure*}

Figure~\ref{distance_pdf}  shows the posterior probability density function obtained for each sample of stars associated with a molecular cloud together with the distance estimates given in Table~\ref{tab_distClusters}. Interestingly, we note that the posterior probability distribution of the various clouds exhibit somewhat different shapes. For example, L~1495, the most populated cloud in the sample, has a very narrow distribution (e.g., compared  with L~1495~NW), which indeed gives the most precise distance estimate in our analysis. Here we report the Bayesian estimates given in Table~\ref{tab_distClusters} as our final results for the distance, because this methodology allows for a proper handling of the uncertainties in our data. 


\begin{landscape}
\renewcommand\thetable{4} 
\begin{table}
\caption{Distance and spatial velocity for the 174 confirmed cluster members. 
\label{tab_distStars}
}
\scriptsize{
\begin{tabular}{lllcccccccccccc}
\hline\hline
2MASS Identifier&Gaia-DR2 Identifier&Other Identifier&$d_{1/\varpi}$&$d$&$X$&$Y$&$Z$&$U$&$V$&$W$&$u$&$v$&$w$&$V_{r}^{LSR}$\\
&&&(pc)&(pc)&(pc)&(pc)&(pc)&(km/s)&(km/s)&(km/s)&(km/s)&(km/s)&(km/s)&(km/s)\\
\hline\hline

2MASS~J04131414+2819108 & Gaia DR2 163246832135164544 & LkCa~1 & $ 128.4 _{ -1.9 }^{+ 1.9 }$& $ 128.5 _{ -2.0 }^{+ 1.8 }$& $ -120.4 _{ -2.0 }^{+ 1.6 }$& $ 25.7 _{ -0.4 }^{+ 0.4 }$& $ -36.3 _{ -0.6 }^{+ 0.5 }$& $ -10.1 _{ -0.2 }^{+ 0.2 }$& $ -12.4 _{ -0.3 }^{+ 0.3 }$& $ -9.1 _{ -0.3 }^{+ 0.3 }$& $ 1.0 _{ -0.8 }^{+ 0.7 }$& $ -0.2 _{ -0.6 }^{+ 0.6 }$& $ -1.8 _{ -0.5 }^{+ 0.5 }$& $ 0.81 \pm 0.12 $\\
2MASS~J04132722+2816247 & Gaia DR2 163233981593016064 & Anon~1 & $ 135.7 _{ -2.2 }^{+ 2.3 }$& $ 135.9 _{ -2.2 }^{+ 2.2 }$& $ -127.6 _{ -2.1 }^{+ 2.1 }$& $ 27.0 _{ -0.4 }^{+ 0.5 }$& $ -38.4 _{ -0.7 }^{+ 0.6 }$& $ -21.6 _{ -0.3 }^{+ 0.3 }$& $ -10.2 _{ -0.4 }^{+ 0.4 }$& $ -12.9 _{ -0.4 }^{+ 0.4 }$& $ -10.5 _{ -0.8 }^{+ 0.8 }$& $ 2.0 _{ -0.6 }^{+ 0.6 }$& $ -5.7 _{ -0.5 }^{+ 0.5 }$& $ 13.13 \pm 0.28 $\\
2MASS~J04141188+2811535 & Gaia DR2 163182888662060928 &  & $ 130.6 _{ -3.2 }^{+ 3.4 }$& $ 130.9 _{ -3.4 }^{+ 3.7 }$& $ -122.7 _{ -3.6 }^{+ 3.0 }$& $ 25.6 _{ -0.6 }^{+ 0.7 }$& $ -36.8 _{ -1.0 }^{+ 0.9 }$& \nodata & \nodata & \nodata & \nodata & \nodata & \nodata & \nodata \\
2MASS~J04141291+2812124 & Gaia DR2 163184366130809984 & V773~Tau~A+B & $ 130.0 _{ -1.4 }^{+ 1.5 }$& $ 130.1 _{ -1.5 }^{+ 1.4 }$& $ -122.1 _{ -1.6 }^{+ 1.2 }$& $ 25.5 _{ -0.3 }^{+ 0.4 }$& $ -36.6 _{ -0.5 }^{+ 0.4 }$& $ -16.5 _{ -2.6 }^{+ 2.6 }$& $ -12.5 _{ -1.2 }^{+ 1.2 }$& $ -10.3 _{ -1.4 }^{+ 1.4 }$& $ -5.4 _{ -2.7 }^{+ 2.7 }$& $ -0.3 _{ -1.3 }^{+ 1.3 }$& $ -3.1 _{ -1.4 }^{+ 1.4 }$& $ 7.18 \pm 2.50 $\\
2MASS~J04141358+2812492 & Gaia DR2 163184366130809472 & FM~Tau & $ 131.4 _{ -1.9 }^{+ 1.9 }$& $ 131.8 _{ -1.9 }^{+ 1.9 }$& $ -123.5 _{ -1.7 }^{+ 1.9 }$& $ 25.8 _{ -0.4 }^{+ 0.4 }$& $ -37.0 _{ -0.5 }^{+ 0.6 }$& \nodata & \nodata & \nodata & \nodata & \nodata & \nodata & \nodata \\
2MASS~J04141458+2827580 & Gaia DR2 164738521519622656 & FN~Tau & $ 130.7 _{ -2.0 }^{+ 2.0 }$& $ 131.0 _{ -2.1 }^{+ 1.8 }$& $ -122.9 _{ -2.1 }^{+ 1.8 }$& $ 26.1 _{ -0.4 }^{+ 0.4 }$& $ -36.4 _{ -0.6 }^{+ 0.5 }$& $ -15.9 _{ -0.2 }^{+ 0.2 }$& $ -11.6 _{ -0.4 }^{+ 0.3 }$& $ -9.9 _{ -0.3 }^{+ 0.4 }$& $ -4.8 _{ -0.8 }^{+ 0.7 }$& $ 0.6 _{ -0.6 }^{+ 0.6 }$& $ -2.6 _{ -0.5 }^{+ 0.5 }$& $ 6.67 \pm 0.09 $\\
2MASS~J04141700+2810578 & Gaia DR2 163184091252903936 & CW~Tau & $ 131.9 _{ -1.8 }^{+ 1.9 }$& $ 132.1 _{ -1.7 }^{+ 2.0 }$& $ -123.9 _{ -2.0 }^{+ 1.6 }$& $ 25.8 _{ -0.3 }^{+ 0.4 }$& $ -37.1 _{ -0.6 }^{+ 0.5 }$& $ -13.9 _{ -0.2 }^{+ 0.2 }$& $ -12.0 _{ -0.3 }^{+ 0.3 }$& $ -10.2 _{ -0.3 }^{+ 0.3 }$& $ -2.8 _{ -0.8 }^{+ 0.7 }$& $ 0.2 _{ -0.6 }^{+ 0.6 }$& $ -3.0 _{ -0.5 }^{+ 0.5 }$& $ 4.77 \pm 0.10 $\\
2MASS~J04141760+2806096 & Gaia DR2 163181342473839744 & CIDA~1 & $ 135.1 _{ -2.4 }^{+ 2.5 }$& $ 135.5 _{ -2.4 }^{+ 2.4 }$& $ -127.0 _{ -2.5 }^{+ 2.0 }$& $ 26.3 _{ -0.4 }^{+ 0.5 }$& $ -38.2 _{ -0.8 }^{+ 0.6 }$& $ -15.5 _{ -0.2 }^{+ 0.2 }$& $ -11.7 _{ -0.4 }^{+ 0.4 }$& $ -10.4 _{ -0.4 }^{+ 0.4 }$& $ -4.4 _{ -0.8 }^{+ 0.7 }$& $ 0.5 _{ -0.6 }^{+ 0.6 }$& $ -3.2 _{ -0.6 }^{+ 0.6 }$& $ 6.37 \pm 0.14 $\\
2MASS~J04142639+2805597 & Gaia DR2 163178353176600448 & MHO~2 & $ 132.1 _{ -4.1 }^{+ 4.3 }$& $ 132.4 _{ -3.8 }^{+ 4.2 }$& $ -124.2 _{ -4.4 }^{+ 3.4 }$& $ 25.7 _{ -0.7 }^{+ 0.9 }$& $ -37.3 _{ -1.4 }^{+ 1.1 }$& \nodata & \nodata & \nodata & \nodata & \nodata & \nodata & \nodata \\
2MASS~J04144739+2803055 & Gaia DR2 163177116226018944 & XEST~20-066 & $ 128.2 _{ -2.0 }^{+ 2.1 }$& $ 128.4 _{ -2.2 }^{+ 2.1 }$& $ -120.5 _{ -2.0 }^{+ 1.9 }$& $ 24.7 _{ -0.4 }^{+ 0.4 }$& $ -36.1 _{ -0.6 }^{+ 0.6 }$& $ -13.2 _{ -0.3 }^{+ 0.3 }$& $ -13.4 _{ -0.4 }^{+ 0.4 }$& $ -9.7 _{ -0.4 }^{+ 0.4 }$& $ -2.1 _{ -0.8 }^{+ 0.8 }$& $ -1.2 _{ -0.6 }^{+ 0.6 }$& $ -2.5 _{ -0.6 }^{+ 0.6 }$& $ 3.71 \pm 0.28 $\\
2MASS~J04144797+2752346 & Gaia DR2 163157325014927360 & LkCa~3A~+B & $ 123.9 _{ -4.6 }^{+ 5.0 }$& $ 125.5 _{ -4.8 }^{+ 5.1 }$& $ -116.8 _{ -4.9 }^{+ 4.0 }$& $ 23.6 _{ -0.8 }^{+ 1.0 }$& $ -35.2 _{ -1.5 }^{+ 1.2 }$& $ -9.8 _{ -0.3 }^{+ 0.3 }$& $ -14.7 _{ -0.8 }^{+ 0.8 }$& $ -8.7 _{ -0.8 }^{+ 0.7 }$& $ 1.3 _{ -0.8 }^{+ 0.8 }$& $ -2.5 _{ -0.9 }^{+ 0.9 }$& $ -1.4 _{ -0.8 }^{+ 0.8 }$& $ -0.03 \pm 0.18 $\\
2MASS~J04145234+2805598 & Gaia DR2 163179006011625088 & XEST~20-071~A & $ 134.7 _{ -2.4 }^{+ 2.5 }$& $ 135.3 _{ -2.5 }^{+ 2.3 }$& $ -126.7 _{ -2.4 }^{+ 2.3 }$& $ 26.0 _{ -0.5 }^{+ 0.5 }$& $ -37.8 _{ -0.7 }^{+ 0.7 }$& $ -17.7 _{ -0.2 }^{+ 0.2 }$& $ -10.5 _{ -0.4 }^{+ 0.4 }$& $ -10.7 _{ -0.4 }^{+ 0.4 }$& $ -6.6 _{ -0.8 }^{+ 0.7 }$& $ 1.7 _{ -0.6 }^{+ 0.6 }$& $ -3.5 _{ -0.5 }^{+ 0.5 }$& $ 8.79 \pm 0.12 $\\
2MASS~J04150651+2728136 & Gaia DR2 162933226506362240 &  & $ 133.1 _{ -2.0 }^{+ 2.1 }$& $ 133.3 _{ -1.9 }^{+ 2.1 }$& $ -125.0 _{ -2.2 }^{+ 1.8 }$& $ 24.5 _{ -0.3 }^{+ 0.4 }$& $ -38.2 _{ -0.7 }^{+ 0.6 }$& \nodata & \nodata & \nodata & \nodata & \nodata & \nodata & \nodata \\
2MASS~J04151471+2800096 & Gaia DR2 163165738856771200 & KPNO~1 & $ 126.4 _{ -10.2 }^{+ 12.2 }$& $ 132.6 _{ -11.5 }^{+ 14.6 }$& $ -120.5 _{ -13.9 }^{+ 8.7 }$& $ 24.4 _{ -2.0 }^{+ 3.1 }$& $ -36.1 _{ -4.7 }^{+ 3.0 }$& \nodata & \nodata & \nodata & \nodata & \nodata & \nodata & \nodata \\
2MASS~J04153916+2818586 & Gaia DR2 164684340508950144 &  & $ 131.0 _{ -2.2 }^{+ 2.2 }$& $ 131.3 _{ -2.1 }^{+ 2.2 }$& $ -123.2 _{ -2.3 }^{+ 1.8 }$& $ 25.4 _{ -0.4 }^{+ 0.5 }$& $ -36.1 _{ -0.7 }^{+ 0.5 }$& $ -16.1 _{ -0.2 }^{+ 0.2 }$& $ -12.1 _{ -0.4 }^{+ 0.4 }$& $ -10.3 _{ -0.4 }^{+ 0.4 }$& $ -5.0 _{ -0.8 }^{+ 0.7 }$& $ 0.2 _{ -0.6 }^{+ 0.6 }$& $ -3.0 _{ -0.5 }^{+ 0.5 }$& $ 6.84 \pm 0.11 $\\
2MASS~J04154131+2915078 & Gaia DR2 164802984685384320 &  & $ 156.3 _{ -4.0 }^{+ 4.2 }$& $ 157.1 _{ -4.0 }^{+ 3.9 }$& $ -147.7 _{ -4.2 }^{+ 3.7 }$& $ 32.3 _{ -0.8 }^{+ 0.9 }$& $ -41.6 _{ -1.2 }^{+ 1.1 }$& \nodata & \nodata & \nodata & \nodata & \nodata & \nodata & \nodata \\
2MASS~J04154269+2909558 & Gaia DR2 164800235906367232 &  & $ 155.8 _{ -4.9 }^{+ 5.3 }$& $ 156.6 _{ -5.1 }^{+ 5.5 }$& $ -147.4 _{ -5.4 }^{+ 4.7 }$& $ 32.0 _{ -1.0 }^{+ 1.2 }$& $ -41.7 _{ -1.7 }^{+ 1.4 }$& \nodata & \nodata & \nodata & \nodata & \nodata & \nodata & \nodata \\
2MASS~J04154278+2909597 & Gaia DR2 164800235906366976 & IRAS~04125+2902 & $ 159.2 _{ -3.0 }^{+ 3.1 }$& $ 159.4 _{ -3.1 }^{+ 3.0 }$& $ -149.9 _{ -3.0 }^{+ 2.6 }$& $ 32.6 _{ -0.6 }^{+ 0.7 }$& $ -42.4 _{ -0.9 }^{+ 0.8 }$& \nodata & \nodata & \nodata & \nodata & \nodata & \nodata & \nodata \\
2MASS~J04155799+2746175 & Gaia DR2 162967384383246336 &  & $ 135.1 _{ -2.6 }^{+ 2.7 }$& $ 135.5 _{ -2.6 }^{+ 2.6 }$& $ -127.5 _{ -2.6 }^{+ 2.3 }$& $ 25.2 _{ -0.5 }^{+ 0.5 }$& $ -38.1 _{ -0.8 }^{+ 0.7 }$& $ -15.3 _{ -0.3 }^{+ 0.2 }$& $ -12.3 _{ -0.5 }^{+ 0.4 }$& $ -11.1 _{ -0.5 }^{+ 0.5 }$& $ -4.2 _{ -0.8 }^{+ 0.7 }$& $ -0.0 _{ -0.7 }^{+ 0.6 }$& $ -3.9 _{ -0.6 }^{+ 0.6 }$& $ 6.30 \pm 0.16 $\\
2MASS~J04161210+2756385 & Gaia DR2 164474986623118592 &  & $ 136.9 _{ -2.8 }^{+ 2.9 }$& $ 137.0 _{ -2.4 }^{+ 2.7 }$& $ -128.8 _{ -3.6 }^{+ 2.1 }$& $ 25.7 _{ -0.4 }^{+ 0.7 }$& $ -38.2 _{ -1.1 }^{+ 0.6 }$& $ -16.6 _{ -0.2 }^{+ 0.2 }$& $ -13.2 _{ -0.5 }^{+ 0.5 }$& $ -11.5 _{ -0.5 }^{+ 0.5 }$& $ -5.5 _{ -0.8 }^{+ 0.7 }$& $ -1.0 _{ -0.7 }^{+ 0.7 }$& $ -4.2 _{ -0.6 }^{+ 0.6 }$& $ 7.42 \pm 0.14 $\\

\hline\hline
\end{tabular}
\tablefoot{We provide for each star the 2MASS and Gaia-DR2 identifiers, distance obtained by inverting the parallax, distance derived from the Bayesian approach (see Sect.~\ref{section5}), $XYZ$ positions, $UVW$ components of the Galactic spatial velocity, $uvw$ components of the peculiar velocity (with respect to the LSR), and radial velocity corrected for the solar motion. (This table will be available in its entirety in machine readable form.) }
}
\end{table}
\end{landscape}
\clearpage

Our results show that the complex of clouds formed by L~1535, L~1529, L~1531, L~1524 and the B~215 clump are the closest structures to the Sun in the Taurus region ($d=129.0^{+0.8}_{-0.8}$ and $d=128.5^{+1.6}_{-1.6}$, respectively). This is consistent with the distance of $d=126.6^{+1.7}_{-1.7}$~pc obtained previously by \citet{Galli2018} for L~1531 based on the VLBI trigonometric parallax of V807~Tau. The GOBELINS survey in Taurus targeted the central and southern clouds of the complex, so \citet{Galli2018} presented L~1536 as the farthest cloud in the region based on the VLBI parallax of HP~Tau~G2 ($d=162.7^{+0.8}_{-0.8}$~pc). This distance estimate is in very good agreement with the result of $d=161.4^{+0.7}_{-0.7}$~pc that we derive in this study for L~1536, but our current analysis in this paper suggests that L~1558 ($d=198.1^{+2.5}_{-2.5}$~pc) should  hereafter be considered as the most remote molecular cloud in Taurus. In general, we see a good agreement between the results reported in the two studies. The only exception is L~1519 for which \citet{Galli2018} used the Gaia-DR1 parallaxes of three stars and the VLBI parallax of HD~282630 (see discussion of this source in Sect.~\ref{section3}) to estimate the distance to the cloud. The reported distance of $d=142.1^{+2.4}_{-2.3}$~pc is not consistent with the new result that we derive in this paper using the more accurate and precise Gaia-DR2 parallaxes. The molecular cloud L~1513 listed in Table~10 of that study is not discussed here, because the only source projected towards this cloud (namely UY~Aur) was flagged as a potential outlier in our clustering analysis presented in Sect.~\ref{section4}. 

\renewcommand\thetable{5} 
\begin{table*}[!h]
\centering
\caption{Distance of the Taurus molecular clouds. 
\label{tab_distClusters}
}
\scriptsize{
\begin{tabular}{lcccccccc}
\hline
\hline
Molecular Cloud&Cluster&$N$&\multicolumn{2}{c}{Distance}&\multicolumn{3}{c}{$X_{0},Y_{0},Z_{0}$}\\
&&&\multicolumn{2}{c}{(pc)}&\multicolumn{3}{c}{(pc)}\\
\hline
&&&Inversion&Bayesian&Mean$\pm$SEM&Median&SD\\
\hline\hline
L~1517, L~1519 & 1 & 14 & $ 159.2 _{ -2.5 }^{+ 2.6 }$& $ 158.5 _{ -1.0 }^{+ 1.0 }$& $( -156.7 , 19.1 , -22.3 )\pm( 1.2 , 0.4 , 0.3 )$&$( -155.6 , 19.4 , -22.1 )$&$( 4.3 , 1.6 , 1.2 )$\\
 L~1544 & 4 & 6 & $ 171.7 _{ -2.9 }^{+ 3.0 }$& $ 171.1 _{ -1.5 }^{+ 1.6 }$& $( -169.5 , 4.9 , -28.4 )\pm( 2.2 , 0.7 , 0.5 )$&$( -168.8 , 4.8 , -28.7 )$&$( 5.4 , 1.7 , 1.1 )$\\
L~1495~NW & 6 & 3 & $ 157.1 _{ -2.4 }^{+ 2.5 }$& $ 157.9 _{ -2.2 }^{+ 2.2 }$& $( -148.3 , 32.3 , -41.9 )\pm( 0.8 , 0.2 , 0.2 )$&$( -147.7 , 32.3 , -41.7 )$&$( 1.4 , 0.3 , 0.4 )$\\
L~1495 & 7 & 39 & $ 130.3 _{ -1.7 }^{+ 1.7 }$& $ 129.9 _{ -0.3 }^{+ 0.4 }$& $( -123.2 , 24.5 , -35.8 )\pm( 0.6 , 0.2 , 0.3 )$&$( -122.7 , 24.5 , -35.9 )$&$( 3.6 , 1.3 , 1.7 )$\\
L~213, B~216 & 8 & 9 & $ 160.7 _{ -2.5 }^{+ 2.6 }$& $ 160.0 _{ -1.2 }^{+ 1.2 }$& $( -153.0 , 24.8 , -44.0 )\pm( 1.7 , 0.7 , 0.7 )$&$( -153.3 , 24.5 , -43.1 )$&$( 5.0 , 2.1 , 2.1 )$\\
B~215 & 9 & 2 & $ 129.2 _{ -1.7 }^{+ 1.8 }$& $ 128.5 _{ -1.6 }^{+ 1.6 }$& $( -122.5 , 16.9 , -37.8 )\pm( 1.9 , 0.4 , 0.5 )$&$( -122.5 , 16.9 , -37.8 )$&$( 2.6 , 0.5 , 0.7 )$\\
Heiles Cloud 2  & 14 & 7 & $ 141.9 _{ -2.0 }^{+ 2.0 }$& $ 140.2 _{ -1.3 }^{+ 1.3 }$& $( -137.5 , 14.3 , -33.4 )\pm( 1.6 , 0.5 , 0.5 )$&$( -136.5 , 14.7 , -32.9 )$&$( 4.3 , 1.3 , 1.3 )$\\
Heiles Cloud 2  & 15 & 5 & $ 138.0 _{ -1.9 }^{+ 1.9 }$& $ 139.9 _{ -1.3 }^{+ 1.3 }$& $( -134.0 , 13.1 , -32.3 )\pm( 1.5 , 0.4 , 0.3 )$&$( -132.1 , 12.8 , -31.9 )$&$( 3.3 , 0.9 , 0.7 )$\\
L~1535, L~1529, L~1531, L~1524 & 16 & 11 & $ 128.7 _{ -1.6 }^{+ 1.7 }$& $ 129.0 _{ -0.8 }^{+ 0.8 }$& $( -123.4 , 12.6 , -35.1 )\pm( 0.7 , 0.2 , 0.3 )$&$( -124.0 , 12.6 , -35.2 )$&$( 2.4 , 0.8 , 1.0 )$\\
L~1536 & 18 & 17 & $ 160.2 _{ -2.5 }^{+ 2.6 }$& $ 161.4 _{ -0.7 }^{+ 0.7 }$& $( -154.6 , 11.9 , -45.8 )\pm( 1.5 , 0.2 , 0.5 )$&$( -153.8 , 12.0 , -45.5 )$&$( 6.3 , 0.8 , 1.9 )$\\
T~Tau cloud & 19 & 4 & $ 144.3 _{ -2.8 }^{+ 2.9 }$& $ 148.1 _{ -1.0 }^{+ 1.0 }$& $( -134.7 , 8.9 , -51.5 )\pm( 2.6 , 0.2 , 1.0 )$&$( -136.3 , 9.0 , -52.1 )$&$( 5.1 , 0.4 , 2.1 )$\\
L~1551 & 20 & 24 & $ 145.3 _{ -2.1 }^{+ 2.1 }$& $ 146.4 _{ -0.5 }^{+ 0.5 }$& $( -136.8 , 1.7 , -49.5 )\pm( 0.6 , 0.2 , 0.2 )$&$( -136.3 , 1.9 , -49.3 )$&$( 3.1 , 1.1 , 1.1 )$\\
L~1558 & 21 & 5 & $ 199.4 _{ -3.9 }^{+ 4.1 }$& $ 198.1 _{ -2.5 }^{+ 2.5 }$& $( -190.4 , -7.3 , -61.0 )\pm( 3.2 , 0.4 , 0.8 )$&$( -188.4 , -7.3 , -60.6 )$&$( 7.1 , 0.8 , 1.7 )$\\
\hline

\end{tabular}
\tablefoot{We provide for each cloud its identifier, corresponding cluster in the HMAC analysis (see Sect.~\ref{section4}), number of stars with measured parallax, distance obtained from the inverse of the mean parallax of the cloud (see Table~\ref{tab_allClusters}), distance obtained from the Bayesian approach (see Sect.~\ref{section5}), mean, standard error of the mean (SEM), median, and standard deviation of the three-dimensional cartesian XYZ coordinates of the cloud center. \vspace{0cm}}
}
\end{table*}

\renewcommand\thetable{6} 
\begin{table*}[!h]
\centering
\caption{Spatial velocity of the Taurus molecular clouds. 
\label{tab_velClusters}
}
\scriptsize{
\begin{tabular}{lcccccccccccccc}
\hline
\hline
Molecular Cloud&Cluster&$N$&\multicolumn{3}{c}{$U$}&\multicolumn{3}{c}{$V$}&\multicolumn{3}{c}{$W$}&\multicolumn{3}{c}{$V_{space}$}\\
&&&\multicolumn{3}{c}{(km/s)}&\multicolumn{3}{c}{(km/s)}&\multicolumn{3}{c}{(km/s)}&\multicolumn{3}{c}{(km/s)}\\
\hline
&&&Mean$\pm$SEM&Median&SD&Mean$\pm$SEM&Median&SD&Mean$\pm$SEM&Median&SD&Mean$\pm$SEM&Median&SD\\
\hline\hline
L~1517, L~1519 & 1 & 2 & $ -14.6 \pm 0.1 $&$ -14.6 $& 0.8 & $ -14.7 \pm 0.3 $&$ -14.7 $& 0.0 & $ -10.3 \pm 0.3 $&$ -10.3 $& 0.2 & $ 23.2 \pm 0.4 $&$ 23.2 $& 0.5 \\
 L~1544 & 4 & 2 & $ -17.6 \pm 0.1 $&$ -17.6 $& 0.4 & $ -12.3 \pm 0.2 $&$ -12.3 $& 0.9 & $ -8.9 \pm 0.2 $&$ -8.9 $& 0.3 & $ 23.2 \pm 0.3 $&$ 23.2 $& 0.7 \\
L~1495~NW & 6 & 0 & \nodata & \nodata & \nodata & \nodata & \nodata & \nodata & \nodata & \nodata & \nodata & \nodata & \nodata & \nodata \\
L~1495 & 7 & 25 & $ -16.1 \pm 0.2 $&$ -15.9 $& 1.1 & $ -12.2 \pm 0.2 $&$ -12.2 $& 0.8 & $ -10.8 \pm 0.1 $&$ -10.8 $& 0.5 & $ 23.0 \pm 0.2 $&$ 22.9 $& 0.9 \\
B~213, B~216 & 8 & 8 & $ -17.9 \pm 0.2 $&$ -18.0 $& 0.5 & $ -13.2 \pm 0.2 $&$ -13.3 $& 0.6 & $ -7.0 \pm 0.2 $&$ -7.1 $& 0.5 & $ 23.4 \pm 0.2 $&$ 23.4 $& 0.4 \\
B~215 & 9 & 2 & $ -16.6 \pm 0.2 $&$ -16.6 $& 0.3 & $ -10.5 \pm 0.3 $&$ -10.5 $& 0.0 & $ -10.3 \pm 0.3 $&$ -10.3 $& 0.0 & $ 22.2 \pm 0.5 $&$ 22.2 $& 0.2 \\
Heiles Cloud 2 & 14 & 3 & $ -15.7 \pm 0.2 $&$ -15.6 $& 0.4 & $ -12.3 \pm 0.2 $&$ -12.4 $& 0.4 & $ -9.1 \pm 0.2 $&$ -9.0 $& 0.3 & $ 21.9 \pm 0.2 $&$ 21.8 $& 0.3 \\
Heiles Cloud 2 & 15 & 5 & $ -15.4 \pm 0.4 $&$ -15.4 $& 0.9 & $ -10.8 \pm 0.2 $&$ -10.5 $& 0.5 & $ -9.2 \pm 0.1 $&$ -9.3 $& 0.3 & $ 20.9 \pm 0.3 $&$ 20.9 $& 0.7 \\
L~1535, L~1529, L~1531, L~1524 & 16 & 10 & $ -16.1 \pm 0.3 $&$ -15.9 $& 1.1 & $ -11.2 \pm 0.2 $&$ -11.1 $& 0.5 & $ -9.5 \pm 0.1 $&$ -9.6 $& 0.2 & $ 21.8 \pm 0.2 $&$ 21.7 $& 0.8 \\
L~1536 & 18 & 14 & $ -16.4 \pm 0.3 $&$ -16.4 $& 1.1 & $ -13.9 \pm 0.3 $&$ -13.7 $& 0.9 & $ -6.8 \pm 0.2 $&$ -6.9 $& 0.8 & $ 22.6 \pm 0.2 $&$ 22.3 $& 0.9 \\
T~Tau cloud  & 19 & 1 & $ -18.0 \pm 0.2 $&$ -18.0 $& \nodata & $ -8.0 \pm 0.6 $&$ -8.0 $& \nodata & $ -8.0 \pm 0.6 $&$ -8.0 $& \nodata & $ 21.3 \pm 0.6 $&$ 21.3 $& \nodata \\
L~1551 & 20 & 12 & $ -15.7 \pm 0.5 $&$ -16.1 $& 1.6 & $ -15.0 \pm 0.3 $&$ -14.9 $& 0.9 & $ -7.3 \pm 0.3 $&$ -7.5 $& 0.9 & $ 22.9 \pm 0.5 $&$ 23.6 $& 1.6 \\
L~1558 & 21 & 0 & \nodata & \nodata & \nodata & \nodata & \nodata & \nodata & \nodata & \nodata & \nodata & \nodata & \nodata & \nodata \\
\hline
Taurus (complex) & all & 92 & $ -16.2 \pm 0.1 $&$ -16.2 $& 1.3 & $ -12.8 \pm 0.2 $&$ -12.7 $& 1.6 & $ -8.9 \pm 0.2 $&$ -9.0 $& 1.7 & $ 22.5 \pm 0.1 $&$ 22.6 $& 1.2 \\
\hline\hline
\end{tabular}
\tablefoot{We provide for each cloud its identifier, corresponding cluster in the HMAC analysis (see Sect.~\ref{section4}), number of stars with measured radial velocity, mean, standard error of the mean (SEM), median, and standard deviation of the Galactic $UVW$ velocity components (not corrected for the solar motion). The standard deviation value given in the table represents the difference between the individual measurements when the molecular cloud has only two representative stars. \vspace{0cm}} 
}
\end{table*}

\subsection{Internal motions, expansion, and rotation} \label{section5.2}

\begin{figure*}[!h]
\begin{center}
\includegraphics[width=0.49\textwidth]{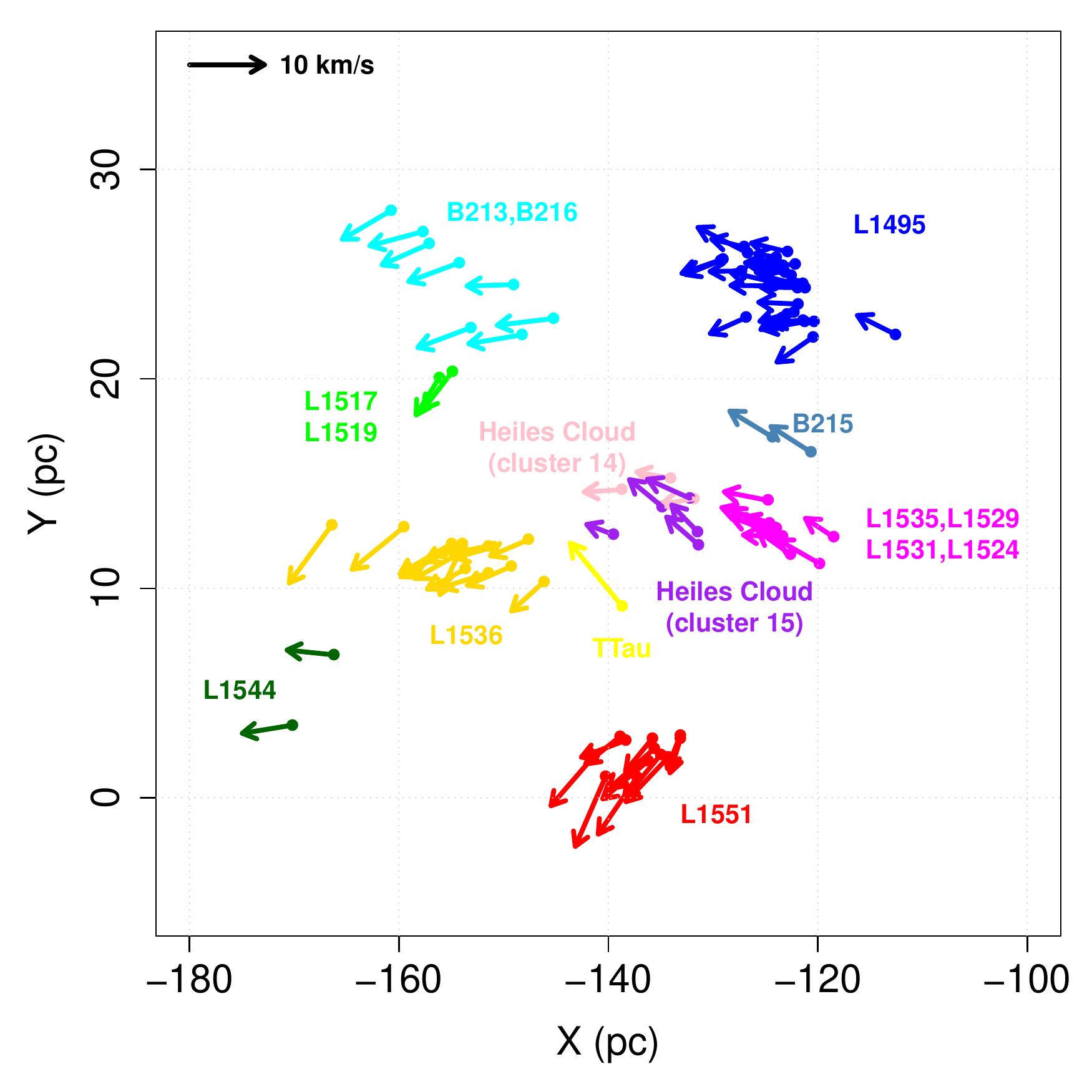}
\includegraphics[width=0.49\textwidth]{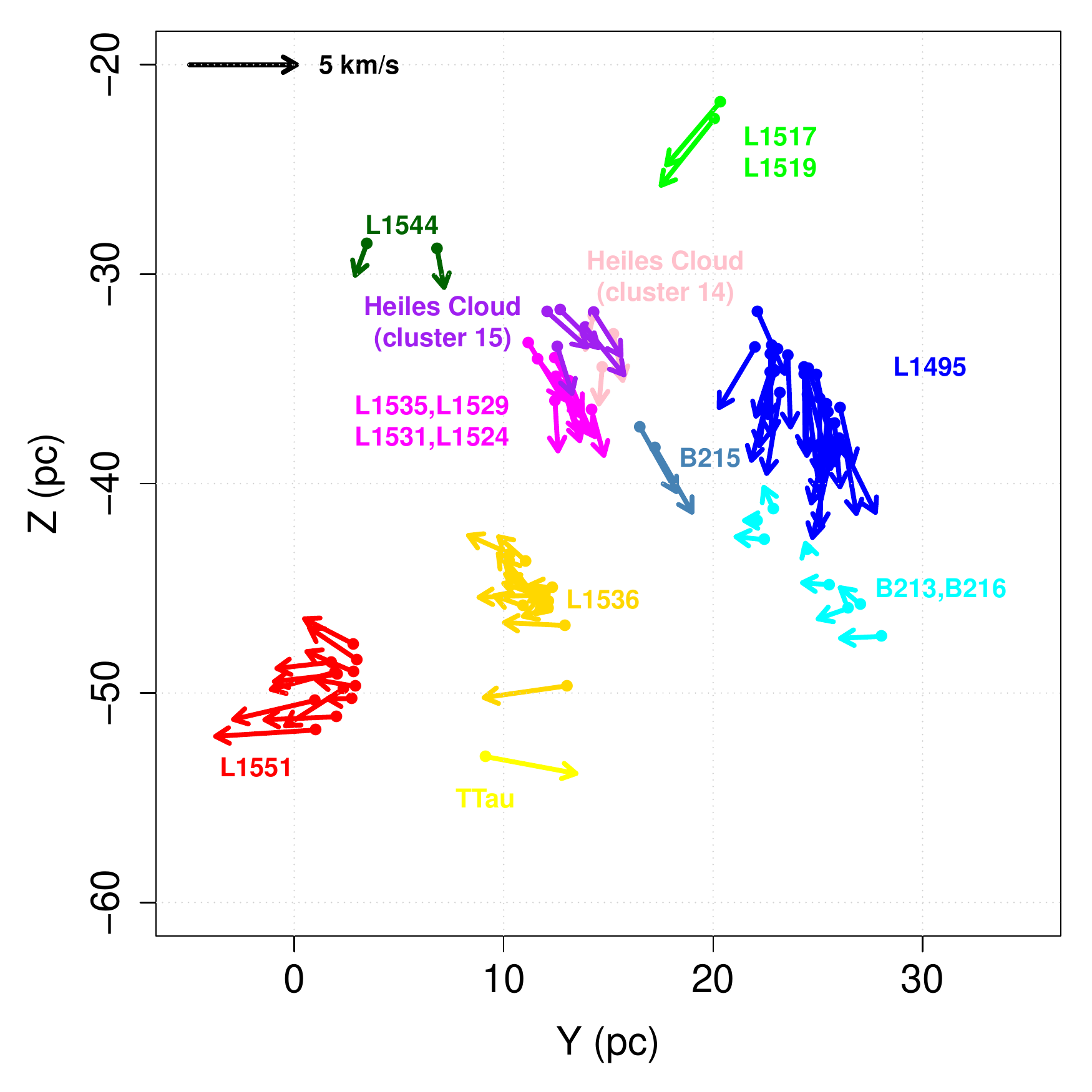}
\includegraphics[width=0.49\textwidth]{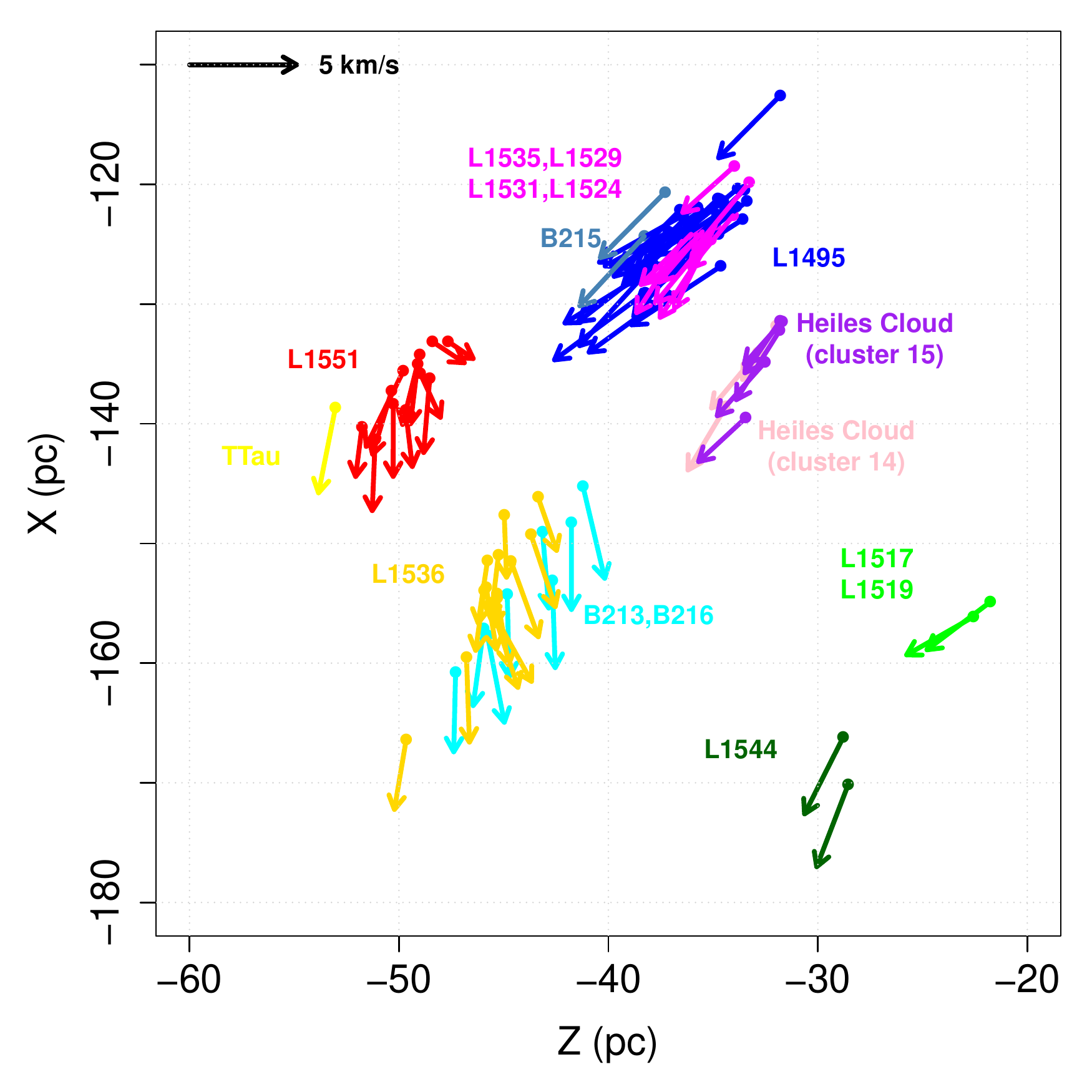}
\caption{Peculiar velocity of the stars  in the various clouds of the Taurus complex projected onto the XY, YZ, and ZX planes.
\label{XYZ_UVW_corr} 
}
\end{center}
\end{figure*}

\begin{figure*}[!h]
\begin{center}
\includegraphics[width=0.49\textwidth]{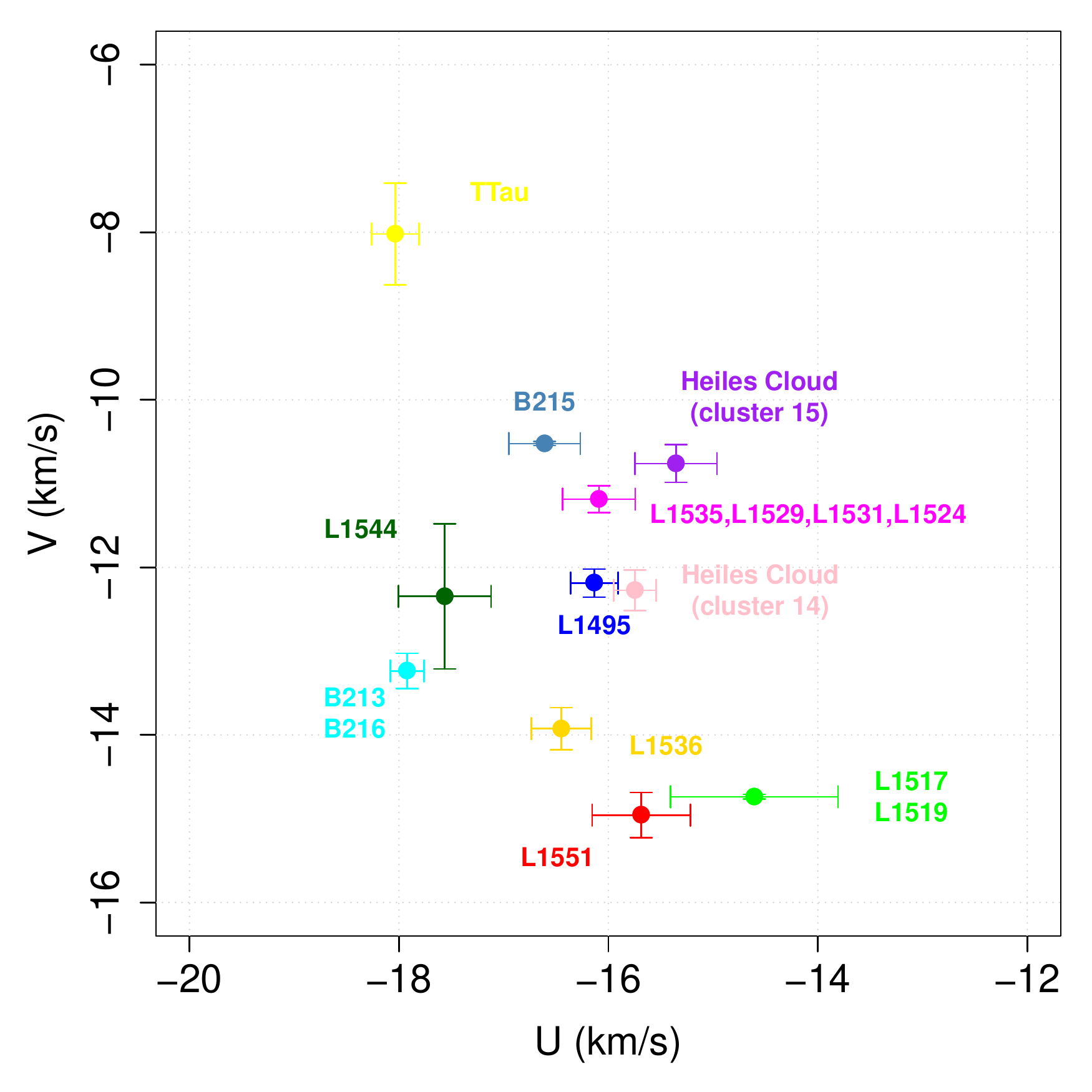}
\includegraphics[width=0.49\textwidth]{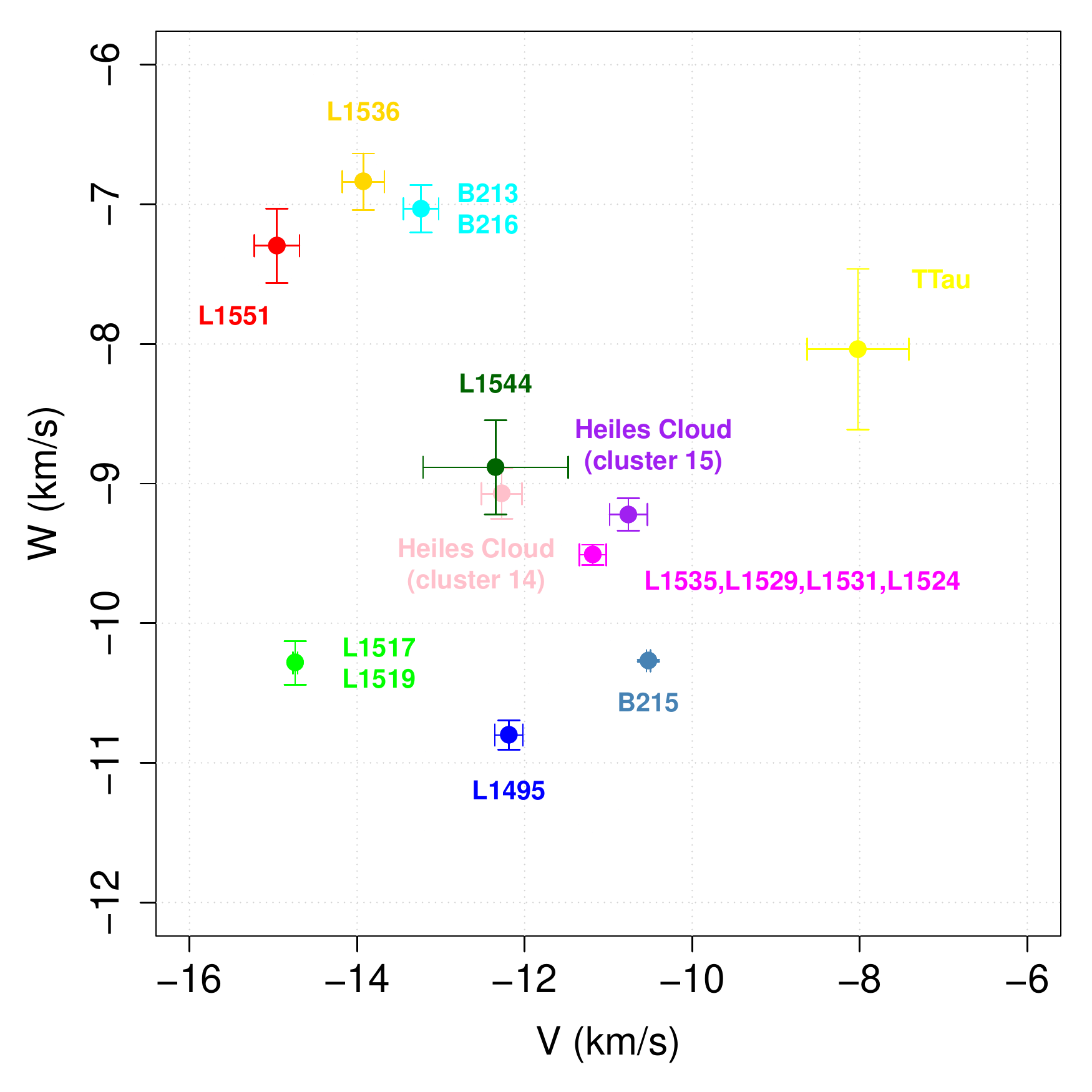}
\includegraphics[width=0.49\textwidth]{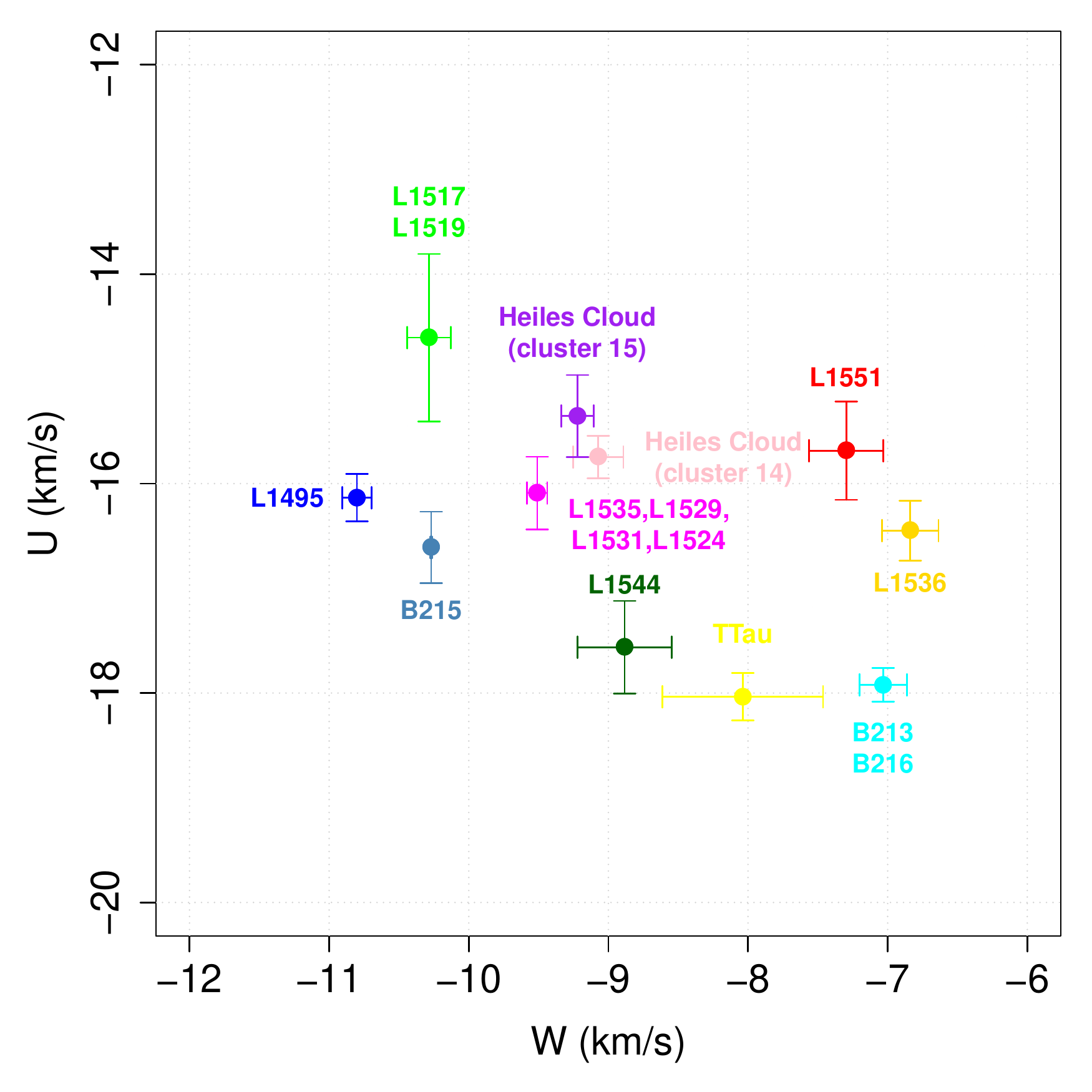}
\caption{Mean spatial velocity of the stars projected towards the various molecular clouds of the Taurus complex.
\label{UVW_scatter} 
}
\end{center}
\end{figure*}

In the following we investigate the internal and relative motions of the stars projected towards the various molecular clouds in the complex. Figure~\ref{XYZ_UVW_corr} shows the spatial velocity of the stars projected onto the XZ, YZ, and ZX planes after correcting for the solar motion. The stellar motions appear less organized when we remove the velocity of the Sun relative to the LSR, but a bulk motion for the various clouds in the complex is still apparent, as illustrated in Figure~\ref{XYZ_UVW_corr}. It is interesting to note that the peculiar velocities of the stars projected onto the XY, YZ, and ZX planes reveal the existence of two groups of molecular clouds with velocity vectors pointing towards different directions. One of these groups is formed by L~1495, L~1535, L~1529, L~1531, L~1524, B~215, and Heiles~Cloud~2 where the velocity vectors point towards the bottom left corner of the ZX plane, for example. Not surprisingly, these clouds (i.e., clusters) are clustered under the same group in the HMAC hierarchical tree at level~12 (see Fig.~\ref{fig_dendrogram}). This suggests a potentially different formation episode for the various clouds in the complex. Interestingly, this effect is also apparent in the three-dimensional space of velocities (see Fig.~\ref{UVW_scatter}). The Taurus subgroups listed above exhibit W velocities that are lower by about 2-3~km/s  compared  to the stars in L~1551, L~1536, B~213, and B~216, among others, whose velocity vectors point towards a different direction.
 
\nopagebreak

We present in Table~\ref{tab_RelMotion} the relative motion of the various clouds in the complex. The T~Tau cloud is not included in this discussion to avoid a biased result based on only one source with measured radial velocity. The relative motion among the various clouds range from about 1 to 5~km/s. The highest value that we observe ($5.4\pm0.5$~km/s) occurs between L~1551 and the B~215 clump. The relative motion between the northernmost cloud (i.e., L~1517 and L~1519) and the southernmost cloud (i.e., L~1551)  is only $3.2\pm0.5$~km/s. We measure a significant relative bulk motion of $4.3\pm0.2$~km/s between the core of L~1495 and its filament (i.e., B~213 and B~216)  confirming that they are indeed independent structures. It is also interesting to note that they exhibit diverging motions in the Z direction (see also Fig.~\ref{XYZ_UVW_corr}). In addition, we also measure a significant non-zero relative motion of $\Delta v=-1.5\pm0.3$~km/s in the $v$ component of the peculiar velocity of the stars in the two subgroups of the Heiles Cloud~2 (i.e., clusters 14 and 15), which justifies our decision to discuss them separately throughout this paper. The high values that we find here for the relative motions between some clouds of the complex \citep[see also][]{Luhman2018} are consistent with the velocity difference among Taurus subgroups reported in the past by \citet{Jones1979} and the velocity dispersion of 6~km/s used by \citet{Bertout2006} in the convergent point search method under the assumption that all stars (independent of the molecular cloud to which they belong) are comoving.      

Let us now assess the quantitative importance of random and organized motions within the complex. We investigate the  potential expansion and rotation effects in the Taurus region following the procedure described by \citet{Rivera2015}. First, we compute the unit position vector $\mathbf{\hat{r}_{*}}=\mathbf{r}_{*}/|\mathbf{r}_{*}|$ for each star that represents the distance of a given star with respect to the center of the corresponding molecular cloud to which it belongs. Second, we compute the velocity $\delta\mathbf{v}_{*}$ of each star relative to its molecular cloud. The dot product between the two quantities ($\mathbf{\hat{r}_{*}} \cdot \delta\mathbf{v}_{*}$) is large and positive (negative) if the group is undergoing expansion (contraction). Analogously, the cross product ($\mathbf{\hat{r}_{*}} \times \delta\mathbf{v}_{*}$) stands as a proxy for the angular momentum of the group and it is large (small) in the case of significant (negligible) rotation effects. We compute the dot and cross product for all stars in our sample with respect to the molecular clouds to which they belong and take the mean of the resulting values as a proxy for the expansion (contraction) and rotation velocities of each group. We run these calculations for all molecular clouds with a minimum of three representative stars with known spatial velocities (i.e., with measured radial velocities). The results of our analysis are presented in Table~\ref{tab_ExpRot}. We note that the resulting quantities are consistent with zero (within $5\sigma$ of the corresponding uncertainties) suggesting that the expansion and rotation effects in the individual molecular clouds are negligible. 

We repeat the same experiment as described above but using the full sample of cluster members with measured radial velocities (92~stars, see Sect.~\ref{section5.1}) to detect large-scale expansion and rotation effects in the Taurus complex. The resulting expansion (contraction) velocity of $0.0\pm0.1$~km/s for the entire complex is consistent with zero. This implies that the internal motions in the radial direction of the complex are dominated by random motions rather than an organized expansion or contraction pattern. On the other hand, the non-zero rotational velocity that we derive here ($|\mathbf{\hat{r}_{*}} \times \delta\mathbf{v}_{*}|=1.5\pm0.1$~km/s, see Table~\ref{tab_ExpRot}) suggests the existence of possible rotation effects in the Taurus complex as a whole. The rotation rate that we derive is nevertheless lower than the result  of $v_{rot}\simeq 2$~km/s obtained previously by \citet{Rivera2015} using a sample of only seven stars with VLBI astrometry. However, it is important to mention that this number is still smaller than the observed three-dimensional velocity dispersion of the stars in our sample ($\sigma=\sqrt{\sigma_{u}^2+\sigma_{v}^2+\sigma_{w}^2}=2.7$~km/s, see Table~\ref{tab_distClusters}). This value suggests that the rotation contributes significantly to the velocity dispersion, but there is also an important contribution from random motions within the complex. 

The relative distances between the Taurus subgroups range from about 4 to 50~pc with a median inter-cloud distance of 25~pc (see Table~\ref{tab_RelMotion}). The crossing time between the various subgroups in this region is on the order of several Myr. For example, if we assume a common origin and birthplace for L~1495 and L~1544, a timescale of about 20~Myr is necessary to explain their present-day positions given their relative distance of $50.9\pm2.1$~pc and bulk motion of $2.3\pm0.4$~km/s. This number greatly exceeds the median age of Taurus stars \citep[$\sim$5~Myr, see, e.g.,][]{Bertout2007}. 

\begin{table*}[!hp]
\renewcommand\thetable{7} 
\centering
\scriptsize{
\caption{Relative space motion among the various clouds of the Taurus complex.   
\label{tab_RelMotion}}
\begin{tabular}{llccccc}
\hline\hline
Molecular Cloud 1&Molecular Cloud 2&$\Delta u$&$\Delta v$&$\Delta w$&$\Delta V_{bulk}$&$\Delta d$\\
&&(km/s)&(km/s)&(km/s)&(km/s)&(pc)\\
\hline\hline
L~1517, L~1519 & L~1544 & $ 3.0 \pm 0.7 $& $ -2.4 \pm 0.6 $& $ -1.4 \pm 0.5 $& $ 4.1 \pm 0.7 $& $ 20.1 \pm 1.7 $\\
L~1517, L~1519 & L~1495  & $ 1.5 \pm 0.6 $& $ -2.6 \pm 0.5 $& $ 0.5 \pm 0.4 $& $ 3.0 \pm 0.5 $& $ 36.5 \pm 1.2 $\\
L~1517, L~1519 & B~213, B~216 & $ 3.3 \pm 0.5 $& $ -1.5 \pm 0.5 $& $ -3.3 \pm 0.4 $& $ 4.9 \pm 0.5 $& $ 22.7 \pm 0.8 $\\
L~1517, L~1519 & B~215 & $ 2.0 \pm 0.8 $& $ -4.2 \pm 0.7 $& $ -0.0 \pm 0.6 $& $ 4.7 \pm 0.7 $& $ 37.6 \pm 2.0 $\\
L~1517, L~1519 & Heiles Cloud 2 (cluster 14) & $ 1.1 \pm 0.6 $& $ -2.5 \pm 0.5 $& $ -1.2 \pm 0.4 $& $ 3.0 \pm 0.5 $& $ 22.7 \pm 1.7 $\\
L~1517, L~1519 & Heiles Cloud 2 (cluster 15) & $ 0.7 \pm 0.6 $& $ -4.0 \pm 0.5 $& $ -1.1 \pm 0.4 $& $ 4.2 \pm 0.5 $& $ 25.5 \pm 1.7 $\\
L~1517, L~1519 & L~1535, L~1529, L~1531, L~1524 & $ 1.5 \pm 0.6 $& $ -3.6 \pm 0.5 $& $ -0.8 \pm 0.4 $& $ 3.9 \pm 0.5 $& $ 36.3 \pm 1.3 $\\
L~1517, L~1519 & L~1536 & $ 1.8 \pm 0.6 $& $ -0.8 \pm 0.5 $& $ -3.4 \pm 0.4 $& $ 4.0 \pm 0.5 $& $ 24.7 \pm 0.6 $\\
L~1517, L~1519 & L~1551 & $ 1.1 \pm 0.7 $& $ 0.2 \pm 0.5 $& $ -3.0 \pm 0.5 $& $ 3.2 \pm 0.5 $& $ 37.9 \pm 0.8 $\\
L~1544 & L~1495  & $ -1.4 \pm 0.6 $& $ -0.2 \pm 0.4 $& $ 1.9 \pm 0.4 $& $ 2.4 \pm 0.4 $& $ 50.9 \pm 2.1 $\\
L~1544 & B~213, B~216 & $ 0.4 \pm 0.5 $& $ 0.9 \pm 0.5 $& $ -1.9 \pm 0.4 $& $ 2.1 \pm 0.4 $& $ 30.2 \pm 1.7 $\\
L~1544 & B~215 & $ -1.0 \pm 0.8 $& $ -1.8 \pm 0.6 $& $ 1.4 \pm 0.5 $& $ 2.5 \pm 0.6 $& $ 49.4 \pm 2.8 $\\
L~1544 & Heiles Cloud 2 (cluster 14) & $ -1.8 \pm 0.6 $& $ -0.1 \pm 0.5 $& $ 0.2 \pm 0.4 $& $ 1.8 \pm 0.5 $& $ 33.7 \pm 2.6 $\\
L~1544 & Heiles Cloud 2 (cluster 15) & $ -2.2 \pm 0.6 $& $ -1.6 \pm 0.5 $& $ 0.3 \pm 0.4 $& $ 2.7 \pm 0.6 $& $ 36.6 \pm 2.6 $\\
L~1544 & L~1535, L~1529, L~1531, L~1524 & $ -1.5 \pm 0.6 $& $ -1.2 \pm 0.4 $& $ 0.6 \pm 0.3 $& $ 2.0 \pm 0.5 $& $ 47.2 \pm 2.3 $\\
L~1544 & L~1536 & $ -1.1 \pm 0.6 $& $ 1.6 \pm 0.5 $& $ -2.0 \pm 0.4 $& $ 2.8 \pm 0.5 $& $ 24.0 \pm 1.8 $\\
L~1544 & L~1551 & $ -1.9 \pm 0.7 $& $ 2.6 \pm 0.5 $& $ -1.6 \pm 0.4 $& $ 3.6 \pm 0.5 $& $ 39.0 \pm 1.9 $\\
L~1495 &B~213, B~216 & $ 1.8 \pm 0.3 $& $ 1.0 \pm 0.3 $& $ -3.8 \pm 0.2 $& $ 4.3 \pm 0.2 $& $ 30.9 \pm 1.7 $\\
L~1495 & B~215 & $ 0.5 \pm 0.6 $& $ -1.7 \pm 0.5 $& $ -0.5 \pm 0.4 $& $ 1.8 \pm 0.5 $& $ 7.9 \pm 0.5 $\\
L~1495 & Heiles Cloud 2 (cluster 14) & $ -0.4 \pm 0.3 $& $ 0.1 \pm 0.3 $& $ -1.7 \pm 0.2 $& $ 1.8 \pm 0.2 $& $ 17.8 \pm 1.4 $\\
L~1495 & Heiles Cloud 2 (cluster 15) & $ -0.8 \pm 0.5 $& $ -1.4 \pm 0.3 $& $ -1.6 \pm 0.2 $& $ 2.3 \pm 0.3 $& $ 16.1 \pm 1.1 $\\
L~1495 & L~1535, L~1529, L~1531, L~1524 & $ -0.0 \pm 0.4 $& $ -1.0 \pm 0.2 $& $ -1.3 \pm 0.1 $& $ 1.6 \pm 0.2 $& $ 11.9 \pm 0.3 $\\
L~1495 & L~1536 & $ 0.3 \pm 0.4 $& $ 1.7 \pm 0.3 $& $ -4.0 \pm 0.2 $& $ 4.3 \pm 0.2 $& $ 35.3 \pm 1.5 $\\
L~1495 & L~1551 & $ -0.4 \pm 0.5 $& $ 2.8 \pm 0.3 $& $ -3.5 \pm 0.3 $& $ 4.5 \pm 0.3 $& $ 30.0 \pm 0.5 $\\
B~213, B~216 & B~215 & $ -1.3 \pm 0.6 $& $ -2.7 \pm 0.5 $& $ 3.2 \pm 0.5 $& $ 4.4 \pm 0.5 $& $ 32.0 \pm 2.4 $\\
B~213, B~216 & Heiles Cloud 2 (cluster 14) & $ -2.2 \pm 0.3 $& $ -1.0 \pm 0.3 $& $ 2.0 \pm 0.2 $& $ 3.1 \pm 0.3 $& $ 21.4 \pm 1.8 $\\
B~213, B~216 & Heiles Cloud 2 (cluster 15) & $ -2.6 \pm 0.4 $& $ -2.5 \pm 0.3 $& $ 2.2 \pm 0.2 $& $ 4.2 \pm 0.3 $& $ 25.1 \pm 1.8 $\\
B~213, B~216 & L~1535, L~1529, L~1531, L~1524 & $ -1.8 \pm 0.4 $& $ -2.1 \pm 0.3 $& $ 2.5 \pm 0.2 $& $ 3.7 \pm 0.3 $& $ 33.2 \pm 1.7 $\\
B~213, B~216 & L~1536 & $ -1.5 \pm 0.3 $& $ 0.7 \pm 0.3 $& $ -0.2 \pm 0.3 $& $ 1.6 \pm 0.3 $& $ 13.1 \pm 0.8 $\\
B~213, B~216 & L~1551 & $ -2.2 \pm 0.5 $& $ 1.7 \pm 0.3 $& $ 0.3 \pm 0.3 $& $ 2.8 \pm 0.4 $& $ 28.7 \pm 1.2 $\\
B~215 & Heiles Cloud 2 (cluster 14) & $ -0.9 \pm 0.6 $& $ 1.8 \pm 0.5 $& $ -1.2 \pm 0.5 $& $ 2.3 \pm 0.5 $& $ 15.9 \pm 2.4 $\\
B~215 & Heiles Cloud 2 (cluster 15) & $ -1.3 \pm 0.7 $& $ 0.2 \pm 0.5 $& $ -1.0 \pm 0.4 $& $ 1.7 \pm 0.6 $& $ 13.3 \pm 2.1 $\\
B~215 & L~1535, L~1529, L~1531, L~1524 & $ -0.5 \pm 0.7 $& $ 0.7 \pm 0.5 $& $ -0.8 \pm 0.4 $& $ 1.1 \pm 0.5 $& $ 5.1 \pm 0.6 $\\
B~215 & L~1536 & $ -0.2 \pm 0.6 $& $ 3.4 \pm 0.5 $& $ -3.4 \pm 0.5 $& $ 4.8 \pm 0.5 $& $ 33.4 \pm 2.3 $\\
B~215 & L~1551 & $ -0.9 \pm 0.7 $& $ 4.4 \pm 0.5 $& $ -3.0 \pm 0.5 $& $ 5.4 \pm 0.5 $& $ 23.9 \pm 1.2 $\\
Heiles Cloud 2 (cluster 14) & Heiles Cloud 2 (cluster 15) & $ -0.4 \pm 0.4 $& $ -1.5 \pm 0.3 $& $ 0.1 \pm 0.2 $& $ 1.6 \pm 0.3 $& $ 3.9 \pm 2.0 $\\
Heiles Cloud 2 (cluster 14) & L~1535, L~1529, L~1531, L~1524 & $ 0.3 \pm 0.4 $& $ -1.1 \pm 0.3 $& $ 0.4 \pm 0.2 $& $ 1.2 \pm 0.3 $& $ 14.4 \pm 1.8 $\\
Heiles Cloud 2 (cluster 14) & L~1536 & $ 0.7 \pm 0.4 $& $ 1.7 \pm 0.3 $& $ -2.2 \pm 0.3 $& $ 2.9 \pm 0.3 $& $ 21.2 \pm 1.8 $\\
Heiles Cloud 2 (cluster 14) & L~1551 & $ -0.1 \pm 0.5 $& $ 2.7 \pm 0.4 $& $ -1.8 \pm 0.3 $& $ 3.2 \pm 0.3 $& $ 20.5 \pm 0.5 $\\
Heiles Cloud 2 (cluster 15) & L~1535, L~1529, L~1531, L~1524 & $ 0.7 \pm 0.5 $& $ 0.4 \pm 0.3 $& $ 0.3 \pm 0.1 $& $ 0.9 \pm 0.5 $& $ 11.0 \pm 1.6 $\\
Heiles Cloud 2 (cluster 15) & L~1536 & $ 1.1 \pm 0.5 $& $ 3.2 \pm 0.3 $& $ -2.4 \pm 0.2 $& $ 4.1 \pm 0.3 $& $ 24.6 \pm 1.8 $\\
Heiles Cloud 2 (cluster 15) & L~1551 & $ 0.3 \pm 0.6 $& $ 4.2 \pm 0.4 $& $ -1.9 \pm 0.3 $& $ 4.6 \pm 0.3 $& $ 20.9 \pm 0.5 $\\
L~1535, L~1529, L~1531, L~1524 & L~1536 & $ 0.4 \pm 0.5 $& $ 2.7 \pm 0.3 $& $ -2.7 \pm 0.2 $& $ 3.8 \pm 0.3 $& $ 33.0 \pm 1.6 $\\
L~1535, L~1529, L~1531, L~1524 & L~1551 & $ -0.4 \pm 0.6 $& $ 3.8 \pm 0.3 $& $ -2.2 \pm 0.3 $& $ 4.4 \pm 0.3 $& $ 22.6 \pm 0.6 $\\
L~1536 & L~1551 & $ -0.8 \pm 0.5 $& $ 1.0 \pm 0.4 $& $ 0.5 \pm 0.3 $& $ 1.4 \pm 0.4 $& $ 20.8 \pm 1.4 $\\
\hline\hline
\end{tabular}
\tablefoot{We provide the relative motion for each component of the spatial velocity  (in the sense molecular cloud~1 minus molecular cloud~2), the resulting bulk motion between the clouds, and their relative distance computed from the $XYZ$ coordinates of the cloud centers (see Table~\ref{tab_distClusters}).}
\vspace{2cm}
}
\end{table*}

\begin{table*}[!hp]
\renewcommand\thetable{8} 
\centering
\scriptsize{
\caption{Results for the expansion and rotation velocity of each molecular cloud and the entire complex.
\label{tab_ExpRot}}
\begin{tabular}{lccc}
\hline
\hline
Molecular Cloud&Cluster&$\mathbf{\hat{r}_{*}} \cdot \delta\mathbf{v}_{*}$&$\mathbf{\hat{r}_{*}} \times \delta\mathbf{v}_{*}$\\
&&(km/s)&(km/s)\\
\hline\hline
L~1495  & 7 & $ -0.1 \pm 0.2 $& $( +0.1 , -0.2 , +0.0 )\pm( 0.1 , 0.1 , 0.1 )$\\
B~213, B~216 & 8 & $ -0.1 \pm 0.1 $& $( -0.1 , -0.2 , +0.3 )\pm( 0.1 , 0.2 , 0.1 )$\\
Heiles Cloud 2 (cluster 14) & 14 & $ +0.2 \pm 0.1 $& $( -0.0 , +0.2 , +0.1 )\pm( 0.1 , 0.2 , 0.2 )$\\
Heiles Cloud 2 (cluster 15) & 15 & $ -0.2 \pm 0.4 $& $( -0.1 , -0.2 , +0.4 )\pm( 0.1 , 0.1 , 0.2 )$\\
L~1535, L~1529, L~1531, L~1524 & 16 & $ +0.0 \pm 0.2 $& $( -0.2 , -0.2 , +0.1 )\pm( 0.1 , 0.2 , 0.2 )$\\
L~1536 & 18 & $ +0.4 \pm 0.2 $& $( -0.1 , +0.3 , +0.6 )\pm( 0.1 , 0.3 , 0.2 )$\\
L~1551 & 20 & $ +0.8 \pm 0.5 $& $( -0.1 , +0.1 , +0.1 )\pm( 0.1 , 0.2 , 0.2 )$\\
\hline
Taurus Complex & all & $ -0.0 \pm 0.1 $& $( -0.8 , +1.0 , +0.8 )\pm( 0.1 , 0.1 , 0.1 )$\\
\hline\hline
\end{tabular}
}
\end{table*}


\clearpage
\subsection{Stellar and molecular gas kinematics}\label{section5.3}

In this section we compare the radial velocities of the stars in our sample with the kinematics of the underlying gaseous clouds. We used the large-scale survey of the Taurus molecular clouds in $^{12}$CO and $^{13}$CO performed by \citet{Goldsmith2008} using the Five College Radio Astronomy Observatory (FCRAO) telescope. The northernmost and southernmost clouds are not included in the surveyed region, so the  analysis is  restricted to the molecular clouds in the central region of the Taurus complex that fall into the FCRAO maps. 

We proceeded as follows to compare the stellar velocities with the kinematics of the molecular gas. First, we convert the (heliocentric) radial velocities of the stars collected from the literature to the LSR. For consistency with our FCRAO data, we deduce the velocity of the Sun with respect to the LSR computed from the solar apex of $(\alpha_{\odot},\delta_{\odot})=(271^{\circ},30^{\circ})$ and $V_{\odot}=20$~km/s \citep[see][]{Jackson2006} rather than the solar motion used in Sect.~\ref{section5.1}. The corrected radial velocities of individual stars are listed in Table~\ref{tab_distStars}. Second, we extract the $^{12}$CO and $^{13}$CO spectra from the FCRAO maps at the position of each star in our sample in a velocity interval from -2 to 14~km/s which conservatively includes the range of observed velocities (with respect to the LSR) in Taurus \citep[see, e.g., Fig.~12 of][]{Goldsmith2008}. Then, we compute the centroid velocity and estimate its uncertainty from the r.m.s. of the spectrum as described by \citet{Dickman1985}.

Two points are worth mentioning here before comparing the stellar radial velocities with the CO velocity. First, the fraction of binaries and multiple systems in Taurus is high \citep[see, e.g.,][]{Leinert1993,Duchene1999} and a complete census of these systems with their  properties (e.g., orbital period, angular separation, and mass ratio) is still lacking in the literature. We reject all known binaries and multiple systems for the current analysis \citep[many of them have been flagged by][]{Joncour2017} to avoid comparing the velocity of the gas with a radial velocity measurement that is variable in time.  Second, a visual inspection of the extracted spectra for the $^{12}$CO molecule reveals that the emission often exhibits complex structures and self-absorbed spectral profiles \citep[see also ][]{Urquhart2007} making it difficult to compute a velocity centroid in such cases. Although the $^{12}$CO molecule is more abundant than its isotopolog $^{13}$CO, the latter is more optically  thin giving access to the full column density that produces the emission \citep[see, e.g.,][]{Cormier2018} and these absorption features are less common in our spectra. We  therefore decided to work with the $^{13}$CO emission to determine the velocity of the molecular gas along the line of sight. Another interesting feature that we observe in some of our spectra is the existence of multiple (overlapping) components for the velocity of the gas as reported previously by \citet{Hacar2013}. It is possible  in principle to compare the stellar radial velocities with the closest component of the gas velocity, but we  decided to discard these spectra from our analysis to avoid a biased correlation between the two velocities.

Table~\ref{tab_StarsGas} lists the individual measurements for the velocity of the stars and the $^{13}$CO molecular gas used in our comparison. This analysis is restricted to 28 stars in our sample that satisfy the conditions described above. We note that three stars (CW~Tau, 2MASS~J04213459+2701388, and 2MASS~J04414825+2534304) have radial velocities that differ by more than 1~km/s with respect to the $^{13}$CO molecular gas velocity. One possibility to explain the different velocities for these sources is the existence of undetected binaries because their proper motions and parallaxes are consistent with membership in the corresponding clouds (as discussed in Sect.~\ref{section4}). In particular, we found three heliocentric radial velocity measurements in the literature for CW~Tau:  $14.5\pm2.0$~km/s \citep{Hartmann1986,HBC}, $13.60\pm0.10$~km/s \citep{Nguyen2012}, and $16.39\pm0.42$~km/s \citep{Kounkel2019}. As explained in Sect.~\ref{section2}, we used the most precise measurement throughout our analysis. The difference between the radial velocity of CW~Tau and the $^{13}$CO molecular gas would still be at the 1~km/s level if, for example,  we used  the most recent measurement of $V_{r}=16.39\pm0.42$~km/s in our comparison. In the case of 2MASS~J04213459+2701388 and 2MASS~J04414825+2534304 we found only one radial velocity measurement in the literature. 

\begin{figure}[!h]
\begin{center}
\includegraphics[width=0.45\textwidth]{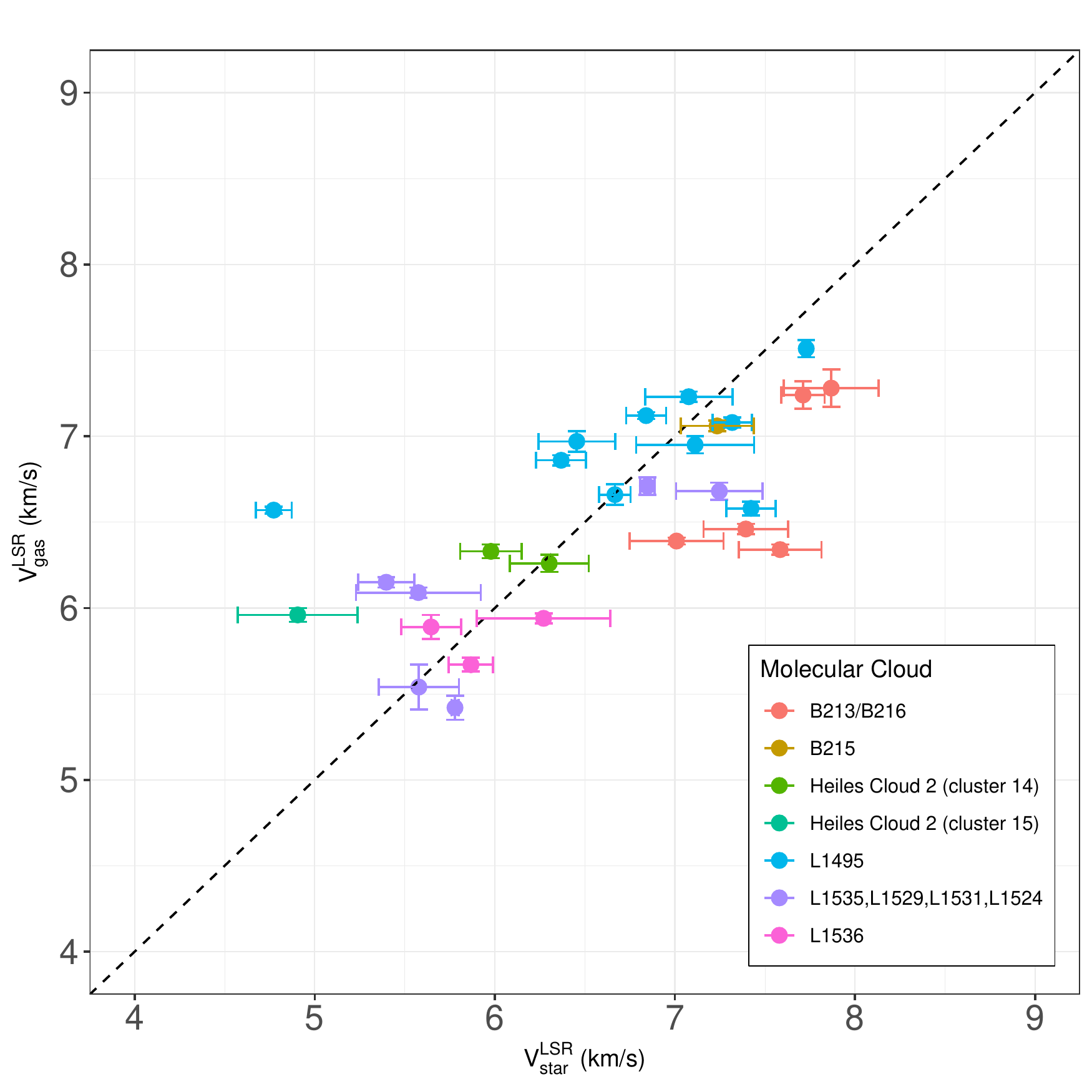}
\caption{Comparison of the radial velocity of the stars (with respect to the LSR) with the centroid velocity of the $^{13}$CO emission extracted from the FCRAO maps at the position of each star. The black dashed line indicates a perfect correlation between the two measurements, and the colors respresent the various molecular clouds to which the stars belong.  
\label{comp_gas_stars} 
}
\end{center}
\end{figure}

As illustrated in Figure~\ref{comp_gas_stars} the correlation between the radial velocity of the stars and the velocity of the $^{13}$CO molecular gas along the line of sight is clearly evident. Here, we report a mean difference between the two velocities of  $0.04\pm0.12$~km/s (in the sense stars minus gas) and r.m.s. of 0.63~km/s. Previous studies in this region performed by \citet{Herbig1977} and \citet{Hartmann1986} reported a mean difference of $0.4\pm 0.5$~km/s (with r.m.s. of 3.9~km/s) and $0.2\pm0.4$~km/s (with r.m.s. of 1.7~km/s), respectively. Our results obtained in this paper reveal that the stars and the gas are even more tightly coupled than previously thought. One reason to explain this result comes from the more precise and accurate radial velocity measurements available to date that have been incorporated in our analysis. In addition, it should also be noted that the sample of Taurus stars used in each study is not the same. 

Our results in this section are consistent with the stars being at the same velocity of the neighboring molecular gas. This finding confirms that the stars in our sample are indeed associated with the various substructures of the complex, and supports our results for the existence of multiple populations, significant depth effects, and internal motions in the Taurus region.  

\begin{table*}
\renewcommand\thetable{9} 
\centering
\caption{Comparison of the velocity of the stars and the $^{13}$CO molecular gas. 
\label{tab_StarsGas}}
\begin{tabular}{llccl}
\hline\hline
2MASS Identifier&Other Identifier&$V_{gas}^{LSR}$&$V_{star}^{LSR}$&Molecular Cloud\\
&&(km/s)&(km/s)&\\
\hline\hline
2MASS~J04141458+2827580 & FN~Tau & $ 6.66 \pm 0.06 $& $ 6.67 \pm 0.09 $& L~1495 \\
2MASS~J04141700+2810578 & CW~Tau & $ 6.57 \pm 0.02 $& $ 4.77 \pm 0.10 $& L~1495  \\
2MASS~J04141760+2806096 & CIDA~1 & $ 6.86 \pm 0.03 $& $ 6.37 \pm 0.14 $& L~1495  \\
2MASS~J04153916+2818586 &  & $ 7.12 \pm 0.02 $& $ 6.84 \pm 0.11 $& L~1495  \\
2MASS~J04161210+2756385 &  & $ 6.58 \pm 0.04 $& $ 7.42 \pm 0.14 $& L~1495  \\
2MASS~J04190110+2819420 & V410~X-ray~6 & $ 7.23 \pm 0.03 $& $ 7.08 \pm 0.24 $& L~1495  \\
2MASS~J04192625+2826142 & V819~Tau & $ 7.51 \pm 0.05 $& $ 7.73 \pm 0.02 $& L~1495  \\
2MASS~J04194819+2750007 &  & $ 7.08 \pm 0.03 $& $ 7.32 \pm 0.11 $& L~1495  \\
2MASS~J04201611+2821325 &  & $ 6.95 \pm 0.05 $& $ 7.11 \pm 0.33 $& L~1495  \\
2MASS~J04213459+2701388 &  & $ 6.34 \pm 0.03 $& $ 7.58 \pm 0.23 $& B~213, B~216 \\
2MASS~J04214013+2814224 & XEST~21-026 & $ 6.97 \pm 0.06 $& $ 6.45 \pm 0.21 $& L~1495 \\
2MASS~J04222404+2646258 & XEST~11-087 & $ 6.46 \pm 0.03 $& $ 7.39 \pm 0.23 $& B~213, B~216 \\
2MASS~J04224786+2645530 & IRAS~04196+2638 & $ 6.39 \pm 0.02 $& $ 7.01 \pm 0.26 $& B~213, B~216 \\
2MASS~J04233919+2456141 & FT~Tau & $ 7.06 \pm 0.03 $& $ 7.23 \pm 0.20 $& B~215 \\
2MASS~J04262939+2624137 & KPNO~3 & $ 7.28 \pm 0.11 $& $ 7.87 \pm 0.26 $& B~213, B~216 \\
2MASS~J04272467+2624199 &  & $ 7.24 \pm 0.08 $& $ 7.71 \pm 0.12 $& B~213, B~216 \\
2MASS~J04295950+2433078 &  & $ 6.68 \pm 0.05 $& $ 7.25 \pm 0.24 $& L~1535, L~1529, L~1531, L~1524 \\
2MASS~J04322329+2403013 &  & $ 5.54 \pm 0.13 $& $ 5.58 \pm 0.22 $& L~1535, L~1529, L~1531, L~1524 \\
2MASS~J04323058+2419572 & FY~Tau & $ 6.15 \pm 0.03 $& $ 5.40 \pm 0.16 $& L~1535, L~1529, L~1531, L~1524 \\
2MASS~J04323176+2420029 & FZ~Tau & $ 6.09 \pm 0.03 $& $ 5.58 \pm 0.35 $& L~1535, L~1529, L~1531, L~1524 \\
2MASS~J04332621+2245293 & XEST~17-036 & $ 5.67 \pm 0.04 $& $ 5.87 \pm 0.12 $& L~1536 \\
2MASS~J04333405+2421170 & GI~Tau & $ 6.71 \pm 0.05 $& $ 6.85 \pm 0.04 $& L~1535, L~1529, L~1531, L~1524 \\
2MASS~J04341099+2251445 & JH~108 & $ 5.89 \pm 0.07 $& $ 5.65 \pm 0.17 $& L~1536 \\
2MASS~J04341527+2250309 & CFHT~1 & $ 5.94 \pm 0.03 $& $ 6.27 \pm 0.37 $& L~1536 \\
2MASS~J04352737+2414589 & DN~Tau & $ 5.42 \pm 0.07 $& $ 5.78 \pm 0.02 $& L~1535, L~1529, L~1531, L~1524 \\
2MASS~J04382858+2610494 & DO~Tau & $ 6.33 \pm 0.04 $& $ 5.98 \pm 0.17 $& Heiles Cloud 2 (cluster 14) \\
2MASS~J04390396+2544264 &  & $ 6.26 \pm 0.05 $& $ 6.30 \pm 0.22 $& Heiles Cloud 2 (cluster 14) \\
2MASS~J04414825+2534304 &  & $ 5.96 \pm 0.04 $& $ 4.91 \pm 0.33 $& Heiles Cloud 2 (cluster 15) \\
\hline\hline

\end{tabular}
\tablefoot{We provide for each star its identifier, velocity of the $^{13}$CO emission at the position of the star, radial velocity of the star converted to the LSR, and the molecular cloud to which it belongs. }
\end{table*}


\section{Conclusions}\label{section6}

We used in this study the best astrometry available to date by combining Gaia-DR2 data with the VLBI results delivered by the GOBELINS project to investigate the three-dimensional structure and kinematics of the Taurus star-forming region. Both projects return consistent results for the targets in common and complement each other in this region of the sky.  

We applied a hierarchical clustering algorithm for partitioning the stars in our sample into groups with similar properties based on the stellar positions, proper motions, and parallaxes. Our methodology allowed us to identify the various substructures of the Taurus region and discuss their relationship (i.e., hierarchy). We found 21 clusters in our sample at the lowest level of the hierarchical tree and a number of outliers that exhibit discrepant properties. Thirteen of these clusters are associated with one molecular cloud of the Taurus complex and have been used to derive the distance and spatial velocity of the corresponding clouds providing the most complete and precise scenario of the six-dimensional structure of this region. 

We confirmed the existence of significant depth effects along the line of sight. The median inter-cloud distance among the various subgroups of the Taurus region is about 25~pc. We report B~215 and L~1558 as the closest ($d=128.5^{+1.6}_{-1.6}$~pc) and most remote ($d=198.1^{+2.5}_{-2.5}$~pc) substructures of the complex, respectively. In addition, we show that the core of the most prominent molecular cloud of the complex L~1495 and the filament connected to it in the plane of the sky are located at significantly different distances ($d=129.9^{+0.4}_{-0.3}$~pc and $d=160.0^{+1.2}_{-1.2}$~pc, respectively) and diverge from each other in the velocity space. 

In a subsequent analysis, we computed the spatial velocities of the stars and the relative bulk motion among the various clouds. The highest values that we derive for the relative motion among the various substructures occur between the B~215 clump in the central region of the complex with the northernmost and southernmost clouds (L~1517, L~1519 and L~1551, respectively) and they reach  about 5~km/s. The one-dimensional velocity dispersion that we obtain from the full sample of Taurus stars with known spatial velocities is on the order of 2~km/s. In addition, we have also investigated the existence of expansion, contraction, and rotation effects. We concluded that these effects are too small (if present at all) in the individual molecular clouds represented in our sample of stars. We do not detect any significant expansion pattern for the Taurus complex as a whole, but we find evidence of potential rotation effects that will require further investigation with different methodologies. 

Finally, we compared the radial velocity of the stars in our sample with the velocity of the underlying gaseous clouds derived from the emission of the $^{13}$CO molecular gas, and showed that they are consistent among themselves. We find a mean difference of $0.04\pm0.12$~km/s (with r.m.s. of 0.63~km/s), which suggests that the stars are indeed following the velocity pattern of the gas in this region.



\begin{acknowledgements}
We thank Alvaro Hacar, Estelle Moraux, and Isabelle Joncour for helpful discussions that improved the manuscript. 
This research has received funding from the European Research Council (ERC) under the European Union’s Horizon 2020 research 
and innovation program (grant agreement No 682903, P.I. H. Bouy), and from the French State in the framework of the 
``Investments for the future” Program, IdEx Bordeaux, reference ANR-10-IDEX-03-02. L.L. acknowledges the financial support from  PAPIIT-UNAM project IN112417, and CONACyT. G.N.O.-L. acknowledges support from the von Humboldt Stiftung. This research has made use of the SIMBAD database, operated at CDS, Strasbourg, France. This work has made use of data from the European Space Agency (ESA) mission {\it Gaia} (\url{https://www.cosmos.esa.int/gaia}), processed by the {\it Gaia} Data Processing and Analysis Consortium (DPAC, \url{https://www.cosmos.esa.int/web/gaia/dpac/consortium}). Funding for the DPAC has been provided by national institutions, in particular the institutions participating in the {\it Gaia} Multilateral Agreement.
\end{acknowledgements}

\bibliographystyle{aa} 
\bibliography{references} 

\begin{appendix}
\section{Performance assessment of the clustering analysis with simulations}\label{appendixA}

Our clustering analysis with HMAC has identified 21 clusters which group 236 stars in our sample, and another 48 outliers with unique properties (see Sect.~\ref{section4.2}). In this section we analyze the robustness and dependence of our previous results on the uncertainties of the astrometric parameters used in the clustering analysis. We investigate the capability of the HMAC algorithm to distinguish between cluster members and outliers in our sample of stars, and we evaluate the clustering of our sample into the 21 clusters derived in our analysis presented in Sect.~\ref{section4.2} (hereafter the true run). The analysis discussed throughout this section refers to the clustering results obtained at the lowest level of the HMAC hierarchical trees that we derived from our simulations, as explained below. 

First, we constructed 1000 synthetic samples of the Taurus association by resampling the five astrometric parameters $(\alpha,\delta,\mu_{\alpha}\cos\delta,\mu_{\delta},\varpi)$ of each star in the true run from a multivariate normal distribution, where mean and standard deviation correspond to the individual measurements and their uncertainties. We used the full $5\times5$ covariance matrix for Gaia-DR2 sources to generate the synthetic data. Then, we ran HMAC on each synthetic sample of the Taurus association using the same sequence of bandwidths as used in the true run (see also Table~\ref{tab_HMAC}), and obtained the cluster membership for the synthetic stars. It is important to mention that the clusters obtained in each realization of this process do not perfectly match the clusters from the true run in regard to the number of stars and their somewhat different location in the five-dimensional parameter space. Thus, to identify the various clusters from the true run in our simulations we first computed their distances to the simulated counterparts, and then assigned the closest cluster in our simulations to each cluster in the true run using Euclidean distances in the five-dimensional space defined by the observables. 

Second, we evaluated the robustness of the clustering analysis in the true run in terms of the reproducibility of these results in our simulations using synthetic data. In each run of our simulations we tracked  the membership status (member vs. outlier) and the cluster membership of the synthetic stars produced in our simulations to compare it with the result given in the true run for each star. In this context, we assigned the classes ``member'' (positive) and ``outlier'' (negative) to describe the membership status of the stars in the true run and in our simulations. It should be noted that the terminology ``outlier'' used throughout the paper refers to the sources that do not belong to any cluster of members with similar properties identified in this study even though they have been identified as YSOs in previous studies and are likely to be associated with the Taurus region. We computed the number of true positives (TP), false positives (FP), false negatives (FN), and true negatives (TN) to quantitatively address the comparison between the actual and predicted classes, which refer to the true run and our simulations, respectively. 

Our simulations allow us to investigate two important points regarding the clustering analysis with HMAC: (i) the dichotomy between cluster members and outliers, and (ii) the possibility of stars being assigned to different clusters in our simulations. In the first case, we do not distinguish between the members that have been assigned to different clusters in our simulations and in the true run, but we investigate the capability of the HMAC algorithm to distinguish between the two classes. In this context, we define the true positive rate (TPR) and the true negative rate (TNR) as follows:
\begin{equation}
TPR=\frac{TP}{TP+FN}\, ,
\end{equation}
\begin{equation}
TNR=\frac{TN}{TN+FP}\,.
\end{equation}
The mean values of TPR and TNR that we obtain after running HMAC for the 1000 synthetic samples as described above are $0.889\pm0.054$ and $0.903\pm0.042$, respectively. The former confirms that a high fraction of the cluster members in the true run are identified as cluster members in our simulations, and the latter shows that  a high fraction of the outliers in the true run are also  identified as outliers in our simulations. The relatively high values that we obtain for both TPR and TNR show that the contamination rate is low, and we therefore confirm the membership status of the sources in the true run.  

We repeated the procedure described above with the individual clusters identified in the true run to investigate the second point of our performance assessment. In this case, we defined the classes ``member'' (positive) and ``non-member'' (negative), which refer to the specific cluster under analysis. In addition to the values of TPR and TNR, we also derived the positive predictive value (PPV) and the negative predictive value (NPV) as follows:
\begin{equation}
PPV=\frac{TP}{TP+FP}\, , 
\end{equation}
\begin{equation}
NPV=\frac{TN}{TN+FN}\, . 
\end{equation}
The PPV shows whether the sample of members in one cluster obtained from our simulations is contaminated by sources identified as non-members in the true run. Analogously, the NPV measures whether our list of non-members (with respect to a given cluster) obtained in the simulations is polluted by sources identified as cluster members in the true run. In addition, we also computed the $F_{1}$ score for the clustering performance within individual clusters which returns the harmonic mean between TPR and PPV. It is given by
\begin{equation}
F_{1}=\frac{2\cdot TPR\cdot PPV}{TPR+PPV}\, . 
\end{equation}
The results of this analysis are shown in Table~\ref{tab_appendix}. We note in particular that clusters 10, 11, and 12 exhibit the lowest performance of  all the clusters (see, e.g., the results for the $F_{1}$ score). However, these numbers are affected by small number statistics (i.e., only two stars in each cluster). The early merging of clusters 11 and 12 with cluster 7 (see Fig.~\ref{fig_dendrogram}) also explains that these stars are often associated with different clusters in our simulations. In the specific case of cluster 10 we note that  one of its members, namely 2MASS~J04312669+2703188, is often classified as an outlier in our simulations due to the large parallax uncertainty ($\varpi=7.019\pm0.893$~mas, see Table~\ref{tab1}) that is used in the resampling procedure described above to generate synthetic stars. Altogether, the results that we obtain in our simulations for the TPR, TNR, PPV, and NPV support the stability and robustness of the clustering results presented in Sect.\ref{section4.2} for the true run with respect to the measurement uncertainties.  

\begin{table*}[!h]
\renewcommand\thetable{A1} 
\centering
\caption{Mean values for the TPR, TNR, PPV, NPV, and $F_{1}$ score obtained for each cluster from our simulations (after 1000 realizations). 
\label{tab_appendix}}
\begin{tabular}{cccccc}
\hline\hline
Cluster &$TPR$&$TNR$&$PPV$&$NPV$&$F_{1}$-score\\
\hline
1 & $ 0.993 \pm 0.019 $& $ 1.000 \pm 0.001 $& $ 0.996 \pm 0.014 $& $ 0.999 \pm 0.001 $& $ 0.995 \pm 0.012 $\\
2 & $ 1.000 \pm 0.011 $& $ 1.000 \pm 0.001 $& $ 0.974 \pm 0.080 $& $ 1.000 \pm 0.000 $& $ 0.987 \pm 0.042 $\\
3 & $ 0.920 \pm 0.098 $& $ 1.000 \pm 0.001 $& $ 0.997 \pm 0.025 $& $ 0.997 \pm 0.003 $& $ 0.957 \pm 0.054 $\\
4 & $ 0.934 \pm 0.136 $& $ 1.000 \pm 0.000 $& $ 1.000 \pm 0.000 $& $ 0.999 \pm 0.003 $& $ 0.966 \pm 0.073 $\\
5 & $ 0.998 \pm 0.035 $& $ 1.000 \pm 0.000 $& $ 1.000 \pm 0.000 $& $ 1.000 \pm 0.000 $& $ 0.999 \pm 0.018 $\\
6 & $ 0.983 \pm 0.081 $& $ 0.999 \pm 0.002 $& $ 0.931 \pm 0.115 $& $ 1.000 \pm 0.001 $& $ 0.956 \pm 0.072 $\\
7 & $ 0.873 \pm 0.189 $& $ 0.992 \pm 0.008 $& $ 0.963 \pm 0.036 $& $ 0.972 \pm 0.038 $& $ 0.915 \pm 0.106 $\\
8 & $ 0.981 \pm 0.075 $& $ 1.000 \pm 0.001 $& $ 0.998 \pm 0.012 $& $ 0.999 \pm 0.004 $& $ 0.989 \pm 0.039 $\\
9 & $ 0.837 \pm 0.326 $& $ 0.998 \pm 0.004 $& $ 0.805 \pm 0.333 $& $ 0.999 \pm 0.002 $& $ 0.821 \pm 0.234 $\\
10 & $ 0.560 \pm 0.163 $& $ 1.000 \pm 0.000 $& $ 1.000 \pm 0.000 $& $ 0.997 \pm 0.001 $& $ 0.718 \pm 0.134 $\\
11 & $ 0.527 \pm 0.479 $& $ 0.997 \pm 0.004 $& $ 0.477 \pm 0.453 $& $ 0.997 \pm 0.003 $& $ 0.501 \pm 0.330 $\\
12 & $ 0.650 \pm 0.477 $& $ 0.997 \pm 0.002 $& $ 0.542 \pm 0.434 $& $ 0.998 \pm 0.003 $& $ 0.591 \pm 0.325 $\\
13 & $ 0.976 \pm 0.106 $& $ 1.000 \pm 0.000 $& $ 0.997 \pm 0.033 $& $ 1.000 \pm 0.001 $& $ 0.986 \pm 0.056 $\\
14 & $ 0.728 \pm 0.227 $& $ 0.998 \pm 0.004 $& $ 0.928 \pm 0.096 $& $ 0.991 \pm 0.007 $& $ 0.816 \pm 0.147 $\\
15 & $ 0.837 \pm 0.180 $& $ 0.996 \pm 0.004 $& $ 0.814 \pm 0.185 $& $ 0.997 \pm 0.003 $& $ 0.825 \pm 0.129 $\\
16 & $ 0.753 \pm 0.134 $& $ 0.996 \pm 0.008 $& $ 0.949 \pm 0.120 $& $ 0.979 \pm 0.011 $& $ 0.840 \pm 0.096 $\\
17 & $ 0.913 \pm 0.188 $& $ 0.997 \pm 0.007 $& $ 0.931 \pm 0.156 $& $ 0.997 \pm 0.007 $& $ 0.922 \pm 0.122 $\\
18 & $ 0.712 \pm 0.199 $& $ 0.997 \pm 0.003 $& $ 0.955 \pm 0.046 $& $ 0.974 \pm 0.017 $& $ 0.816 \pm 0.132 $\\
19 & $ 0.794 \pm 0.226 $& $ 0.999 \pm 0.001 $& $ 0.960 \pm 0.088 $& $ 0.997 \pm 0.003 $& $ 0.869 \pm 0.140 $\\
20 & $ 0.951 \pm 0.037 $& $ 0.999 \pm 0.002 $& $ 0.993 \pm 0.016 $& $ 0.993 \pm 0.005 $& $ 0.972 \pm 0.021 $\\
21 & $ 0.971 \pm 0.072 $& $ 1.000 \pm 0.001 $& $ 0.985 \pm 0.048 $& $ 0.999 \pm 0.001 $& $ 0.978 \pm 0.043 $\\

\hline\hline
\end{tabular}
\tablefoot{The results listed in the table with zero uncertainty arise from the fact that either FP or FN is zero in all realizations of our simulations.}
\end{table*}
\end{appendix}


\end{document}